\documentclass[12pt]{article}
\pdfoutput=1
\usepackage{amsmath,amssymb,amsthm,bm,graphicx,float,array,multirow,multicol,rotfloat,caption,subcaption,hyperref,cleveref,enumerate,geometry,mathdots,adjustbox,booktabs,parskip,mathtools,tikz,tikz-cd,pdflscape,csquotes,lscape,rotating,empheq,mathrsfs,longtable,stackengine,afterpage}
\usepackage[para]{threeparttable}
\usepackage[all]{xy}
\usepackage[normalem]{ulem}
\usepackage[aligntableaux=center]{ytableau}
\usepackage[numbers,sort&compress]{natbib}
\numberwithin{equation}{section}
\numberwithin{figure}{section}
\allowdisplaybreaks
\makeatletter

\usepackage{paralist}

\usetikzlibrary{arrows,shapes.misc,positioning,decorations.pathmorphing,decorations.markings,decorations.pathreplacing,matrix,patterns,backgrounds}
\tikzset{
node/.style={circle, thick, draw=black!100,fill=white!100,  minimum size=3mm, inner sep=0pt},
sqnode/.style={rectangle
, thick, draw=black!100,fill=white!100,  minimum size=2mm, inner sep=0pt
},
sonode/.style={circle, thick, draw=black!100,fill=red!100,  minimum size=3mm, inner sep=0pt},
spnode/.style={circle, thick, draw=black!100,fill=blue!100,  minimum size=3mm, inner sep=0pt},
spnode/.style={circle, thick, draw=black!100,fill=blue!100,  minimum size=3mm, inner sep=0pt},
fnode/.style={rectangle, thick, draw=black!100,fill=white!100,  minimum size=3mm, inner sep=0pt},
tnode/.style={rounded rectangle, outer sep=0pt, thick, minimum size=5mm}
}

\definecolor{myGreen}{RGB}{54, 150, 45}
\newcommand{\Lpagenumber}{\ifdim\textwidth=\linewidth\else\bgroup
	\dimendef\margin=0 %use \margin instead of \dimen0
	\ifodd\value{page}\margin=\oddsidemargin
	\else\margin=\evensidemargin
	\fi
	\raisebox{\dimexpr +2 \topmargin-\headheight-\headsep-0.5\linewidth}[0pt][0pt]{%
		\rlap{\hspace{\dimexpr \margin+\textheight+2\footskip}%
			\llap{\rotatebox{90}{\hfill\thepage\hfill}}}}%
	\egroup\fi}
\AddToHook{shipout/background}{\Lpagenumber}%

\usepackage[many]{tcolorbox}

\usepackage{placeins}
\usepackage{oubraces}
\usepackage{tikz}
\usetikzlibrary{shapes.geometric}
\tikzset{gauge1/.style={draw=none,minimum size=0.4cm,fill=white,circle, draw}}
\tikzset{gauge5/.style={draw=none,minimum size=0.6cm,fill=white,circle, draw}}
\tikzset{supergauge/.style={draw=none,minimum size=0.9cm,fill=white,circle, draw}}
\tikzset{bluegauge/.style={draw=none,minimum size=0.4cm,fill=blue,circle, draw}}
\tikzset{redgauge/.style={draw=none,minimum size=0.4cm,fill=red,circle, draw}}
\tikzset{gauge3/.style={draw=none,minimum size=0.4cm,fill=white,circle, draw}}
\tikzset{dotsize/.style={draw=none,minimum size=0.6pt,fill=black,circle,inner sep=1pt, draw}}
\tikzset{mini/.style={draw=none,minimum size=1pt,fill=white,circle,inner sep=3pt, draw}}
\tikzset{miniG/.style={draw=none,minimum size=1pt,fill=black,circle,inner sep=3pt, draw}}
\tikzset{cyane/.style={draw=none,minimum size=0.4cm,fill=cyan,circle, draw}}
\tikzset{pinklinet/.style={draw=none,minimum size=0.4cm,fill=magenta,circle, draw}}
\tikzset{greenlinet/.style={draw=none,minimum size=0.4cm,fill=green,circle, draw}}
\tikzset{blacknode/.style={draw=none,minimum size=0.4cm,fill=black,circle, draw}}
\tikzset{brownlinet/.style={draw=none,minimum size=0.4cm,fill=olive,circle, draw}}
\tikzset{magicmintlinet/.style={draw=none,minimum size=0.4cm,fill=red,circle, draw}}
\tikzset{orangeet/.style={draw=none,minimum size=0.4cm,fill=orange,circle, draw}}
\tikzset{grayet/.style={draw=none,minimum size=0.4cm,fill=gray,circle, draw}}
\tikzset{blueet/.style={draw=none,minimum size=0.4cm,fill=blue,circle, draw}}
\tikzset{flavour2/.style={draw=none,minimum size=0.8cm,fill=white, regular polygon,regular polygon sides=4,draw}}
\tikzset{flavour2/.style={draw=none,minimum size=0.6cm,fill=white, regular polygon,regular polygon sides=4,draw}}
\tikzset{redflavour/.style={draw=none,minimum size=0.6cm,fill=red, regular polygon,regular polygon sides=4,draw}}
\tikzset{blueflavour/.style={draw=none,minimum size=0.6cm,fill=blue, regular polygon,regular polygon sides=4,draw}}
\tikzset{greenflavour/.style={draw=none,minimum size=0.6cm,fill=green, regular polygon,regular polygon sides=4,draw}}
\tikzset{brownflavour/.style={draw=none,minimum size=0.6cm,fill=brown, regular polygon,regular polygon sides=4,draw}}
\tikzset{pinkflavour/.style={draw=none,minimum size=0.6cm,fill=magenta, regular polygon,regular polygon sides=4,draw}}
\tikzset{grayflavour/.style={draw=none,minimum size=0.6cm,fill=gray, regular polygon,regular polygon sides=4,draw}}
\tikzset{none/.style={draw=none}}
\tikzset{new edge style 1/.style={dashed}}
\tikzset{dashedline/.style={dashed}}
\tikzset{brace1/.style={decorate,decoration={brace,amplitude=5pt,mirror}}}
\tikzset{bluee/.style={line width=0.5mm,blue}}
\tikzset{orangee/.style={line width=0.5mm,orange}}
\tikzset{magentae/.style={line width=0.5mm,magenta}}
\tikzset{rede/.style={line width=0.5mm,red}}
\tikzset{thickred/.style={line width=5mm,red}}
\tikzset{greene/.style={line width=0.5mm,green}}
\tikzset{darke/.style={line width=0.5mm,black}}
\tikzset{cyaneX/.style={line width=0.5mm,cyan}}
\tikzset{new edge style 3/.style={dashed,red}}
\tikzset{magicmintline/.style={line width=0.5mm,gray}}
\tikzset{brownline/.style={line width=0.5mm,brown}}
\tikzset{greenline/.style={line width=0.5mm,green}}
\tikzset{oliveline/.style={line width=0.5mm,green}}
\tikzset{darkgreenline/.style={line width=0.5mm,olive}}
\tikzset{pinkline/.style={line width=0.5mm,magenta}}
\tikzset{dottedz/.style={line width=0.5mm,black,dotted}}
\tikzset{pinkline2/.style={line width=0.5mm,magenta,dotted}}
\tikzset{brace2/.style={decorate,decoration={brace,amplitude=5pt}}}
\tikzset{reddotted/.style={line width=0.5mm,red,dotted}}
\tikzset{bluedotted/.style={line width=0.5mm,blue,dotted}}
\tikzset{magicmintdotted/.style={line width=0.5mm,gray,dotted}}
\tikzset{greendotted/.style={line width=0.5mm,green,dotted}}
\tikzset{browndotted/.style={line width=0.5mm,brown,dotted}}
\tikzset{arrowed/.style={line width=0.5mm,->}}
\pgfdeclarelayer{edgelayer}
\pgfdeclarelayer{nodelayer}
\usetikzlibrary{decorations.pathreplacing}
\pgfsetlayers{edgelayer,nodelayer,main}

\usepackage{xcolor,colortbl}

\usepackage{ytableau}
\tikzstyle{brane}=[draw]
\tikzset{D7/.style={circle, draw=black, inner sep=0pt, fill=white, minimum size=3mm}}
\tikzset{hasse/.style={circle, fill,inner sep=2pt}}
\tikzset{flavour/.style={regular polygon,regular polygon sides=4,inner sep=2.5pt, draw}}
\tikzset{gauge/.style={circle, draw,inner sep=2.5pt}}
\tikzset{gaugeb/.style={circle, draw,fill=black,inner sep=2.5pt}}
\tikzset{gaugered/.style={circle, draw,fill=red,inner sep=2.5pt}}
\tikzset{gaugeblue/.style={circle, draw,fill=blue,inner sep=2.5pt}}
\tikzset{gaugegreen/.style={circle, draw,fill=green,inner sep=2.5pt}}
\tikzset{bd/.style={circle, draw=black, inner sep=0pt, fill=black, minimum size=2mm}}
\tikzset{wd/.style={circle, draw=black, inner sep=0pt, fill=white, minimum size=2mm}}
\tikzset{Dynkin/.style={circle, draw=black, inner sep=0pt, fill=white, minimum size=2mm}}
\tikzstyle{ligne}=[draw, thick] 
\tikzset{doublearrow/.style={ draw=black!75, color=black!75, thick, double distance=3pt, }} 
\tikzset{bluedouble/.style={ draw=blue!75, color=blue!75, thick, double distance=3pt, }}

\tikzset{reddashed/.style={ draw=red, color=red, thick }}
\tikzset{redarrow/.style={ draw=red, -> , thick }}

\theoremstyle{plain}
\newtheorem*{thm*}{Theorem}

\theoremstyle{definition}

\newtheorem*{defn*}{Definition}

\newcommand{\Ncal}{\mathcal{N}}
\makeatother

\graphicspath{{figures/}}

\begin{document}

\begin{titlepage}
% Report number
\vspace*{-3cm} 
\begin{flushright}
{\tt DESY-24-185}\\
{\tt UWThPh 2024-26}\\
\end{flushright}
\begin{center}
\vspace{2.2cm}
{\LARGE\bfseries A Pathway to Decay and Fission of \\[0.3em] Orthosymplectic Quiver Theories}
\vspace{1.2cm}

{\large
Craig Lawrie,$^{1}$ Lorenzo Mansi,$^1$ Marcus Sperling,$^2$ and Zhenghao Zhong$^3$\\}
\vspace{.7cm}
{ $^1$ Deutsches Elektronen-Synchrotron DESY,\\
Notkestr.~85, 22607 Hamburg, Germany}\par
\vspace{.2cm}
{ $^2$ Fakultät für Physik, Universität Wien, \\Boltzmanngasse 5, 1090 Wien, Austria}\par
\vspace{.2cm}
{ $^3$ University of Oxford, Radcliffe Observatory, Andrew Wiles Building, \\Woodstock Rd, Oxford OX2 6GG}\par
\vspace{.2cm}

\vspace{.3cm}

\scalebox{.65}{\tt craig.lawrie1729@gmail.com, lorenzo.mansi@desy.de, marcus.sperling@univie.ac.at, zhenghao.zhong@maths.ox.ac.uk}\par
\vspace{1.2cm}
\textbf{Abstract}
\end{center}

We present an algorithm to extract the Coulomb branch Hasse diagram of orthosymplectic 3d $\mathcal{N}=4$ quiver gauge theories. The algorithm systematically predicts all descendant theories arising from Coulomb branch Higgsing, thereby detailing the stratification of the symplectic singularity defined by the initial Coulomb branch. Leveraging the Lie algebra isomorphism $\mathfrak{su}(4) \cong \mathfrak{so}(6)$, we validate our algorithm via the 3d mirror of 4d theories of class $\mathcal{S}$ of such type. This comparison involves moduli spaces that admit both orthosymplectic and unitary quiver realisations, the latter being well-understood via standard techniques such as Decay and Fission. Higgsing on the Coulomb branch of the 3d mirror or magnetic quiver translates to Higgs branch renormalization group flows of the corresponding higher-dimensional SCFTs. Thus,
we benchmark our method via Higgsing 6d $\mathcal{N}=(1,0)$ D-type orbi-instanton theories, predicting novel Higgsing patterns involving products of interacting fixed points, and class $\mathcal{S}$ theories of type $\mathfrak{so}(2N)$, demonstrating Higgsing to products of theories of types specified by Levi subalgebras of $\mathfrak{so}(2N)$.

\vfill 
\end{titlepage}

\tableofcontents
\newpage

\section{Introduction}\label{sec:Intro}

The connection between the Higgsing patterns on the Higgs branch of supersymmetric quantum field theories (SQFTs) with eight supercharges and the stratification of symplectic singularities has been elucidated in greater depth following \cite{Bourget:2019aer}. Physically, the Englert--Brout--Higgs--Guralnik--Hagen--Kibble mechanism \cite{Englert:1964et,Higgs:1964pj,Guralnik:1964eu,Kibble:1967sv}, commonly referred to as the Higgs mechanism, describes the process of a scalar field acquiring a vacuum expectation value (VEV), breaking the gauge group $G$ to a subgroup $H \subset G$. 

In SQFTs with eight supercharges, the hypermultiplet contains four scalars that are rotated under an $SU(2)$ factor of the R-symmetry. The region of the moduli space where these scalars acquire a VEV, called the Higgs branch, defines a variety which is both hyper-K\"ahler \cite{cmp/1104116624} and a symplectic singularity \cite{Beauville_2000}. Mathematically, a symplectic singularity $X$ admits a finite stratification \cite{Kaledin_2006} in \emph{symplectic leaves}, $X_i$, partially ordered by inclusion $X_0 \subset X_1 \subset \cdots \subset X_n$, with $X_0=\{\varnothing \}$ and $X_n=X$. Each stratum is such that the singular part of $X_i$ is in $X_{i-1}$ and its normalisation is a symplectic singularity. Moreover, near each stratum, $X_i$, the variety admits a local decomposition in the product of the closed strata itself and a transverse slice $T_i$, which is itself a symplectic singularity. The physical and mathematical description interpolates the Higgsing pattern with the stratification in symplectic leaves: in a physical theory, each leaf $X_i$ is spanned by the distinct set of VEVs that Higgs the initial theory to a specific residual theory. The transverse slice $T_i$ (to the entire space $X$) is then the moduli space of the residual theory.
This information can be effectively encoded in a Hasse diagram, which captures the structure of a partially ordered set after transitive reduction.

While in three and four dimensions, the Higgs branch $\mathcal{H}$ is ``classical'', or at most well behaved, and can be computed directly once given the physical content of the theory; in five and six dimensions, this space receives infinite corrections from instantons, therefore it is necessary to resort to a method that can resum all these contributions. A solution is to utilize what is known as the magnetic quiver for the Higgs branch \cite{Cabrera:2018jxt,Cabrera:2019izd}, where the demanding problem of summing instantons is traded for the much better under-control task of resuming monopoles in 3d $\mathcal{N}=4$ Lagrangian theories. This technique consists of engineering a set of 3d $\mathcal{N}=4$ \emph{magnetic} theories $\mathrm{M}_i$ whose Coulomb branch $\mathcal{C}$ is isomorphic to the Higgs branch $\mathcal{H}$ of the original \emph{electric} theory $\mathrm{E}$:
\begin{align}
\mathcal{H}_\text{3/4/5/6d} (\mathrm{E}) \,\cong\, \bigcup_i \,\mathcal{C}_{3d} (\mathrm{M}_i) \,.
\end{align}
In fact, due to Poincarè duality in 3d, the Abelian gauge bosons can be dualised to (periodic) scalars, thus giving the space spanned by the vacuum expectation values of the vector multiplet scalar fields, known as the Coulomb branch, also the structure of a symplectic singularity. This variety is parameterised by monopole operators that get ``dressed'' by gauge invariant operators of the residual massless theory in a non-trivial monopole background; all of which is repackaged in the monopole formula \cite{Cremonesi:2013lqa}. 

Such monopole operators \cite{Aharony:1997bx,Borokhov:2002ib,Borokhov:2002cg} have emerged as crucial class of half-BPS operators that are utilised to provide a quantum-corrected description of the 3d $\mathcal{N}=4$ Coulomb branch. Given a UV $G$ gauge theory, (bare) monopole operators are in one-to-one correspondence with highest weight representations of the GNO-dual group $G^\vee$ \cite{Goddard:1976qe} due to a quantisation condition \cite{Englert:1976ng}. A given monopole operator is, moreover, labelled by its conformal dimension (equal to the R-symmetry charge) and its charges under Coulomb branch symmetries, provided such are at least partially manifest in the UV description. Most notably, monopole operators play a key role in the phenomenon of the enhancement of the Coulomb branch global symmetry in the infrared, see \cite{Borokhov:2002cg,Gaiotto:2008ak,Bashkirov:2010hj} and subsequent works. The rapid development of Coulomb branch techniques has laid the foundation for study of Higgs branches via magnetic quivers.

Although the magnetic quiver technique has been fruitfully applied to a plethora of cases, studying the stratification of the variety, and thus the Higgsing pattern of the theory, requires an additional ingredient. The key point is defining a \emph{quiver algorithm}, which is a generic term for any procedure that allows for the extraction of both the leaves and the transverse slices that characterise the singularity resorting only to its formulation in terms of quiver gauge theory.
While various algorithms have been proposed, \cite{Cabrera:2018ann,Cabrera:2019dob,Bourget:2023dkj,Bourget:2024mgn,Bennett:2024loi}, a full solution has been found only for unitary quiver gauge theories. In \cite{Bourget:2023dkj,Bourget:2024mgn} it was shown that, studying the stratification of the Coulomb branch moduli space, a 3d $\mathcal{N}=4$ quiver theory $Q$, with only unitary gauge nodes, can only undergo two possible processes:
\begin{compactitem}
\item \emph{Decay} to a quiver $Q'$ having the same shape as $Q$, but gauge nodes of lower ranks.
\item \emph{Fission} to a product of quivers $Q_1$ and $Q_2$.
\end{compactitem}
The resulting quivers correspond to magnetic quivers for the partially Higgsed theory (i.e., they keep track of the transverse slices) and are all \emph{good} theories in the sense of \cite{Gaiotto:2008ak}: that is, they all have monopole operators whose conformal dimension sits above the unitarity bound, guaranteeing that the UV and IR R-symmetries are the same.

In this paper, we propose an extension of this algorithm to the realm of less-understood orthosymplectic quiver gauge theories. Framed orthosymplectic quivers (i.e., quivers with flavour nodes) played a pivotal role in realising 3d mirror symmetry via S-duality on Type IIB brane configurations with orientifold 3-planes \cite{Uranga:1998uj,Gimon:1996rq,Hanany:1997gh,Kapustin:1998fa,Hanany:1999sj,Feng:2000eq,Gaiotto:2008ak}. In contrast, unframed orthosymplectic quivers (the class relevant for this work) appeared, for example, as 3d mirrors of class $\mathcal{S}$ theories \cite{Benini:2009gi,Benini:2010uu} and, more generally, as magnetic quivers of higher-dimensional theories \cite{Cabrera:2019dob,Bourget:2020gzi,Akhond:2020vhc,Akhond:2021knl,Sperling:2021fcf,Hanany:2022itc,Bao:2024eoq}. Similar to their unitary counterparts, foliating the Coulomb branch forces the orthosymplectic quiver to undergo either decay or fission. Special to this case is that an orthosymplectic quiver does not need to remain of this type: it can, in fact, undergo \emph{unitarisation} and become a unitary theory, or more generically, fission to a product of an orthosymplectic quiver and a unitary one. Moreover, as expected from \cite{Sperling:2021fcf}, an orthosymplectic quiver can decay to an orthosymplectic quiver whose Coulomb branch is a product.

The paper is structured as follows.
In Section \ref{sec:Algorithm}, we explain the Orthosymplectic Decay and Fission algorithm that we propose and provide a practical example of the algorithm and its range of applicability.
In Section \ref{sec:Class_S}, we introduce class $\mathcal{S}$ theories of types $\mathfrak{su}(4)\cong\mathfrak{so}(6)$ and show that the algorithm predicts the same result for the orthosymplectic realization as the unitary one. Next, in Section \ref{sec:Rules}, we provide some deeper explanation for the physics underpinning the proposed algorithm. Then, in Section \ref{sec:Benchmark}, we benchmark the algorithm against quivers and Hasse diagrams obtained independently, focusing in particular on a new result in 6d $\Ncal=(1,0)$ D-Type orbi-instanton theories. Finally, we summarize our results and discuss some future avenues for research in Section \ref{sec:discussion}.

\section{The Orthosymplectic Decay and Fission Algorithm}\label{sec:Algorithm}

In the introduction, we have stressed the importance of having an algorithm that is able to predict the stratification of the Coulomb branch of a 3d $\mathcal{N}=4$ quiver theory. In the remainder of this work, we refer to a quiver theory by the following standard dictionary:
\begin{itemize}
    \item Circular nodes are gauge nodes, they are depicted in a different colour corresponding to the associated gauge algebra. In this paper, we fix all the global forms and we implicitly use the colour coding to symbolise the gauge group:\footnote{Different choices of the global form of a gauge group affect the weight lattice of the GNO-dual group, which labels the monopole operators parameterising the Coulomb branch. See, for example,  \cite{Hanany:2016ezz,Bourget:2020xdz}.}
\begin{align}
\begin{gathered}
    \begin{tikzpicture}
        \node[sonode,label=below:{$k$}] at (0,0) () {};
    \end{tikzpicture} 
\end{gathered} \leftrightarrow SO(k) \quad , \quad \begin{gathered}
    \begin{tikzpicture}
        \node[spnode,label=below:{$2k$}] at (0,0) () {};
    \end{tikzpicture} 
\end{gathered} \leftrightarrow USp(2k) \quad , \quad \begin{gathered}
    \begin{tikzpicture}
        \node[node,label=below:{$k$}] at (0,0) () {};
    \end{tikzpicture} 
\end{gathered} \leftrightarrow U(k) \,.
\end{align}
\item Square nodes denote the presence of a flavour symmetry algebra. The same colour coding used for gauge nodes is employed for flavour algebras.
\item An edge between a square and a circular node signals the presence of hypermultiplets in the bifundamental representation of the gauge and flavour symmetry algebras.
\end{itemize}

From \cite{Gaiotto:2008ak}, we have a prescription, based only on the quiver's shape, to assess whether the conformal dimension of the theory's monopole operators falls within the unitarity bound. The quantity that controls this property is called the \emph{balance}, usually denoted by $b$; it depends on the flavours $N_f$, i.e., the amount of (full) hypermultiplets, a certain gauge node $G$ sees:
\begin{equation}\label{eqn:balance}
    b=\begin{cases}
        N_f-2N & \mathrm{iff} \ G=U(N)\,,\\
        N_f-N+1 & \mathrm{iff} \ G=SO(N)\,,\\
        N_f-2N-1 & \mathrm{iff} \ G=USp(2N)\,.
    \end{cases}
\end{equation}
When $b\geq 0$, the node is \emph{good}: in this case, all the monopole operators are above the unitarity bound. When $b=0$, the node is referred to as \emph{balanced}: such a node is still good, but it admits an operator contributing to the flavour symmetry, leading to an enhancement. The case $b=-1$ is called \emph{ugly}; here, in the unitary case, a monopole operator exists at the unitarity bound, implying that this is a decoupled twisted hypermultiplet in disguise. For orthosymplectic nodes, $b=-1$ already falls within the \emph{bad} category. Lastly, when $b<-1$, the node is \emph{bad} and admits monopole operators below the unitarity bound: this signals that the $R$-symmetry in the UV is not the one in the IR. Therefore, we are counting operators under the wrong symmetry. The good, ugly, and bad classification, albeit being a local statement for each gauge node, is then inherited by the full theory: if no bad or ugly nodes are encountered, the theory is good. However, when making global statements on the goodness of a quiver theory, it may happen that extended monopoles, i.e., monopoles valued in the magnetic lattice of multiple gauge group factors, may render an apparently good theory bad: this is a known case of affine Dynkin-shaped unitary quiver theories whose Coulomb branch is related to the moduli space of instantons of the respective Dynkin algebra in flat space \cite{Cremonesi:2014xha}. To elucidate this discussion with an example, we depict the following good orthosymplectic  quiver:
\begin{equation}\label{eqn:FirstQuiver}
    \raisebox{-0.5\height}{\scalebox{0.8}{\includegraphics[width=0.55\linewidth,page=1]{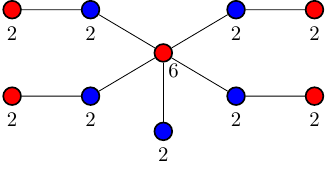}}
    } \,.
\end{equation}
Here, we can see that the $USp(2)$ gauge node at the bottom of the quiver, as well as the four $SO(2)$ nodes, are balanced, and the remaining gauge nodes are all good. We return to this quiver as an example anon.

The orthosymplectic quivers we discuss in this paper have a special feature compared to their unitary counterparts: the monopole operators can take charges both in an integer lattice \emph{and} a shifted lattice where all the charges are half-integer. In the unitary case, the goodness condition on each gauge node is equivalent to the statement that no monopole charged in the integer lattice has R-charge below the unitarity bound. In particular, in the orthosymplectic case, monopoles charged in the half-integer lattice can fall below the unitarity bound in a way that is invisible to the balance of the individual gauge nodes. A more accurate notion of the scope of applicability of our algorithm is when the Coulomb branch is such that it contains exactly one singular point.\footnote{Higgs branch algorithms such as \cite{Bennett:2024loi} are intrinsically free from this shortcoming due to the ``classical'' nature of the space.} Although this usually corresponds to the goodness of the theory in the Gaiotto-Witten sense, i.e., the balance requirement, this definition is more general: extended monopole operators with non-trivial fluxes under multiple gauge group factors, half-integer lattice effects not visible from the balance, and other subtleties are all allowed under this alternative definition of applicability. However, since a notion of balance is useful in stating the algorithm, we postulate a re-adjustment of the balance notion for $SO(6)$ gauge nodes to partially be sensitive to the half-integer monopole charges. This is captured in the following rule, for which a justification is given in Section \ref{sec:Rules}:
\begin{enumerate}
    \item[1.] [\textbf{$USp(2)$ contribution}] A stand-alone $USp(2)$ gauge node next to an $SO(6)$ gauge node effectively contributes to $N_f$ as $\frac{5}{8}$, instead of $1$. A $USp(2)$ in the between $SO(6)$ and $SO(2)$ gauge nodes effectively contributes to $N_f$ as $\frac{5}{4}$ towards the balance of the $SO(6)$ node.
\end{enumerate}
Once the goodness of the theory is correctly assessed, it is possible to move forward and decay the quiver in accordance with the following prescription:
\begin{enumerate}
    \item[2.] [\textbf{Decay existence}] A decay is admissible only if it leads to an orthosymplectic quiver with nodes of balance greater than or equal to $-1$. In the $SO(6)$ where fractional balances between $-1$ and $0$ are allowed, these cases are not admissible as decay products.
    \item[3a.] [\textbf{Allowed Decays}] A quiver with subsets of balanced nodes decays only by subtracting the maximal set of labels allowed in the list of Table \ref{tab:admitted decays} from the balanced nodes. In this case, the transverse slice can be uniquely determined from the quiver's shape and the subtracted labels.
    Whereas nodes with $b>0$ can decay to gauge nodes of the same group but with lowered rank.
    \item[3b.] [\textbf{Allowed Decays}] Gauge nodes with $b=-1$ can only decay to nodes of the same kind with rank lowered by $1$.
\end{enumerate}

Another option that must be considered at any stage is the possibility of fissioning the quiver into a product of two theories, with one possibly being degenerate.

\begin{enumerate}
    \item[4a.] [\textbf{Fission}] An $\mathfrak{so}({2N})$ algebra can fission to any $\mathfrak{so}({2N-2K})\oplus \mathfrak{u}({K})$ algebras for $K\le N$, similarly a $\mathfrak{usp}({2N})$ algebra can fission to any $\mathfrak{usp}({2N-2K})\oplus \mathrm{u}(K)$ algebras; these fission processes uplift from direct sum of algebras to product groups. Said fission process is possible only if the resulting unitary and orthosymplectic quivers are good. 
    
    \item[4b.] [\textbf{Unitarisation}] The extreme case of the fission product when $N=K$ for all nodes in the quiver is called \emph{unitarisation}:  in this process, differently from a standard fission, the obtained unitary gauge groups may need to be lowered such that the resulting unitary quiver is globally good.
\end{enumerate}
After a fission operation, the unitary quiver that is generated can be Higgsed further via standard Decay and Fission.

To get a feeling of a practical application of the Orthosymplectic Decay and Fission algorithm, we consider the orthosymplectic quiver theory $\mathcal{Q}$ of equation \eqref{eqn:FirstQuiver} and derive the Hasse diagram of the Coulomb branch. This is depicted in Figure \ref{fig:Algorithm_Appliation}. 

We comment on each derivation step:
\begin{itemize}
    \item \textbf{Step 0}. Checking the overall balance of the quiver according to the new prescription, we can start from the central $SO(6)$ node of $\mathcal{Q}$ and compute the balance
    \begin{equation}
        b=4\cdot\frac{5}{4}+5/8-5=5/8 \,.
    \end{equation}
    Therefore, this node is good; here it was important that we used the new balance-contribution prescription given in {\tt Rule 1}. The stand-alone $USp(2)$ is balanced ($b=3-2-1=0$), as are all the other $SO(2)$ gauge nodes ($b=1-2+1=0$). The remaining $USp(2)$ nodes are all good with balance $b=1+3-2-1=1$.
    \item \textbf{Step 1}. We now analyse all the possible decay options.  The stand-alone $USp(2)$ node, according to Table \ref{tab:admitted decays}, can be decayed; giving rise to the leftmost quiver in the first row from the bottom of Figure \ref{fig:Algorithm_Appliation}. Any of the $SO(2)$ nodes can similarly be decayed, leading to the central quiver of the same row. Both these options lead to good quivers. A third option consists of decaying the central node to $SO(4)$: this leads to the rightmost quiver in the first row; in this case, the stand-alone $USp(2)$ has balance $b=-1$, therefore the quiver is bad, but we allow it due to {\tt Rule 2}. Any other subtraction of the allowed labels in Table \ref{tab:admitted decays} leads to theories with at least one node of badness less than $-1$, hence they are not admissible. 
    
   It is now time to turn our attention to fission processes. The central $\mathfrak{so}(6)$ algebra can fission to $\mathfrak{so}(4) \oplus \mathfrak{u}(1)$.\footnote{The idea behind this fission can be observed from brane set-ups with orientifold planes. A Higgsing related to fission takes a subset of dynamical branes and moves them away from the orientifold planes. Brane dynamics tells us this exactly leads to the decomposition of the gauge algebras $\mathfrak{so}({2n})\rightarrow \mathfrak{so}({2l})\oplus \mathfrak{u}(k)$ and $\mathfrak{usp}(2n)\rightarrow \mathfrak{usp}(2l)\oplus \mathfrak{u}(k)$ with $n=k+l$. In the case where $l=0$, this leads to unitarisation which is akin to moving \emph{all} the branes away from the orientifold plane. This explanation is discussed in more detail in Section \ref{unitarization}.} Due to the Lie algebra isomorphism between $\mathfrak{so}_4$ and $\mathfrak{su}_2 \oplus \mathfrak{su}_2$, one of the product factors can recombine with the $\mathfrak{u}_1$ leading to the following fission of the central node
    \begin{equation}
        \mathfrak{so}(6) \rightarrow \mathfrak{usp}(2) \oplus \mathfrak{u}(2) \,,
    \end{equation}
    where we implicitly used the algebra isomorphism $\mathfrak{su}(2) \cong \mathfrak{usp}(2)$. The $\mathfrak{usp}(2)$ algebras attached to the central node can fission to $\mathfrak{u}(1)$, which can be thought of as either an $\mathfrak{so}(2)$ or as product algebra of the ``trivial''  orthosymplectic algebra $\mathfrak{\varnothing}$ and $\mathfrak{u}(1)$. Explicitly in the case of the leftmost quiver:
    \begin{equation}
        \begin{gathered}
            \includegraphics[page=14,width=0.9\linewidth]{DecayAndFission_figures.pdf}
        \end{gathered}\,,
    \end{equation}
    where we have conveniently written the product algebras such that the leftmost would be associated with the orthosymplectic quiver theory and the rightmost to the unitary one. No minimal unitarisation processes are available at this stage.
    
    \item \textbf{Step 2}. 
    Starting again from studying the possible decays, the only option consistent with the rules for the leftmost quiver in the first row is decaying the central $SO(6)$ gauge node. All tentative decays on the central theory would result in a bad quiver theory. Likewise, the rightmost non-product quiver in the first row can decay the $b=-1$ node, leading to the leftmost quiver in the second row.
    It is now time to turn our attention to fission processes: both the leftmost and the central theories in the first row can fission. For instance, in the case of the leftmost quiver, we have:
    \begin{equation}
        \begin{gathered}
            \includegraphics[page=12,width=0.9\linewidth]{DecayAndFission_figures.pdf}
        \end{gathered}\,,
    \end{equation}
    where we have adopted the previous employed convention of denoting the product algebras such that the leftmost would be associated with the orthosymplectic quiver theory and the rightmost to the unitary one. 
    
    Alongside fission, it is possible to consider quiver unitarisations: these occur for both the central and the rightmost quiver of the first row. For example:
    \begin{equation}
                \begin{gathered}
            \includegraphics[page=13,width=0.9\linewidth]{DecayAndFission_figures.pdf}
        \end{gathered}\,.
    \end{equation}

    \item \textbf{Step 3}. Apart from the leftmost quiver and the orthosymplectic quiver in the central position in the second row of Figure \ref{fig:Algorithm_Appliation}, which undergo unitarisation (albeit the latter unitarises to the trivial theory) from here on it is possible to resort to standard Decay and Fission as all the quivers are of unitary type. 
\end{itemize}

\begin{landscape}
\pagestyle{empty}
    \begin{figure}
    \begin{subfigure}[b]{0.47\linewidth}
        \centering
        \includegraphics[width=\linewidth,page=2]{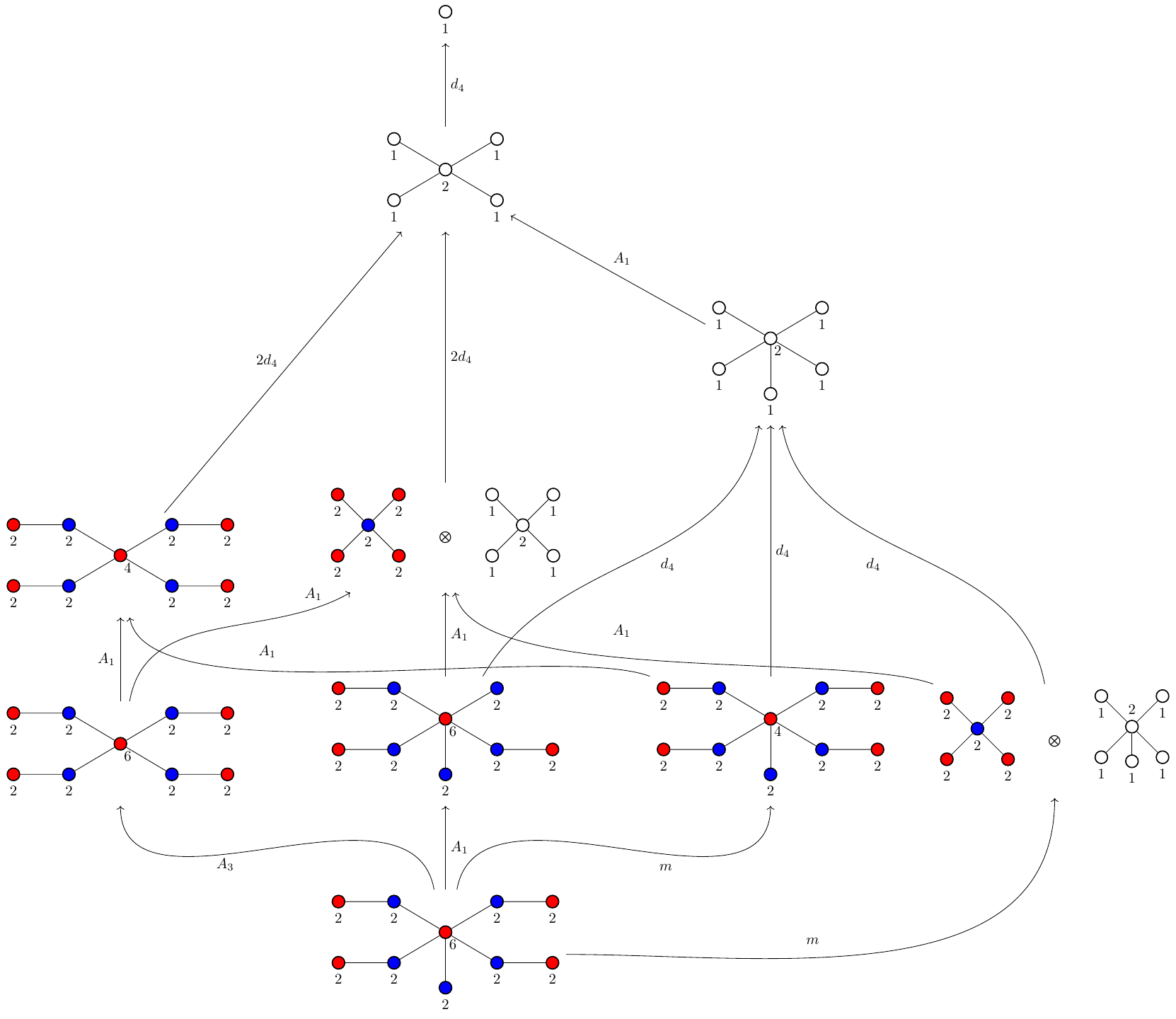}
        \caption{}
        \label{fig:Algorithm_Appliation}
        \end{subfigure} \hspace{0.5cm}
    \begin{subfigure}[b]{0.47\linewidth}
        \centering
        \includegraphics[width=0.9\linewidth,page=15]{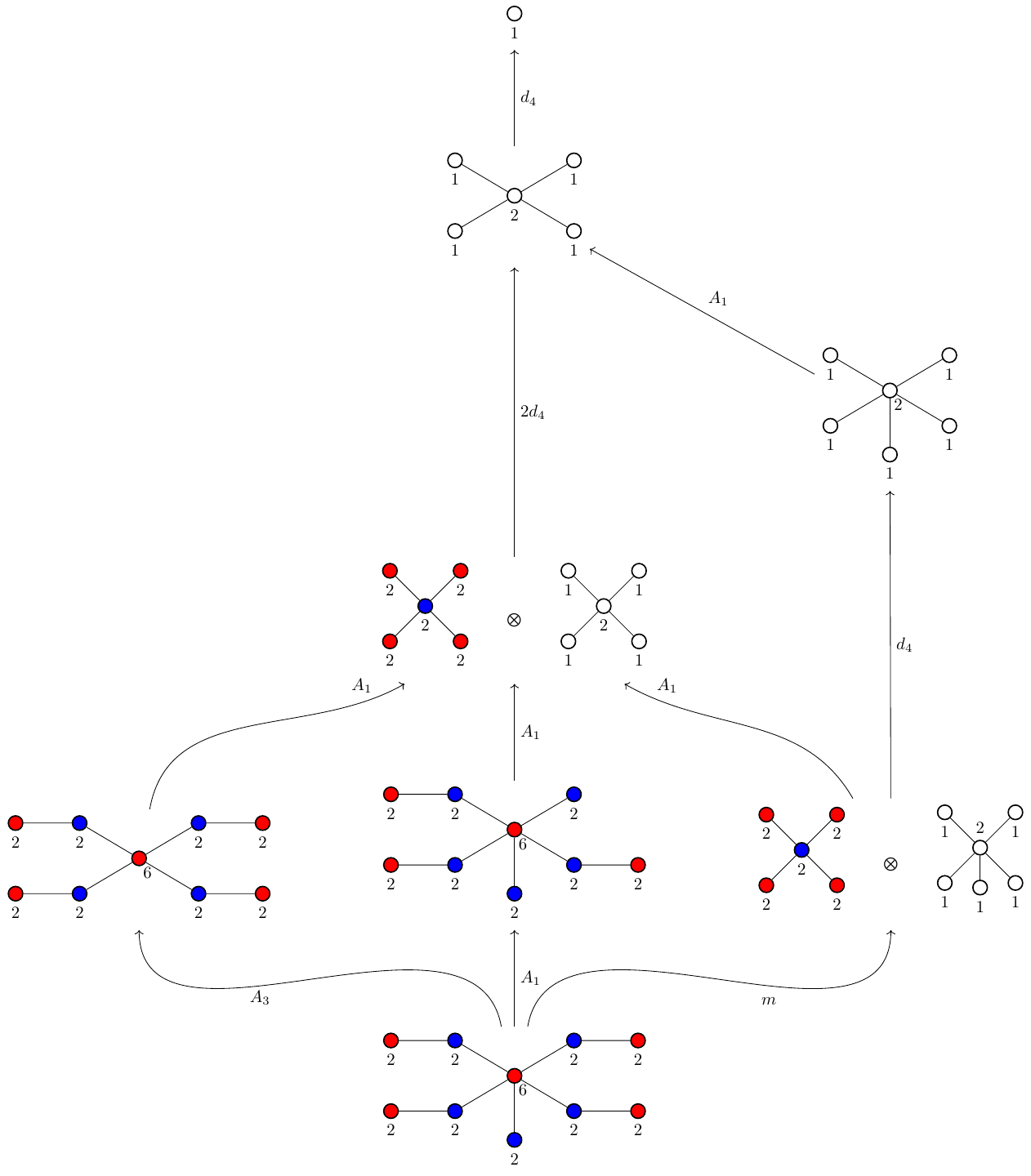}
        \caption{}
        \label{fig:Algorithm_Appliation_Clean Hasse}
    \end{subfigure}
    \caption{(a) Tentative Hasse diagram for the Coulomb branch of the quiver 3d $\mathcal{N}=4$  orthosymplectic quiver $\mathcal{Q}$ of equation \eqref{eqn:FirstQuiver} extracted via the Orthosymplectic Decay and Fission algorithm. The transverse slices are inferred and double-checked via standard Decay and Fission on an alternative unitary realisation of the same moduli space. (b) Hasse diagram for the Coulomb branch of the quiver 3d $\mathcal{N}=4$  orthosymplectic quiver $\mathcal{Q}$ of equation \eqref{eqn:FirstQuiver} obtained by removing the redundant quivers obtained via the Orthosymplectic Decay and Fission. }
    \end{figure}
    
\end{landscape}

The diagram of Figure \ref{fig:Algorithm_Appliation} does not qualify as a Hasse diagram in the standard sense.
Consider the two rightmost quivers belonging to the first row from the bottom and the two leftmost quivers belonging to the second row from the bottom of Figure \ref{fig:Algorithm_Appliation}. Both pairs have the same Coulomb branch dimension, the same descendants obtained via Higgsing, and in the latter case, where the Coulomb branch Hilbert series is computable, the same Coulomb branch (two copies of $d_4$ intersecting at the origin). In particular, we have shown that
\begin{equation}
    \operatorname{HS}_{\mathcal{C}}\left( \begin{gathered}
        \includegraphics[page=22,width=0.3\linewidth]{DecayAndFission_figures.pdf}
    \end{gathered}\right) = \operatorname{HS}_{\mathcal{C}}\left( \begin{gathered}
        \includegraphics[page=23,width=0.1\linewidth]{DecayAndFission_figures.pdf}
    \end{gathered}\right) \times \operatorname{HS}_{\mathcal{C}}\left(\begin{gathered}
        \includegraphics[page=24,width=0.1\linewidth]{DecayAndFission_figures.pdf}
    \end{gathered} \right)\,,
\end{equation}
where $\operatorname{HS}_{\mathcal{C}}$ denotes the Coulomb branch Hilbert series.

We conjecture that, in general, when a pair of theories, either good or bad, engineers the same Hasse diagram, those two theories are actually different presentations of the same theory. However, it is impossible for the algorithm alone to correctly identify them: the algorithm thus predicts the Decay and Fission products of an orthosymplectic quiver, but with the caveat of possibly engineering a plurality of theories in between that must be identified in order to extract the correct Hasse diagram, see Figure~\ref{fig:Algorithm_Appliation_Clean Hasse}. Moreover, the algorithm is only applicable to simply laced theories, for which the Coulomb branch symmetry is composed of algebras other than $\mathfrak{g}_2$ and $\mathfrak{f}_4$. In the latter case, this limitation intrinsically reflects the absence of a simply laced linear quiver gauge theory encoding in its Coulomb branch, Slodowy slices from the regular orbit to a generic orbit $O$ or from $O$ to the trivial orbit.

Having highlighted the range of applicability of Orthosymplectic Decay and Fission, we can now turn our attention to justifying the rules for the algorithm. In order to do as proposed, we need to introduce a further ingredient: the 4d $\mathcal{N}=2$ class $\mathcal{S}$ theories.

\section{Lessons from \texorpdfstring{\boldmath{$\mathfrak{su}(4) \cong \mathfrak{so}(6)$}}{su(4) = so(6)} \texorpdfstring{Class $\mathcal{S}$}{Class S}}\label{sec:Class_S}

In Section~\ref{sec:Algorithm}, we proposed an algorithm by which the structure
of the Coulomb branch of a 3d $\mathcal{N}=4$ orthosymplectic quiver gauge
theory can be elucidated. Of course, this algorithm is not something that Moses
carried down the mountain in his other hand, but is instead motivated initially
by reference to myriad examples. In particular, we can exploit the Lie algebra 
isomorphism 
\begin{equation}\label{eqn:iso}
	\mathfrak{su}(4) \cong \mathfrak{so}(6) \,,
\end{equation}
and compare the results of Orthosymplectic Decay and Fission with the structure 
for the Coulomb branch for the corresponding unitary theory obtained via 
unitary Decay and Fission. In this section, we consider the Higgs branch of 
class $\mathcal{S}$ theories of types $\mathfrak{su}(4)$ and $\mathfrak{so}(6)$, 
and demonstrate that the algorithm of Section \ref{sec:Algorithm} produces the 
expected result.

Class $\mathcal{S}$ \cite{Gaiotto:2009hg,Gaiotto:2009we} is a construction of
4d $\mathcal{N}=2$ SCFTs that starts from the 6d $(2,0)$ SCFT of type
$\mathfrak{g}$, where $\mathfrak{g}$ is a simple and simply-laced Lie algebra, i.e., of ADE type,
that is further compactified on a $n$-punctured genus $g$ Riemann surface with
a topological twist to preserve eight supercharges. We denote such a 4d theory
as
\begin{equation}\label{eqn:classS}
    \mathcal{S}_\mathfrak{g}\langle C_{g,n} \rangle \{ O_1, \cdots, O_n \} \,,
\end{equation}
where $O_i$ capture the data defining the $n$ punctures. In this paper, we are
concerned only with regular untwisted punctures, in which case each $O_i$
corresponds to a choice of nilpotent orbit of $\mathfrak{g}$
\cite{Chacaltana:2012zy}. See \cite{Couzens:2023kyf} for a recent summary of
the pertinent details and terminology of the class $\mathcal{S}$ construction. An 
important point about the class $\mathcal{S}$ construction is that the flavour symmetry algebra
of the resulting 4d $\mathcal{N}=2$ SCFT decomposes over the set of punctures; we have
\begin{equation}\label{eqn:flavS}
	\mathfrak{f} \supseteq \bigoplus_{i=1}^n \mathfrak{f}(O_i) \,,
\end{equation}
where $\mathfrak{f}(O_i)$ is the flavour locally associated to the puncture $O_i$ by 
the choice of nilpotent orbit. The generic case is for the inequality in equation 
\eqref{eqn:flavS} to be saturated; when it is not saturated the theory is said to 
have ``enhanced'' flavour symmetry.

\begin{figure}[t]
    \centering
    \begin{tikzpicture}
        % \node[circle,draw=none] at (2,2) (1) {};
            \node at (1.2,1.2) (1a) {$T_{O_1}(G)$};
        % \node[circle,draw] at (3.14,0) (2) {};
            \node at (1.92,0) (2a) {$T_{O_2}(G)$};
        % \node[circle,draw] at (2,-2) (3) {};
            \node at (1.2,-1.2) (3a) {$T_{O_3}(G)$};
        % \node[circle,draw] at (-3.34,0) (n-1) {};
            \node at (-2.14,0) (n-1a) {$T_{O_{n{-}1}}(G)$};
        % \node[circle,draw] at (-2,2) (n) {};
            \node at (-1.2,1.2) (na) {$T_{O_n}(G)$};

        % \filldraw (1,-1) circle (1pt);
        % \filldraw (0,-1.41) circle (1pt);
        % \filldraw (-1,-1) circle (1pt);
        % \node at (-1,-1) {$\ddots$};
            
        \node[circle,draw, minimum size=40pt, label={[yshift=-28pt]$G$}] at (0,0) (c) {};
    \draw[dotted,very thick] (-1,-1) to[out=315,in=180]  (0,-1.41);
    
        % \draw (1)--(1a) (1a)--(c) (2)--(2a) (2a)--(c) (3)--(3a) (3a)--(c) (n)--(na) (na)--(c) (n-1)--(n-1a) (n-1a)--(c);
        \draw (1a)--(c) (2a)--(c) (3a)--(c) (na)--(c) (n-1a)--(c);

        \node at (4.0,0.0) () {\Huge$/$\footnotesize$\, Z(G)$};
        \end{tikzpicture}
    \caption{The star-shaped mirror dual quiver for the class $\mathcal{S}$ theory in equation \eqref{eqn:classS} compactified on a circle. The $T_\rho(G)$, which are the 3d $\mathcal{N}=4$ theories, have their $G$ (Higgs branch) flavour symmetries diagonally gauged. We explicitly write the quotient action by the center of the group $G$, denoted as $Z(G)$, that acts diagonally on all the gauge groups in the theory; the subtlety of this extra
quotient was observed in \cite{Cremonesi:2014vla}.}
    \label{fig:mirror_classS}
\end{figure}
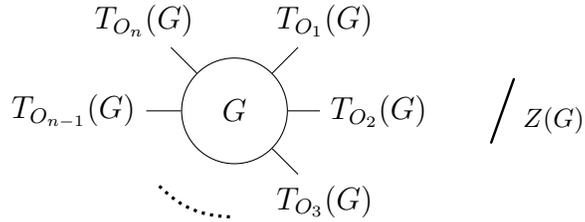

Typically class $\mathcal{S}$ theories are non-Lagrangian, and understanding
their Higgs branches is a priori challenging. However, when $\mathfrak{g}$ is a
classical Lie algebra, one can compactify on an $S^1$ and perform the mirror
duality to obtain a 3d $\mathcal{N}=4$ Lagrangian quiver, which is known as the
3d mirror. The Coulomb branch of this 3d mirror is isomorphic to the Higgs
branch of the class $\mathcal{S}$ theory, and since the 3d mirror is Lagrangian
we can utilize the well-developed toolkits to understand the Higgs branch of
the non-Lagrangian 4d SCFT.\footnote{However, one can also study the 4d Higgs
branches of class $\mathcal{S}$ directly, without exploiting the 3d mirror,
see, for example, \cite{DKL} for details.} For a class $\mathcal{S}$ theory
of the form in equation \eqref{eqn:classS} with only regular untwisted
punctures, constructed on genus zero Riemann surfaces,\footnote{The 3d mirror
is given in \cite{Benini:2010uu} for generic genus, however, for simplicity, we focus only on $g = 0$.} the 3d mirror was given in
\cite{Benini:2010uu} and is schematically shown in Figure
\ref{fig:mirror_classS}.\footnote{The 3d mirrors for class $\mathcal{S}$ with
twisted regular punctures are also known, see \cite{Giacomelli:2020ryy,Kang:2022zsl} for the case
where $\mathfrak{g} = \mathfrak{su}(2n+1)$.  Such quivers are often
(special-)unitary-orthosymplectic; such theories would be a productive test
for a further extension of the Decay and Fission algorithm to these more
general quivers.}

As we can see from Figure \ref{fig:mirror_classS}, the 3d mirrors involve the
diagonal gauging of the flavour symmetry of certain 3d $\mathcal{N}=4$ SCFTs 
labelled as $T_\rho(G)$ \cite{Gaiotto:2008ak}, where $O_\rho$ is a nilpotent orbit of $G$. These theories have a Higgs branch which is the closure of nilpotent orbit $O_{\rho^\vee}$ of the Langlands dual of $G$\footnote{The orbit $\rho^\vee$ of $G^\vee$ is determined by the action of the Spaltenstein map $\vee:G\rightarrow G^\vee$ \cite{Spaltenstein1982} exchanging the nilpotent orbit, specified by $\rho$, of $G$ with a nilpotent orbit of $G^\vee$.}, and a Coulomb branch which is the Slodowy slice, $S_{\rho}$, transverse to $O_\rho$ inside of the nilpotent cone of $G$. 

For $\mathfrak{g}$ a classical Lie algebra, the 3d mirror of the class $\mathcal{S}$ theories are Lagrangian as the $T_{\rho}(G)$ theories are themselves linear Lagrangian quivers. To determine these quiver gauge theories, we first note that nilpotent orbits of classical Lie algebras are associated with integer partitions;\footnote{For a textbook treatment of nilpotent orbits of (semi-)simple Lie algebras, see \cite{Collingwood_1993}.} in terms of this partition, $\rho$, an algorithm is given to construct the $T_{\rho}(G)$ in \cite{Gaiotto:2008ak}.\footnote{We often use $\rho$ to refer to both the nilpotent orbit and the associated integer partition.} We write the partitions associated with nilpotent orbits and the corresponding linear quivers in Table \ref{tbl:A3vsD3}; in particular, we see that $\mathfrak{su}(4)$ nilpotent orbits are in one-to-one correspondence with $\mathfrak{so}(6)$ nilpotent orbits, consistent with the Lie algebra isomorphism. We conclude that the 3d mirrors of the class $\mathcal{S}$ theories we are considering are star-shaped quivers where the tails are $T_\rho$ theories associated with the nilpotent orbits describing the punctures.

We notice that the class $\mathcal{S}$ description, and thus the Coulomb branch
of the 3d mirror, is completely specified by the Lie algebra $\mathfrak{g}$ as
well as the collection of nilpotent orbits capturing the punctures. As such, by
the isomorphism in equation \eqref{eqn:iso}, we expect that the
$\mathfrak{su}(4)$ and $\mathfrak{so}(6)$ theories are identical, where we also
map the nilpotent orbits as in Table \ref{tbl:A3vsD3}. A priori, this is not
obvious when looking at the Lagrangian quivers describing the 3d mirror, as the
$T_\rho(SO(6))$ can involve gauge nodes which do not themselves have a dual
unitary description. 

Thus, the central point of this construction is having equivalent unitary and orthosymplectic interpretations
of the class $\mathcal{S}$ theories $\mathcal{S}_{\mathfrak{so}_6}\langle
C_{g,n} \rangle \{ O_1, \cdots, O_n \}$. From the unitary 3d mirror, via the unitary
Decay and Fission algorithm, we can derive the Higgsing pattern for the unitary
4d theory and map each partially Higgsed theory to its orthosymplectic
counterpart. In this way, we can extensively test the Orthosymplectic Decay and
Fission algorithm presented in Section \ref{sec:Algorithm}.

\begin{table}[t]
    \centering
    \renewcommand{\arraystretch}{1.2}
        \begin{tabular}{c|c|c|c}
           \toprule
            $O_{A_3}$ & $T_{O_{A_3}}(SU(4))$ & $O_{D_3}$ & $T_{O_{D_3}}(SO(6))$ \\ \midrule
            $[1^4]$ & $\begin{gathered}
    \begin{tikzpicture}[baseline=0,font=\footnotesize]
        \node[node, label=below:{$1$}] (A3) {};
      \node[node, label=below:{$2$}] (A4) [right=8mm of A3] {};
        \node[node, label=below:{$3$}] (A6) [right=8mm of A4] {};
        \node[flavour, label=below:{$4$}] (A7) [right=8mm of A6] {};
    
        \draw (A3.east) -- (A4.west);
      \draw (A4.east) -- (A6.west);
      \draw (A6.east) -- (A7.west);
    \end{tikzpicture} 
 \end{gathered}$ 
 & $[1^6]$ & $\begin{gathered}
      \begin{tikzpicture}[baseline=0,font=\footnotesize]
        \node[node, label=below:{$2$},fill=red] (A3) {};
      \node[node, label=below:{$2$},fill=blue] (A4) [right=8mm of A3] {};
        \node[node, label=below:{$4$},fill=red] (A5) [right=8mm of A4] {};
        \node[node, label=below:{$4$},fill=blue] (A6) [right=8mm of A5] {};
        \node[flavour, label=below:{$6$},fill=red] (A7) [right=8mm of A6] {};
    
        \draw (A3.east) -- (A4.west);
      \draw (A4.east) -- (A5.west);
      \draw (A5.east) -- (A6.west);
      \draw (A6.east) -- (A7.west);
    \end{tikzpicture} 
  \end{gathered}$ 
  \\%\hline
            $[2, 1^2]$ & $\begin{gathered}
    \begin{tikzpicture}[baseline=0,font=\footnotesize]
        \node[node, label=below:{$1$}] (A3) {};
      \node[node, label=below:{$2$}] (A4) [right=8mm of A3] {};
        \node[flavour, label=below:{$4$}] (A7) [right=8mm of A4] {};
    
        \draw (A3.east) -- (A4.west);
      \draw (A4.east) -- (A7.west);
    \end{tikzpicture} 
  \end{gathered}$ 
  & $[2^2, 1^2]$ & $\begin{gathered}
      \begin{tikzpicture}[baseline=0,font=\footnotesize]
        \node[node, label=below:{$2$},fill=red] (A5) {};
        \node[node, label=below:{$4$},fill=blue] (A6) [right=8mm of A5] {};
        \node[flavour, label=below:{$6$},fill=red] (A7) [right=8mm of A6] {};
    
      \draw (A5.east) -- (A6.west);
      \draw (A6.east) -- (A7.west);
    \end{tikzpicture} 
 \end{gathered}$ 
 \\%\hline
            $[2^2]$ & $\begin{gathered}
    \begin{tikzpicture}[baseline=0,font=\footnotesize]
        \node[node, label=below:{$2$}] (A3) {};
        \node[flavour, label=below:{$4$}] (A7) [right=8mm of A3] {};
    
        \draw (A3.east)  -- (A7.west);
    \end{tikzpicture} 
  \end{gathered}$ 
  &  $[3, 1^3]$ & $\begin{gathered}
      \begin{tikzpicture}[baseline=0,font=\footnotesize]
        \node[node, label=below:{$2$},fill=red] (A5) {};
        \node[node, label=below:{$2$},fill=blue] (A6) [right=8mm of A5] {};
        \node[flavour, label=below:{$6$},fill=red] (A7) [right=8mm of A6] {};
    
      \draw (A5.east) -- (A6.west);
      \draw (A6.east) -- (A7.west);
    \end{tikzpicture} 
  \end{gathered}$ 
  \\%\hline
            $[3, 1]$ & $\begin{gathered}
    \begin{tikzpicture}[baseline=0,font=\footnotesize]
        \node[node, label=below:{$1$}] (A3) {};
        \node[flavour, label=below:{$4$}] (A7) [right=8mm of A3] {};
    
        \draw (A3.east)  -- (A7.west);
    \end{tikzpicture} 
  \end{gathered}$ 
  & $[3^2]$ & $\begin{gathered}
      \begin{tikzpicture}[baseline=0,font=\footnotesize]
        \node[node, label=below:{$2$},fill=blue] (A6) {};
        \node[flavour, label=below:{$6$},fill=red] (A7) [right=8mm of A6] {};
    
      \draw (A6.east) -- (A7.west);
    \end{tikzpicture} 
  \end{gathered}$ 
  \\%\hline
            $[4]$ & $\begin{gathered}
    \begin{tikzpicture}[baseline=0,font=\footnotesize]
        \node[flavour, label=below:{$4$}] (A7)  {};
    \end{tikzpicture} 
  \end{gathered}$ 
 & $[5, 1]$ & $\begin{gathered}
      \begin{tikzpicture}[baseline=0,font=\footnotesize]
        \node[flavour, label=below:{$6$},fill=red] (A7) {};
    \end{tikzpicture} 
 \end{gathered}$  \\ \bottomrule
        \end{tabular}
    \caption{The nilpotent orbits of $\mathfrak{su}(4)$ and $\mathfrak{so}(6)$ written as integer partitions of $4$ and $6$, respectively. We also write the linear quiver descriptions of the $T_\rho$ theories associated to each nilpotent orbit. The nilpotent orbits written in the same row are isomorphic under the $\mathfrak{su}(4) \cong \mathfrak{so}(6)$ isomorphism. We write the partitions using multiplicative notation throughout this paper.}
    \label{tbl:A3vsD3}
\end{table}

The generic Higgsing pattern in class $\mathcal{S}$ is known as partial puncture closure. This occurs when there is a Higgs branch renormalization group flow between theories
\begin{equation}
    \mathcal{S}_\mathfrak{g}\langle C_{g,n}\rangle\{ O_1, \cdots \} \quad \rightarrow \quad \mathcal{S}_\mathfrak{g}\langle C_{g,n}\rangle\{ O_1', \cdots \} \,,
\end{equation}
where the punctures in the $\cdots$ remain the same; i.e., only one puncture is modified. It is believed that there exists such a flow whenever $O' < O$ under the partial ordering on nilpotent orbits of $\mathfrak{g}$.\footnote{It is, however, not entirely clear that all such Higgsings exist from the four-dimensional perspective. See \cite{Distler:2022nsn} for some of the subtleties.} 

Eventually, after sufficient partial puncture closure, the resulting class $\mathcal{S}$ theory develops an enhanced flavour symmetry; at this point partial puncture closure can be obstructed as half-BPS operators that enter the spectrum from one puncture can mix with those of another puncture to form representations of the enhanced flavour symmetry. Thus, giving vacuum expectation values to operators arising from one puncture can affect the other punctures. Since the enhanced symmetry is not manifest in the class $\mathcal{S}$ geometry, and thus does not have a geometric interpretation like partial puncture closure, the Higgs branch of such theories requires individual analysis. However, enhanced symmetry only occurs in a small number of cases, and always occurs near the end of the Higgs branch Hasse diagram for a generic class $\mathcal{S}$ theory.

These Higgsings obtained from partially closing punctures do not cover all possible Higgsings of a class $\mathcal{S}$ theory. The missing Higgsings are precisely the fission, and therefore unitarisation, processes in the Orthosymplectic Decay and Fission algorithm. 

However, they are sufficient to describe all possible Higgsing when $\mathfrak{g} = \mathfrak{su}(4) \cong \mathfrak{so}(6)$. More general possibilities, in fact these are the fissions and the unitarizations, which can be associated to the generators of the Higgs branch chiral ring \cite{Maruyoshi:2013hja} that are not moment-map operators, are discussed from the perspective of the 4d Higgs branch in \cite{DKL}, and can partially be related to the complex deformation theory of the $\mathbb{C}^2/\Gamma$ orbifolds that engineer the 6d $(2,0)$ SCFTs in Type IIB string theory, c.f., \cite{Bourget:2022ehw,Lawrie:2024zon}.

\begin{figure}[t]
    \centering
    \begin{subfigure}[b]{0.47\textwidth}
    \centering
    \begin{tikzpicture}
\node[align=left] at (0,0) (null) [label=left:{\footnotesize{$[1^4]$:}}] {
 \raisebox{-0.5\height}{
    \begin{tikzpicture}
\node[fnode,label=below:{\footnotesize $4$}] at (0,0) (4) {};
\node[node,label=below:{\footnotesize $3$}] [right=5mm of 4 ] (3) {};
\node[node,label=below:{\footnotesize $2$}] [right=5mm of 3 ] (2) {};
\node[node,label=below:{\footnotesize $1$}] [right=5mm of 2 ] (1) {};
\draw (4)--(3)--(2)--(1);
\end{tikzpicture}}
};
\node[align=left] at (0,+1.5) (minimal)  [label=left:{\footnotesize{$[2,1^2]$:}}]{
\raisebox{-.5\height}{
    \begin{tikzpicture}
\node[fnode,label=below:{\footnotesize $4$}] at (0,0) (4) {};
\node[node,label=below:{\footnotesize $2$}] [right=5mm of 4 ] (3) {};
\node[node,label=below:{\footnotesize $1$}] [right=5mm of 3 ] (2) {};
\draw (4)--(3)--(2);
\end{tikzpicture} }
};
\node[align=left] at (0,+3) (orbit) [label=left:{\footnotesize{$[2^2]$:}}]{
    \raisebox{-.5\height}{\begin{tikzpicture}
\node[fnode,label=below:{\footnotesize $4$}] at (0,0) (4) {};
\node[node,label=below:{\footnotesize $2$}] [right=5mm of 4 ] (3) {};
\draw (4)--(3);
\end{tikzpicture}}
};
\node[align=left] at (0,+4.5) (subregular)[label=left:{\footnotesize{$[3,1]$:}}] {
    \raisebox{-.5\height}{\begin{tikzpicture}
\node[fnode,label=below:{\footnotesize $4$}] at (0,0) (4) {};
\node[node,label=below:{\footnotesize $1$}] [right=5mm of 4 ] (3) {};
\draw (4)--(3);
\end{tikzpicture}
}
};
\node[align=left] at (0,+6) (regular) [label=left:{\footnotesize{$[4]$:}}]{
    \raisebox{-.5\height}{\begin{tikzpicture}
\node[fnode,label=below:{\footnotesize $4$}] at (0,0) (4) {};
\end{tikzpicture}}
};

\draw (null.north)--(minimal.south) node[midway,right] () {\footnotesize $a_3$};
\draw (minimal)--(orbit) node[midway,right] () {\footnotesize $a_1$};
\draw (orbit)--(subregular) node[midway,right] () {\footnotesize $A_1$};
\draw (subregular)--(regular) node[midway,right] () {\footnotesize $A_3$};

\end{tikzpicture}
    \caption{$T_\rho (SU(4))$.}
    \label{eqn:conesu_4+trho}
\end{subfigure}
\quad
\begin{subfigure}[b]{0.47\textwidth}
    \centering
    \begin{tikzpicture}
\node[align=left] at (0,0) (null) [label=left:{\footnotesize{$[1^4]$:}}] {
    \begin{tikzpicture}
\node[fnode,fill=red, label=below:{\footnotesize $6$}] at (0,0) (4) {};
\node[spnode,label=below:{\footnotesize $4$}] [right=5mm of 4 ] (3) {};
\node[sonode,label=below:{\footnotesize $4$}] [right=5mm of 3 ] (2) {};
\node[spnode,label=below:{\footnotesize $2$}] [right=5mm of 2 ] (1) {};
\node[sonode,label=below:{\footnotesize $2$}] [right=5mm of 1 ] (0) {};
\draw (4)--(3)--(2)--(1)--(0);
\end{tikzpicture}
};
\node[align=left] at (0,+1.5) (minimal) [label=left:{\footnotesize{$[2,1^2]\sim [2^2,1^2]$:}}]{
    \begin{tikzpicture}
\node[fnode,fill=red, label=below:{\footnotesize $6$}] at (0,0) (4) {};
\node[spnode,label=below:{\footnotesize $4$}] [right=5mm of 4 ] (3) {};
\node[sonode,label=below:{\footnotesize $2$}] [right=5mm of 3 ] (2) {};
\draw (4)--(3)--(2);
\end{tikzpicture}
};
\node[align=left] at (0,+3) (orbit) [label=left:{\footnotesize{$[2^2]\sim [3,1^3]$:}}] {
    \begin{tikzpicture}
\node[fnode,fill=red, label=below:{\footnotesize $6$}] at (0,0) (4) {};
\node[spnode,label=below:{\footnotesize $2$}] [right=5mm of 4 ] (3) {};
\node[sonode,label=below:{\footnotesize $2$}] [right=5mm of 3 ] (2) {};
\draw (4)--(3)--(2);
\end{tikzpicture}
};
\node[align=left] at (0,+4.5) (subregular) [label=left:{\footnotesize{$[3,1]\sim[3^2]$:}}]{
    \begin{tikzpicture}
\node[fnode,fill=red, label=below:{\footnotesize $6$}] at (0,0) (4) {};
\node[spnode,label=below:{\footnotesize $2$}] [right=5mm of 4 ] (3) {};
\draw (4)--(3);
\end{tikzpicture}
};
\node[align=left] at (0,+6) (regular) [label=left:{\footnotesize{$[4]\sim [5,1]$:}}] {
    \begin{tikzpicture}
\node[fnode,fill=red, label=below:{\footnotesize $6$}] at (0,0) (4) {};
\end{tikzpicture}
};

\draw (null.north)--(minimal.south) node[midway,right] () {\footnotesize $a_3$};
\draw (minimal)--(orbit) node[midway,right] () {\footnotesize $a_1$};
\draw (orbit)--(subregular) node[midway,right] () {\footnotesize $A_1$};
\draw (subregular)--(regular) node[midway,right] () {\footnotesize $A_3$};

\end{tikzpicture}
    \caption{$T_\rho (SO(6))$.}
    \label{eqn:cone_so_6+trho}
    \end{subfigure}
    \caption{The Hasse diagram under the dominance ordering on partitions of the nilpotent orbits of $\mathfrak{su}(4)$ and $\mathfrak{so}(6)$. We also depict the associated $T_\rho$ theories and we can see that there exists both unitary and orthosymplectic decays, respectively, which reproduce the Hasse diagram from the dominance ordering.}
\end{figure}
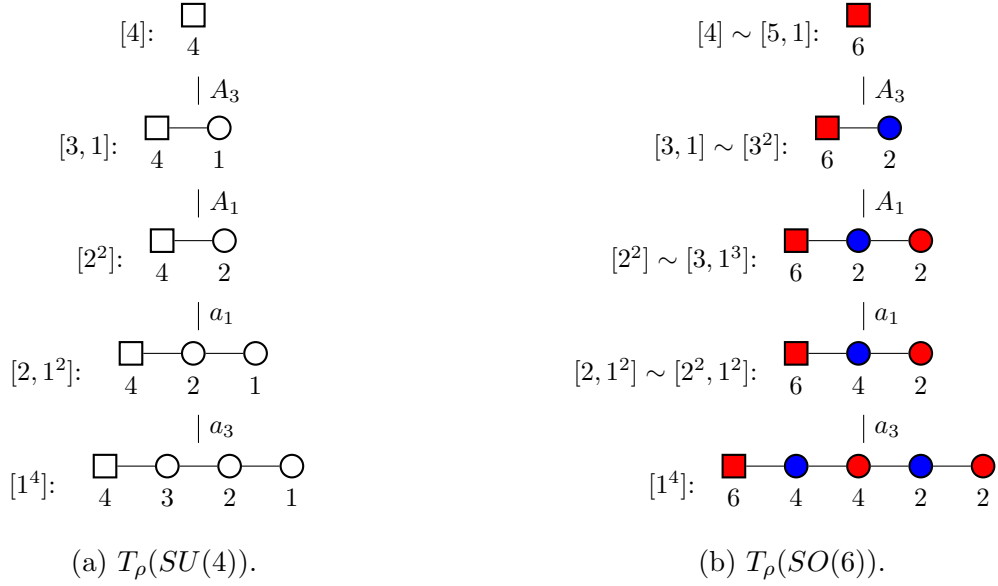

Let us focus first on just one tail of the star-shaped quiver and assume that the balance of the central $\mathfrak{so}(6)$ node is sufficiently positive that there does not exist any enhanced flavour symmetry. We assume that the puncture that we are interested in is $O_1$, and let $O_1' < O_1$ such that the balance of the $\mathfrak{so}(6)$ after replacing $O_1$ with $O_1'$ is such that the $\mathfrak{so}(6)$ node has non-negative balance. Then it is clear that the Decay and Fission algorithm of Section \ref{sec:Algorithm} allows one to decay $T_{O_1}(SO(6))$ to $T_{O_1'}(SO(6))$ following the Hasse diagram presented in Figure \ref{eqn:cone_so_6+trho}. This directly matches with the unitary quivers, as the Hasse diagrams for $\mathfrak{su}(4)$ and $\mathfrak{so}(6)$ nilpotent orbits match via the isomorphism.

\begin{figure}[p]
    \centering
        \begin{subfigure}[b]{0.45\linewidth}
        \centering
        \begin{tikzpicture}
        \node[node,label=below:{$4$}] at (0,0) (c) {}; 
        % \node[node,label=above right:{$1$}] [above right=0.5cm and 0.5cm of c] (l1r) {}; 
        % \node[node,label=above left:{$1$}] [above left=0.5cm and 0.5cm of c] (l1l) {};
        % \node[node,label=right:{$2$}] [above =0.7cm  of c] (l1t) {};
        \node[node,label=below:{$3$}] [left= 0.7cm of c] (l1ll) {};
        \node[node,label=below:{$2$}] [left= 0.7cm of l1ll] (l1lll) {};
        \node[node,label=below:{$1$}] [left= 0.7cm of l1lll] (l1llll) {};
        \node[node,label=below:{$3$}] [right= 0.7cm of c] (l1rr) {};
        \node[node,label=below:{$2$}] [right= 0.7cm of l1rr] (l1rrr) {};
        \node[node,label=below:{$1$}] [right= 0.7cm of l1rrr] (l1rrrr) {};
        \node[node,label=right:{$2$}] [above =0.7cm  of c] (l1t) {};
        \node[node,label=right:{$1$}] [above= 0.7cm of l1t] (l1tt) {};
        % \node[node,label=right:{$1$}] [above= 0.7cm of l1t] (l1tt) {};
% \node[node,label=below right:{$1$}] [below right=0.5cm and 0.5cm of c] (l1br) {}; 
        % \node[node,label=below left:{$1$}] [below left=0.5cm and 0.5cm of c] (l1bl) {};
        \draw (l1rr)--(c)--(l1t);
        \draw (c)--(l1ll);
        \draw (l1rr)--(l1rrr)--(l1rrrr) (l1ll)--(l1lll)--(l1llll);
        \draw[dashed] (l1t)--(l1tt);
    \end{tikzpicture}
        \caption{}
    \end{subfigure}
    \quad
    \begin{subfigure}[b]{0.45\linewidth}
        \centering
        \begin{tikzpicture}
        \node[node,label=below:{$4$}] at (0,0) (c) {}; 
        \node[node,label=above right:{$1$}] [above right=0.5cm and 0.5cm of c] (l1r) {}; 
        \node[node,label=above left:{$1$}] [above left=0.5cm and 0.5cm of c] (l1l) {};
        % \node[node,label=right:{$2$}] [above =0.7cm  of c] (l1t) {};
        \node[node,label=below:{$3$}] [left= 0.7cm of c] (l1ll) {};
        \node[node,label=below:{$2$}] [left= 0.7cm of l1ll] (l1lll) {};
        \node[node,label=below:{$1$}] [left= 0.7cm of l1lll] (l1llll) {};
        \node[node,label=below:{$3$}] [right= 0.7cm of c] (l1rr) {};
        \node[node,label=below:{$2$}] [right= 0.7cm of l1rr] (l1rrr) {};
        \node[node,label=below:{$1$}] [right= 0.7cm of l1rrr] (l1rrrr) {};
        % \node[node,label=right:{$1$}] [above= 0.7cm of l1t] (l1tt) {};
% \node[node,label=below right:{$1$}] [below right=0.5cm and 0.5cm of c] (l1br) {}; 
        % \node[node,label=below left:{$1$}] [below left=0.5cm and 0.5cm of c] (l1bl) {};
        \draw (l1r)--(c)--(l1l) (l1rr)--(c);
        \draw (c)--(l1ll);
        \draw (l1rr)--(l1rrr)--(l1rrrr) (l1ll)--(l1lll)--(l1llll);
    \end{tikzpicture}
        \caption{}
    \end{subfigure}
    
    \vspace{0.5cm}

    \begin{subfigure}[b]{0.45\linewidth}
        \centering
        \begin{tikzpicture}
        \node[node,label=below:{$4$}] at (0,0) (c) {}; 
        \node[node,label=above:{$2$}] [above right=0.5cm and 0.5cm of c] (l1r) {}; 
        \node[node,label=above right:{$1$}] [above right=0.5cm and 0.5cm of l1r] (l1rXXX) {}; 
        \node[node,label=above:{$2$}] [above left=0.5cm and 0.5cm of c] (l1l) {};
        \node[node,label=above left:{$1$}] [above left=0.5cm and 0.5cm of l1l] (l1lXXX) {};
        % \node[node,label=right:{$2$}] [above =0.7cm  of c] (l1t) {};
        \node[node,label=below:{$2$}] [left= 0.7cm of c] (l1ll) {};
        \node[node,label=below:{$1$}] [left= 0.7cm of l1ll] (l1lll) {};
        \node[node,label=below:{$2$}] [right= 0.7cm of c] (l1rr) {};
        \node[node,label=below:{$1$}] [right= 0.7cm of l1rr] (l1rrr) {};
        % \node[node,label=right:{$1$}] [above= 0.7cm of l1t] (l1tt) {};
% \node[node,label=below right:{$1$}] [below right=0.5cm and 0.5cm of c] (l1br) {}; 
        % \node[node,label=below left:{$1$}] [below left=0.5cm and 0.5cm of c] (l1bl) {};
        \draw (l1r)--(c)--(l1l) (l1rr)--(c);
        \draw (c)--(l1ll);
        \draw[dashed] (l1rr)--(l1rrr) (l1ll)--(l1lll) (l1lXXX)--(l1l) (l1r)--(l1rXXX);
    \end{tikzpicture}
        \caption{}
    \end{subfigure}
    \quad
    \begin{subfigure}[b]{0.45\linewidth}
        \centering
        \begin{tikzpicture}
        \node[node,label=below:{$4$}] at (0,0) (c) {}; 
        \node[node,label=above right:{$1$}] [above right=0.5cm and 0.5cm of c] (l1r) {}; 
        \node[node,label=above left:{$1$}] [above left=0.5cm and 0.5cm of c] (l1l) {};
        \node[node,label=right:{$2$}] [above =0.7cm  of c] (l1t) {};
        \node[node,label=below:{$2$}] [left= 0.7cm of c] (l1ll) {};
        \node[node,label=below:{$1$}] [left= 0.7cm of l1ll] (l1lll) {};
        \node[node,label=below:{$2$}] [right= 0.7cm of c] (l1rr) {};
        \node[node,label=below:{$1$}] [right= 0.7cm of l1rr] (l1rrr) {};
        \node[node,label=right:{$1$}] [above= 0.7cm of l1t] (l1tt) {};
% \node[node,label=below right:{$1$}] [below right=0.5cm and 0.5cm of c] (l1br) {}; 
        % \node[node,label=below left:{$1$}] [below left=0.5cm and 0.5cm of c] (l1bl) {};
        \draw (l1r)--(c)--(l1l) (l1rr)--(c)--(l1t);
        \draw (c)--(l1ll);
        \draw[dashed] (l1rr)--(l1rrr) (l1ll)--(l1lll) (l1t)--(l1tt);
    \end{tikzpicture}
        \caption{}
    \end{subfigure}

    \vspace{0.5cm}
    
        \begin{subfigure}[b]{0.45\linewidth}
        \centering
        \begin{tikzpicture}
        \node[node,label=below:{$4$}] at (0,0) (c) {}; 
        \node[node,label=above right:{$1$}] [above right=0.5cm and 0.5cm of c] (l1r) {}; 
        \node[node,label=above left:{$1$}] [above left=0.5cm and 0.5cm of c] (l1l) {};
        \node[node,label=above:{$1$}] [above =0.7cm  of c] (l1t) {};
        \node[node,label=below:{$2$}] [left= 0.7cm of c] (l1ll) {};
        \node[node,label=below:{$1$}] [left= 0.7cm of l1ll] (l1lll) {};
        \node[node,label=below:{$3$}] [right= 0.7cm of c] (l1rr) {};
        \node[node,label=below:{$2$}] [right= 0.7cm of l1rr] (l1rrr) {};
        \node[node,label=below:{$1$}] [right= 0.7cm of l1rrr] (l1rrrr) {};
% \node[node,label=below right:{$1$}] [below right=0.5cm and 0.5cm of c] (l1br) {}; 
        % \node[node,label=below left:{$1$}] [below left=0.5cm and 0.5cm of c] (l1bl) {};
        \draw (l1r)--(c)--(l1l) (l1rrrr)--(l1rrr)--(l1rr)--(c)--(l1t);
        \draw (c)--(l1ll);
        \draw[dashed] (l1ll)--(l1lll);
    \end{tikzpicture}
        \caption{}
    \end{subfigure}
        \quad
    \begin{subfigure}[b]{0.45\linewidth}
        \centering
        \begin{tikzpicture}
        \node[node,label={[xshift=-5pt,yshift=-25pt]$4$}] at (0,0) (c) {}; 
        \node[node,label=above right:{$1$}] [above right=0.5cm and 0.5cm of c] (l1r) {}; 
        \node[node,label=above left:{$1$}] [above left=0.5cm and 0.5cm of c] (l1l) {};
        \node[node,label=above:{$1$}] [above =0.7cm  of c] (l1t) {};
        \node[node,label=below:{$1$}] [left= 0.7cm of c] (l1ll) {};
        \node[node,label=below:{$1$}] [below =0.7cm  of c] (l1b) {};
        \node[node,label=below:{$3$}] [right= 0.7cm of c] (l1rr) {};
        \node[node,label=below:{$2$}] [right= 0.7cm of l1rr] (l1rrr) {};
        \node[node,label=below:{$1$}] [right= 0.7cm of l1rrr] (l1rrrr) {};
% \node[node,label=below right:{$1$}] [below right=0.5cm and 0.5cm of c] (l1br) {}; 
        % \node[node,label=below left:{$1$}] [below left=0.5cm and 0.5cm of c] (l1bl) {};
        \draw (l1r)--(c)--(l1l) (l1rrrr)--(l1rrr)--(l1rr)--(c)--(l1t);
        \draw (l1b)--(c)--(l1ll);
    \end{tikzpicture}
        \caption{}
    \end{subfigure}

    \vspace{0.5cm}
    
    \begin{subfigure}[b]{0.3\linewidth}
        \centering
        \begin{tikzpicture}
        \node[node,label={[xshift=-5pt,yshift=-25pt]$4$}] at (0,0) (c) {}; 
        \node[node,label=above right:{$1$}] [above right=0.5cm and 0.5cm of c] (l1r) {}; 
        \node[node,label=above left:{$1$}] [above left=0.5cm and 0.5cm of c] (l1l) {};
        \node[node,label=above:{$1$}] [above =0.7cm  of c] (l1t) {};
        \node[node,label=below:{$2$}] [left= 0.7cm of c] (l1ll) {};
        \node[node,label=below:{$1$}] [left= 0.7cm of l1ll] (l1lll) {};
        \node[node,label=below:{$1$}] [below =0.7cm  of c] (l1b) {};
        \node[node,label=below:{$2$}] [right= 0.7cm of c] (l1rr) {};
        \node[node,label=below:{$1$}] [right= 0.7cm of l1rr] (l1rrr) {};
        % \node[node,label=below right:{$1$}] [below right=0.5cm and 0.5cm of c] (l1br) {}; 
        % \node[node,label=below left:{$1$}] [below left=0.5cm and 0.5cm of c] (l1bl) {};
        \draw (l1r)--(c)--(l1l) (l1rr)--(c)--(l1t);
        \draw (l1b)--(c)--(l1ll);
        \draw[dashed] (l1ll)--(l1lll) (l1rrr)--(l1rr);
    \end{tikzpicture}
        \caption{}
    \end{subfigure}
    \quad
    \begin{subfigure}[b]{0.3\linewidth}
        \centering
        \begin{tikzpicture}
        \node[node,label={[xshift=-5pt,yshift=-25pt]$4$}] at (0,0) (c) {}; 
        \node[node,label=above right:{$1$}] [above right=0.5cm and 0.5cm of c] (l1r) {}; 
        \node[node,label=above left:{$1$}] [above left=0.5cm and 0.5cm of c] (l1l) {};
        \node[node,label=above:{$1$}] [above =0.7cm  of c] (l1t) {};
        \node[node,label=left:{$1$}] [left= 0.7cm of c] (l1ll) {};
        \node[node,label=below:{$1$}] [below =0.7cm  of c] (l1b) {};
        \node[node,label=below:{$2$}] [right= 0.7cm of c] (l1rr) {};
        \node[node,label=below:{$1$}] [right= 0.7cm of l1rr] (l1rrr) {};
        % \node[node,label=below right:{$1$}] [below right=0.5cm and 0.5cm of c] (l1br) {}; 
        \node[node,label=below left:{$1$}] [below left=0.5cm and 0.5cm of c] (l1bl) {};
        \draw (l1r)--(c)--(l1l) (l1rr)--(c)--(l1t);
        \draw (c)--(l1bl) (l1b)--(c)--(l1ll);
        \draw[dashed] (l1rrr)--(l1rr);
    \end{tikzpicture}
        \caption{}
    \end{subfigure}
    \quad
    \begin{subfigure}[b]{0.3\linewidth}
        \centering
        \begin{tikzpicture}
        \node[node,label={[xshift=-5pt,yshift=-25pt]$4$}] at (0,0) (c) {}; 
        \node[node,label=above right:{$1$}] [above right=0.5cm and 0.5cm of c] (l1r) {}; 
        \node[node,label=above left:{$1$}] [above left=0.5cm and 0.5cm of c] (l1l) {};
        \node[node,label=above:{$1$}] [above =0.7cm  of c] (l1t) {};
        \node[node,label=left:{$1$}] [left= 0.7cm of c] (l1ll) {};
        \node[node,label=below:{$1$}] [below =0.7cm  of c] (l1b) {};
        \node[node,label=right:{$1$}] [right= 0.7cm of c] (l1rr) {};
        \node[node,label=below right:{$1$}] [below right=0.5cm and 0.5cm of c] (l1br) {}; 
        \node[node,label=below left:{$1$}] [below left=0.5cm and 0.5cm of c] (l1bl) {};
        \draw (l1r)--(c)--(l1l) (l1rr)--(c)--(l1t);
        \draw (l1br)--(c)--(l1bl) (l1b)--(c)--(l1ll);
    \end{tikzpicture}
        \caption{}
    \end{subfigure}
    \caption{All unitary 3d mirrors for the class $\mathcal{S}$ theories of type $\mathfrak{su}(4)$ on a sphere with regular untwisted punctures and enhanced flavour symmetry. The partitions associated to the punctures can be read off from Table \ref{tbl:A3vsD3}. To prevent an unnecessary proliferation, we have used a dashed line to indicate that the attached tail may be either that of $T_{[2,1^2]}(SU(4))$ or $T_{[2^2]}(SU(4))$.}\label{fig:all_enc}
\end{figure}
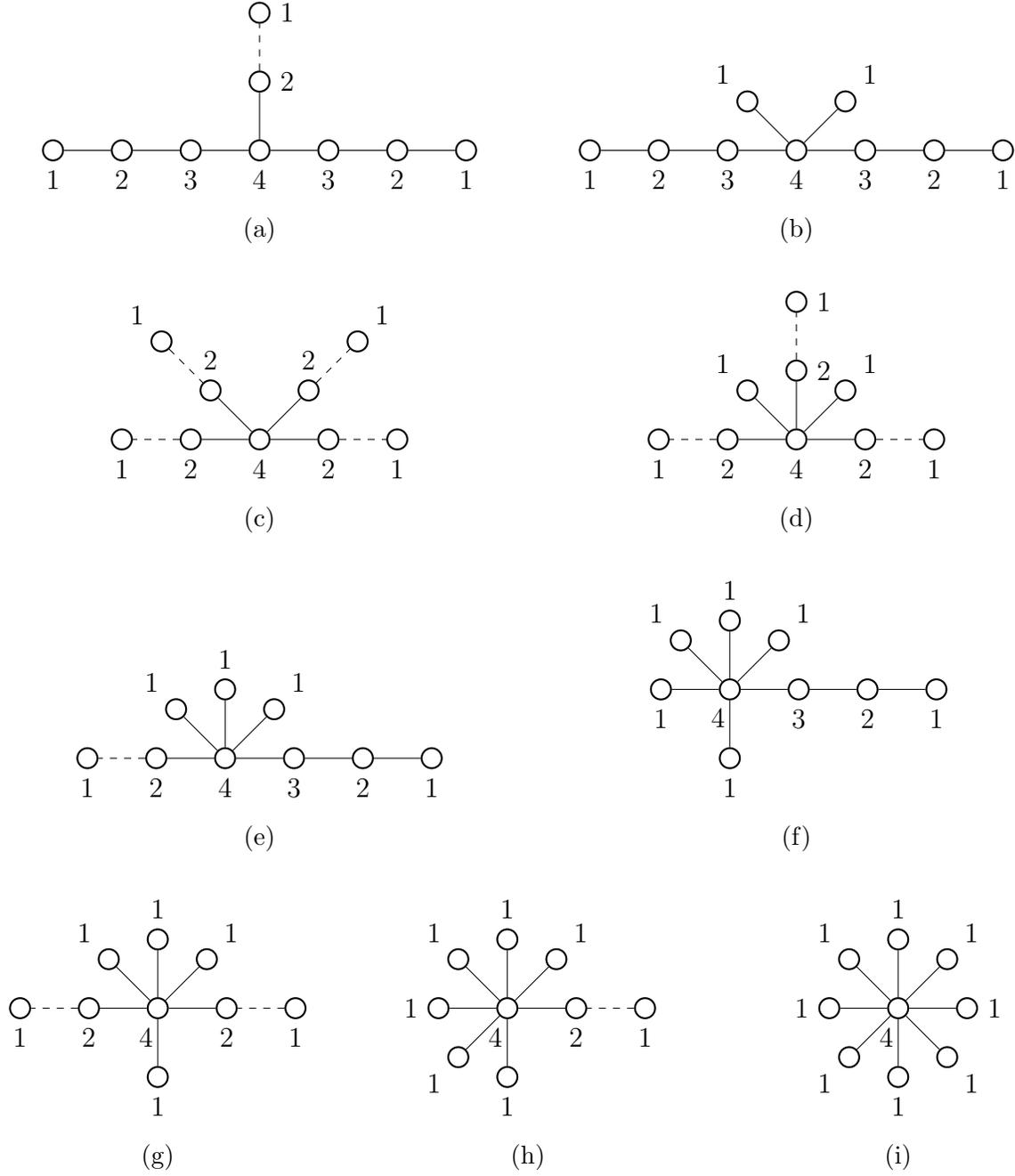

Let us now consider the cases of $\mathfrak{su}(4) \cong \mathfrak{so}(6)$
class $\mathcal{S}$ theories where there is an enhanced flavour symmetry, that
is, where the inequality in equation \eqref{eqn:flavS} is strict.
In the unitary 3d mirror of such theories, the central
$U(4)$ gauge node of the quiver becomes balanced, which leads to the additional
Coulomb symmetries. There are only a small number of examples of gaugings of
$T_{O_i}(SU(4))$ such that the $U(4)$ node is balanced, and thus we are able to
consider the cases exhaustively here. We depict each of these quivers in Figure \ref{fig:all_enc}.

\begin{figure}
    \centering
    \includegraphics[width=0.7\linewidth,page=16]{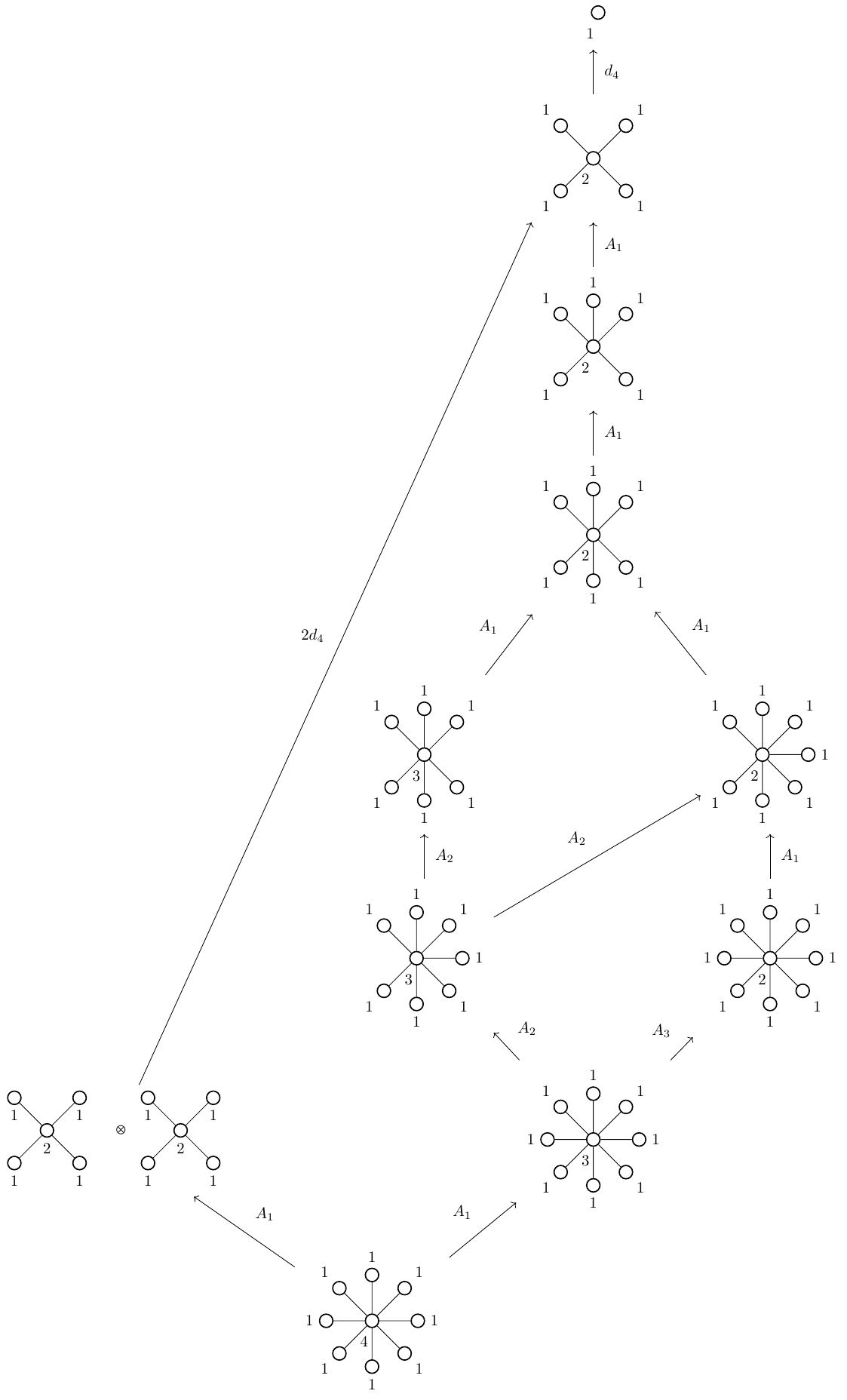}
    \caption{Applying the unitary Decay and Fission algorithm to the quiver in equation \eqref{eqn:weedle} generates this Coulomb branch Hasse diagram. This structure is reproduced by applying the Orthosymplectic Decay and Fission algorithm to the quiver in equation \eqref{eqn:pikachu}.}
    \label{fig:U(4)with8U(1)s-Hasse}
\end{figure}

First, we consider class $\mathcal{S}$ of type $\mathfrak{su}(4)$ on a sphere with eight punctures associated to the nilpotent orbits $[3,1]$. The unitary 3d mirror of this theory is
\begin{equation}\label{eqn:weedle}
    \begin{gathered}\begin{tikzpicture}
        \node[node,label={[xshift=-5pt,yshift=-25pt]$4$}] at (0,0) (c) {}; 
        \node[node,label=above right:{$1$}] [above right=0.5cm and 0.5cm of c] (l1r) {}; 
        \node[node,label=above left:{$1$}] [above left=0.5cm and 0.5cm of c] (l1l) {};
        \node[node,label=above:{$1$}] [above =0.7cm  of c] (l1t) {};
        \node[node,label=left:{$1$}] [left= 0.7cm of c] (l1ll) {};
        \node[node,label=below:{$1$}] [below =0.7cm  of c] (l1b) {};
        \node[node,label=right:{$1$}] [right= 0.7cm of c] (l1rr) {};
        \node[node,label=below right:{$1$}] [below right=0.5cm and 0.5cm of c] (l1br) {}; 
        \node[node,label=below left:{$1$}] [below left=0.5cm and 0.5cm of c] (l1bl) {};
        \draw (l1r)--(c)--(l1l) (l1rr)--(c)--(l1t);
        \draw (l1br)--(c)--(l1bl) (l1b)--(c)--(l1ll);
    \end{tikzpicture}\end{gathered} \quad\,,
\end{equation}
and the structure of the Higgs branch, as obtained from performing Decay and Fission on the unitary 3d mirror, is given in Figure \ref{fig:U(4)with8U(1)s-Hasse}. Using the isomorphism between $\mathfrak{su}(4)$ and $\mathfrak{so}(6)$, we can see that there also exists an orthosymplectic 3d mirror of this class $\mathcal{S}$ theory, which is
\begin{equation}\label{eqn:pikachu}
    \begin{gathered}\begin{tikzpicture}
        \node[sonode,label={[xshift=-5pt,yshift=-25pt]$6$}] at (0,0) (c) {}; 
        \node[spnode,label=above right:{$2$}] [above right=0.5cm and 0.5cm of c] (l1r) {}; 
        \node[spnode,label=above left:{$2$}] [above left=0.5cm and 0.5cm of c] (l1l) {};
        \node[spnode,label=above:{$2$}] [above =0.7cm  of c] (l1t) {};
        \node[spnode,label=left:{$2$}] [left= 0.7cm of c] (l1ll) {};
        \node[spnode,label=below:{$2$}] [below =0.7cm  of c] (l1b) {};
        \node[spnode,label=right:{$2$}] [right= 0.7cm of c] (l1rr) {};
        \node[spnode,label=below right:{$2$}] [below right=0.5cm and 0.5cm of c] (l1br) {}; 
        \node[spnode,label=below left:{$2$}] [below left=0.5cm and 0.5cm of c] (l1bl) {};
        \draw (l1r)--(c)--(l1l) (l1rr)--(c)--(l1t);
        \draw (l1br)--(c)--(l1bl) (l1b)--(c)--(l1ll);
    \end{tikzpicture}\end{gathered} \quad\,.
\end{equation}

The first transition in the Coulomb branch of the quiver in equation \eqref{eqn:pikachu} involves unitarization, where
\begin{equation}
    \mathfrak{so}(6) \,\, \rightarrow \,\, \mathfrak{u}(3) \,, \qquad
    \mathfrak{usp}(2) \,\, \rightarrow \,\, \mathfrak{u}(1) \,.
\end{equation}
This leads to
\begin{equation}
    \begin{gathered}\begin{tikzpicture}
        \node[sonode,label={[xshift=-5pt,yshift=-25pt]$6$}] at (0,0) (c) {}; 
        \node[spnode,label=above right:{$2$}] [above right=0.5cm and 0.5cm of c] (l1r) {}; 
        \node[spnode,label=above left:{$2$}] [above left=0.5cm and 0.5cm of c] (l1l) {};
        \node[spnode,label=above:{$2$}] [above =0.7cm  of c] (l1t) {};
        \node[spnode,label=left:{$2$}] [left= 0.7cm of c] (l1ll) {};
        \node[spnode,label=below:{$2$}] [below =0.7cm  of c] (l1b) {};
        \node[spnode,label=right:{$2$}] [right= 0.7cm of c] (l1rr) {};
        \node[spnode,label=below right:{$2$}] [below right=0.5cm and 0.5cm of c] (l1br) {}; 
        \node[spnode,label=below left:{$2$}] [below left=0.5cm and 0.5cm of c] (l1bl) {};
        \draw (l1r)--(c)--(l1l) (l1rr)--(c)--(l1t);
        \draw (l1br)--(c)--(l1bl) (l1b)--(c)--(l1ll);
        \end{tikzpicture}\end{gathered}
        \quad\qquad \longrightarrow \quad\qquad 
        \begin{gathered}\begin{tikzpicture}
        \node[node,label={[xshift=-5pt,yshift=-25pt]$3$}] at (0,0) (c) {}; 
        \node[node,label=above right:{$1$}] [above right=0.5cm and 0.5cm of c] (l1r) {}; 
        \node[node,label=above left:{$1$}] [above left=0.5cm and 0.5cm of c] (l1l) {};
        \node[node,label=above:{$1$}] [above =0.7cm  of c] (l1t) {};
        \node[node,label=left:{$1$}] [left= 0.7cm of c] (l1ll) {};
        \node[node,label=below:{$1$}] [below =0.7cm  of c] (l1b) {};
        \node[node,label=right:{$1$}] [right= 0.7cm of c] (l1rr) {};
        \node[node,label=below right:{$1$}] [below right=0.5cm and 0.5cm of c] (l1br) {}; 
        \node[node,label=below left:{$1$}] [below left=0.5cm and 0.5cm of c] (l1bl) {};
        \draw (l1r)--(c)--(l1l) (l1rr)--(c)--(l1t);
        \draw (l1br)--(c)--(l1bl) (l1b)--(c)--(l1ll);
    \end{tikzpicture}\end{gathered} \qquad\,,
\end{equation}
which manifestly reproduces one of the two first steps in Figure \ref{fig:U(4)with8U(1)s-Hasse}, and since the result of the unitarization is a unitary quiver, the further Decay and Fission evidently leads to an identical structure. The second allowed Decay and Fission of equation \eqref{eqn:pikachu} again involves a fission step, in particular:
\begin{equation}
    \mathfrak{so}(6) \,\, \rightarrow \,\, \mathfrak{usp}(2) \oplus \mathfrak{u}(2) \,, \qquad
    \mathfrak{usp}(2) \,\, \rightarrow \,\, \mathfrak{so}(2) \cong \mathfrak{u}(1) \,.
\end{equation}
It is easy to see that there is only one way to attach the eight $SO(2) \cong U(1)$ nodes to the $USp(2)$ and $U(2)$ central nodes such that the resulting quivers are sufficiently good, following the notions of goodness that were explained in Section \ref{sec:Algorithm}.\footnote{Note that the star-shaped quiver with $\mathfrak{usp}(2)$ in the center and three $\mathfrak{so}(2)$ legs has a non-local monopole operator that decouples, despite each of the four gauge nodes being locally good. As such, it is ruled out.} Therefore, we obtain the following transition:
\begin{equation}\label{eqn:raichu}
    \begin{gathered}\begin{tikzpicture}
        \node[sonode,label={[xshift=-5pt,yshift=-25pt]$6$}] at (0,0) (c) {}; 
        \node[spnode,label=above right:{$2$}] [above right=0.5cm and 0.5cm of c] (l1r) {}; 
        \node[spnode,label=above left:{$2$}] [above left=0.5cm and 0.5cm of c] (l1l) {};
        \node[spnode,label=above:{$2$}] [above =0.7cm  of c] (l1t) {};
        \node[spnode,label=left:{$2$}] [left= 0.7cm of c] (l1ll) {};
        \node[spnode,label=below:{$2$}] [below =0.7cm  of c] (l1b) {};
        \node[spnode,label=right:{$2$}] [right= 0.7cm of c] (l1rr) {};
        \node[spnode,label=below right:{$2$}] [below right=0.5cm and 0.5cm of c] (l1br) {}; 
        \node[spnode,label=below left:{$2$}] [below left=0.5cm and 0.5cm of c] (l1bl) {};
        \draw (l1r)--(c)--(l1l) (l1rr)--(c)--(l1t);
        \draw (l1br)--(c)--(l1bl) (l1b)--(c)--(l1ll);
        \end{tikzpicture}\end{gathered}
        \qquad \longrightarrow \qquad 
        \begin{gathered}\begin{tikzpicture}
        \node[node,label=below:{$2$}] at (0,0) (c) {}; 
        \node[node,label=above right:{$1$}] [above right=0.5cm and 0.5cm of c] (l1r) {}; 
        \node[node,label=above left:{$1$}] [above left=0.5cm and 0.5cm of c] (l1l) {};
        \node[node,label=below right:{$1$}] [below right=0.5cm and 0.5cm of c] (l1br) {}; 
        \node[node,label=below left:{$1$}] [below left=0.5cm and 0.5cm of c] (l1bl) {};
        \draw (l1r)--(c)--(l1l);
        \draw (l1br)--(c)--(l1bl);
    \end{tikzpicture}\end{gathered} \quad \otimes \quad \begin{gathered}\begin{tikzpicture}
        \node[spnode,label=below:{$2$}] at (0,0) (c) {}; 
        \node[sonode,label=above right:{$2$}] [above right=0.5cm and 0.5cm of c] (l1r) {}; 
        \node[sonode,label=above left:{$2$}] [above left=0.5cm and 0.5cm of c] (l1l) {};
        \node[sonode,label=below right:{$2$}] [below right=0.5cm and 0.5cm of c] (l1br) {}; 
        \node[sonode,label=below left:{$2$}] [below left=0.5cm and 0.5cm of c] (l1bl) {};
        \draw (l1r)--(c)--(l1l);
        \draw (l1br)--(c)--(l1bl);
    \end{tikzpicture}\end{gathered} \quad \,.
\end{equation}
It is already well-known that both quivers on the right-hand side have
identical Coulomb branches, which, as symplectic singularities, are just the
closure of the minimal nilpotent orbit of $\mathfrak{so}(8)$. It remains to
argue that the rightmost quiver in equation \eqref{eqn:raichu}, under the
Orthosymplectic Decay and Fission algorithm, has a Coulomb branch consisting of
only a single slice which leads to a trivial quiver. We can see this, for
example, by noting that a unitarization step would lead to a quiver with no
remaining gauge nodes, and that none of the other rules presented in Section
\ref{sec:Algorithm} lead to a non-trivial quiver. Therefore, we have reproduced
the full structure (given in Figure \ref{fig:U(4)with8U(1)s-Hasse}) of the
Coulomb branch of the unitary quiver in equation \eqref{eqn:weedle} starting
from the orthosymplectic quiver in equation \eqref{eqn:pikachu}, and applying
the algorithm for Orthosymplectic Decay and Fission. 

\begin{figure}
    \centering
    \includegraphics[width=0.45\linewidth,page=17]{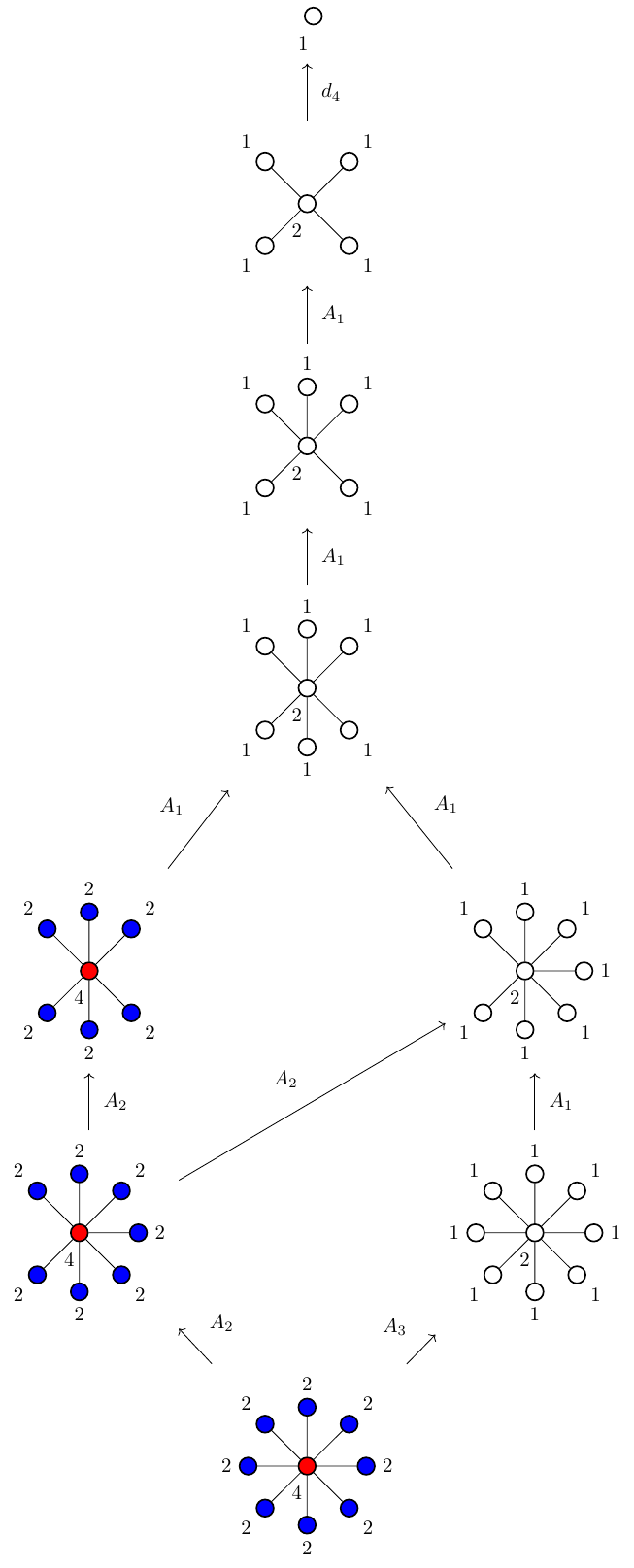}
    \caption{Applying the Orthosymplectic Decay and Fission algorithm to the quiver on the right in equation \eqref{eqn:ditto} we find this Coulomb branch Hasse diagram. This is identical to the subdiagram of Figure \ref{fig:U(4)with8U(1)s-Hasse} starting from the quiver on the left in equation \eqref{eqn:ditto}.}
    \label{fig:SO(4)with8USp(2)s-Hasse}
\end{figure}

Applying the Orthosymplectic Decay and Fission algorithm of Section
\ref{sec:Algorithm}, it would appear that there exists a third consistent
transition:
\begin{equation}\label{eqn:ditto}
    \begin{gathered}\begin{tikzpicture}
        \node[sonode,label={[xshift=-5pt,yshift=-25pt]$6$}] at (0,0) (c) {}; 
        \node[spnode,label=above right:{$2$}] [above right=0.5cm and 0.5cm of c] (l1r) {}; 
        \node[spnode,label=above left:{$2$}] [above left=0.5cm and 0.5cm of c] (l1l) {};
        \node[spnode,label=above:{$2$}] [above =0.7cm  of c] (l1t) {};
        \node[spnode,label=left:{$2$}] [left= 0.7cm of c] (l1ll) {};
        \node[spnode,label=below:{$2$}] [below =0.7cm  of c] (l1b) {};
        \node[spnode,label=right:{$2$}] [right= 0.7cm of c] (l1rr) {};
        \node[spnode,label=below right:{$2$}] [below right=0.5cm and 0.5cm of c] (l1br) {}; 
        \node[spnode,label=below left:{$2$}] [below left=0.5cm and 0.5cm of c] (l1bl) {};
        \draw (l1r)--(c)--(l1l) (l1rr)--(c)--(l1t);
        \draw (l1br)--(c)--(l1bl) (l1b)--(c)--(l1ll);
        \end{tikzpicture}\end{gathered}
        \qquad \longrightarrow \qquad 
        \begin{gathered}\begin{tikzpicture}
        \node[sonode,label={[xshift=-5pt,yshift=-25pt]$4$}] at (0,0) (c) {}; 
        \node[spnode,label=above right:{$2$}] [above right=0.5cm and 0.5cm of c] (l1r) {}; 
        \node[spnode,label=above left:{$2$}] [above left=0.5cm and 0.5cm of c] (l1l) {};
        \node[spnode,label=above:{$2$}] [above =0.7cm  of c] (l1t) {};
        \node[spnode,label=left:{$2$}] [left= 0.7cm of c] (l1ll) {};
        \node[spnode,label=below:{$2$}] [below =0.7cm  of c] (l1b) {};
        \node[spnode,label=right:{$2$}] [right= 0.7cm of c] (l1rr) {};
        \node[spnode,label=below right:{$2$}] [below right=0.5cm and 0.5cm of c] (l1br) {}; 
        \node[spnode,label=below left:{$2$}] [below left=0.5cm and 0.5cm of c] (l1bl) {};
        \draw (l1r)--(c)--(l1l) (l1rr)--(c)--(l1t);
        \draw (l1br)--(c)--(l1bl) (l1b)--(c)--(l1ll);
        \end{tikzpicture}\end{gathered} \quad\,.
\end{equation}
How can we explain the presence of this transition when we know from the
unitary quiver description that there exists only two inequivalent elementary
transitions from the theory in equation \eqref{eqn:pikachu}? In fact, this is
an example of one of the properties of the Orthosymplectic Decay and Fission
algorithm highlighted in Section \ref{sec:Algorithm}: specifically, the
algorithm produces sometimes two different quivers describing the same
sublocus in the Coulomb branch. In this case, the Coulomb branches of 
\begin{equation}\label{eqn:dittoCB}
    \begin{gathered}\begin{tikzpicture}
        \node[node,label={[xshift=-5pt,yshift=-25pt]$3$}] at (0,0) (c) {}; 
        \node[node,label=above right:{$1$}] [above right=0.5cm and 0.5cm of c] (l1r) {}; 
        \node[node,label=above left:{$1$}] [above left=0.5cm and 0.5cm of c] (l1l) {};
        \node[node,label=above:{$1$}] [above =0.7cm  of c] (l1t) {};
        \node[node,label=left:{$1$}] [left= 0.7cm of c] (l1ll) {};
        \node[node,label=below:{$1$}] [below =0.7cm  of c] (l1b) {};
        \node[node,label=right:{$1$}] [right= 0.7cm of c] (l1rr) {};
        \node[node,label=below right:{$1$}] [below right=0.5cm and 0.5cm of c] (l1br) {}; 
        \node[node,label=below left:{$1$}] [below left=0.5cm and 0.5cm of c] (l1bl) {};
        \draw (l1r)--(c)--(l1l) (l1rr)--(c)--(l1t);
        \draw (l1br)--(c)--(l1bl) (l1b)--(c)--(l1ll);
    \end{tikzpicture}\end{gathered}
        \qquad \text{ and } \qquad 
        \begin{gathered}\begin{tikzpicture}
        \node[sonode,label={[xshift=-5pt,yshift=-25pt]$4$}] at (0,0) (c) {}; 
        \node[spnode,label=above right:{$2$}] [above right=0.5cm and 0.5cm of c] (l1r) {}; 
        \node[spnode,label=above left:{$2$}] [above left=0.5cm and 0.5cm of c] (l1l) {};
        \node[spnode,label=above:{$2$}] [above =0.7cm  of c] (l1t) {};
        \node[spnode,label=left:{$2$}] [left= 0.7cm of c] (l1ll) {};
        \node[spnode,label=below:{$2$}] [below =0.7cm  of c] (l1b) {};
        \node[spnode,label=right:{$2$}] [right= 0.7cm of c] (l1rr) {};
        \node[spnode,label=below right:{$2$}] [below right=0.5cm and 0.5cm of c] (l1br) {}; 
        \node[spnode,label=below left:{$2$}] [below left=0.5cm and 0.5cm of c] (l1bl) {};
        \draw (l1r)--(c)--(l1l) (l1rr)--(c)--(l1t);
        \draw (l1br)--(c)--(l1bl) (l1b)--(c)--(l1ll);
        \end{tikzpicture}\end{gathered} \quad \,,
\end{equation}
are identical.\footnote{In this case, the orthosymplectic quiver has bad
$USp(2)$ nodes, and thus we cannot use the monopole formula to
compute the Coulomb branch Hilbert series and compare it to that of the unitary
quiver. Here we use the weaker condition that the Coulomb branch Hasse diagrams
are identical, though this is not in general sufficient to demonstrate that the
Coulomb branches themselves are identical.} It is straightforward to observe
this by comparing the sub-graph of Figure \ref{fig:U(4)with8U(1)s-Hasse}
starting from the unitary quiver on the left (generated by the unitary decay
and fission algorithm) with the Coulomb branch Hasse diagram of the
orthosymplectic quiver on the right (generated by the orthosymplectic decay and
fission algorithm) which we depict in Figure \ref{fig:SO(4)with8USp(2)s-Hasse}.

There is one further subtlety in the derivation of the Hasse diagram in Figure \ref{fig:SO(4)with8USp(2)s-Hasse} that we briefly highlight here. Since we allow $USp(2)$ gauge nodes with balance $-1$ in the algorithm in Section \ref{sec:Algorithm}, we can no longer use standard notions of badness to know when a quiver should not appear in the Decay and Fission algorithm. Instead, we ask when the corresponding 4d class $\mathcal{S}$ theory is good and use that as a diagnostic to rule out certain 3d quivers. Consider class $\mathcal{S}$ of type $\mathfrak{so}(4)$. Since the partition $[2^2]$ of $4$ is very even,\footnote{A very even partition has only even entries, each with even multiplicity.} it is associated to two distinct nilpotent orbits, which we label as
\begin{equation}
    [2^2]_I \qquad \text { and } \qquad [2^2]_{II} \,.
\end{equation}
While these two different nilpotent orbits generally lead to different class $\mathcal{S}$ theories with different Higgs branch chiral rings (see \cite{Distler:2022yse} for the explicit difference in the half-BPS operator spectrum), the naive 3d mirror appears to be agnostic to this, since
\begin{equation}
    T_{[2^2]_I}(SO(4)) = T_{[2^2]_{II}}(SO(4)) = \quad \begin{gathered}
       \raisebox{-0.5\height}{ \begin{tikzpicture}
\node[fnode,fill=red, label=below:{\footnotesize $4$}] at (0,0) (4) {};
\node[spnode,label=below:{\footnotesize $2$}] [right=5mm of 4 ] (3) {};
\draw (4)--(3);
\end{tikzpicture}} \quad \,.
    \end{gathered} 
\end{equation}
From the class $\mathcal{S}$ perspective one can compute the Hall--Littlewood index \cite{Gadde:2011uv,Kinney:2005ej,Gadde:2009kb,Gadde:2011ik,Lemos:2012ph}  which, for genus zero with regular untwisted punctures \cite{Kang:2022zsl}, counts Higgs branch operators. Schematically, it is determined as 
\begin{equation}
    \operatorname{HL}(\tau) \,\,=\,\, \operatorname{Tr}_{\mathcal{H}_{\operatorname{HL}}} \tau^{2(\Delta - R)} (-1)^F \,\,=\,\, \sum_{\Lambda} \cdots  \,,
\end{equation}
where the sum runs over all irreducible representations of $\mathfrak{g}$, and $\cdots$ represents a complicated product of Hall--Littlewood polynomials over the punctures. For the details we refer to \cite{Chacaltana:2018vhp,Distler:2022yse} where a straightforward prescription was given to determine the lowest order $n_\Lambda$ that the representation $\Lambda$ contributes to the Hall--Littlewood index; i.e., Hall--Littlewood index contains a term $\tau^{n_\Lambda}$. If $n_\Lambda < 1$, for any $\Lambda$, then the theory is bad. Consider now $\mathfrak{so}(4)$ class $\mathcal{S}$ on a sphere with $n_R$ punctures $[2^2]_I$ and $n_B$ punctures $[2^2]_{II}$; the two spinor representations of $\mathfrak{so}(4)$ contribute to the index at the following orders:
\begin{equation}\label{eqn:balQ}
    n_{\bm{2}} = -2 + n_B \qquad \text{ and } \qquad n_{\bm{2}'} = -2 + n_R \,.
\end{equation}
As we can see, it is necessary to have at least six punctures, three $[2^2]_I$ and three $[2^2]_{II}$ for the class $\mathcal{S}$ theory to be not bad. Therefore, we propose that the 3d mirror of such a bad 4d theory is also bad, and should be removed from the results of the Decay and Fission algorithm.
With this computation in hand, it is tempting the suggest an interpretation directly in the 3d Lagrangian quiver: if we consider 
\begin{equation}
    \mathfrak{so}(4) \cong \mathfrak{su}(2) \oplus \mathfrak{su}(2) \,,
\end{equation}
and let $n_R$ and $n_B$ denote the number of fundamental flavour attached to the two $\mathfrak{su}(2)$ factors, then equation \eqref{eqn:balQ} is precisely the balances of the $\mathfrak{su}(2)$ factors, respectively. It is not clear why this \emph{should} be the correct interpretation, and since it is orthogonal to the analysis of this paper, we leave a further exploration for future work.

In the example, we might have expected, from the orthosymplectic quiver, that there exists another transition:
\begin{equation}\label{eqn:spearow}
    \begin{gathered}\begin{tikzpicture}
        \node[sonode,label={[xshift=-5pt,yshift=-25pt]$6$}] at (0,0) (c) {}; 
        \node[spnode,label=above right:{$2$}] [above right=0.5cm and 0.5cm of c] (l1r) {}; 
        \node[spnode,label=above left:{$2$}] [above left=0.5cm and 0.5cm of c] (l1l) {};
        \node[spnode,label=above:{$2$}] [above =0.7cm  of c] (l1t) {};
        \node[spnode,label=left:{$2$}] [left= 0.7cm of c] (l1ll) {};
        \node[spnode,label=below:{$2$}] [below =0.7cm  of c] (l1b) {};
        \node[spnode,label=right:{$2$}] [right= 0.7cm of c] (l1rr) {};
        \node[spnode,label=below right:{$2$}] [below right=0.5cm and 0.5cm of c] (l1br) {}; 
        \node[spnode,label=below left:{$2$}] [below left=0.5cm and 0.5cm of c] (l1bl) {};
        \draw (l1r)--(c)--(l1l) (l1rr)--(c)--(l1t);
        \draw (l1br)--(c)--(l1bl) (l1b)--(c)--(l1ll);
        \end{tikzpicture}\end{gathered}
        \qquad \longrightarrow \qquad 
        \begin{gathered}\begin{tikzpicture}
        \node[sonode,label={[xshift=-5pt,yshift=-25pt]$6$}] at (0,0) (c) {}; 
        \node[spnode,label=above right:{$2$}] [above right=0.5cm and 0.5cm of c] (l1r) {}; 
        \node[spnode,label=above left:{$2$}] [above left=0.5cm and 0.5cm of c] (l1l) {};
        \node[spnode,label=above:{$2$}] [above =0.7cm  of c] (l1t) {};
        % \node[spnode,label=left:{$2$}] [left= 0.7cm of c] (l1ll) {};
        \node[spnode,label=below:{$2$}] [below =0.7cm  of c] (l1b) {};
        \node[spnode,label=right:{$2$}] [right= 0.7cm of c] (l1rr) {};
        \node[spnode,label=below right:{$2$}] [below right=0.5cm and 0.5cm of c] (l1br) {}; 
        \node[spnode,label=below left:{$2$}] [below left=0.5cm and 0.5cm of c] (l1bl) {};
        \draw (l1r)--(c)--(l1l) (l1rr)--(c)--(l1t);
        \draw (l1br)--(c)--(l1bl) (l1b)--(c);
        \end{tikzpicture}\end{gathered} \,.
\end{equation}
On the right-hand-side of equation \eqref{eqn:spearow}, each of the
$\mathfrak{usp}(2)$ and $\mathfrak{so}(6)$ gauge nodes have non-negative
balance, in the conventional sense of \cite{Gaiotto:2008ak}. Instead, if we use
the rules from Section \ref{sec:Algorithm}, each isolated $USp(2)$
should contribute to the balance of the $SO(6)$ gauge node as $5/8$
full hypermultiplets in the $\bm{6}$ of $\mathfrak{so}(6)$ (i.e., \texttt{Rule
1}). Thus, the effective balance of the $SO(6)$ node should be
considered as
\begin{equation}
    b = 7 \times \frac{5}{8} - 6 + 1 = - \frac{5}{8} < 0 \,.
\end{equation}
Since the modified balance of the special orthogonal node is negative, it is
treated as bad and this configuration is ruled out. Indeed, this is a good
opportunity to explain the motivation for the modified balance contribution in
\texttt{Rule 1}. While each of the nodes is individually good, there still
exists a monopole operator which crosses the unitarity bound; this operator
arises from the half-integer lattice.\footnote{Notice that although it stays true for $n\le 7$ that a monopole operator arising from the half-integer lattice crosses the unitarity bound, only for $n=7$, the theory we get is, in fact, free.} We consider first the general theory of a
$SO(2K)$ gauge node attached to $n$ $USp(2)$ gauge node: 
\begin{equation}
\begin{gathered}
    \begin{tikzpicture}
        \node[sonode,label=below:{$2K$}] at (0,0) (c) {}; 
        \node[spnode,label=above:{$2$}] [above right=0.5cm and 1cm of c] (l1r) {};
        \node[label=right:{$\vdots$}] [right=0.6cm of c] (dots) {};
        \node[spnode,label=below:{$2$}] [below right=0.5cm and 1cm of c] (l1l) {};
\draw[decorate,decoration={brace, raise=1ex}] (l1r.east)--(l1l.east) node[midway,label={[xshift=30pt,yshift=-5pt]$n$ nodes}] () {};
        \draw (l1r)--(c)--(l1l);

    \end{tikzpicture} 
    \end{gathered}\,.
\end{equation}
The minimally-charged monopole operator inside of the half-integer lattice
crosses the unitarity bound if $n < 8$ for any $K \geq 3$, except $K = 4$; in
the latter case the bound is $n < 7$. We can see that the condition for the
monopole not to decouple when $\mathfrak{so}(6)$ is imposed by \texttt{rule 1}.

We now consider another $\mathfrak{su}(4) \cong \mathfrak{so}(6)$ class
$\mathcal{S}$ theory with enhanced flavour symmetry from Figure
\ref{fig:all_enc}. We turn to the theory
\begin{equation}
	\mathcal{S}_{\mathfrak{su}_4}\langle C_{0,3} \rangle \{ [1^4], [1^4],[2,1^2] \} \,,
\end{equation}
which has the following unitary 3d mirror:
\begin{equation}\label{eqn:ClassSHiggsableE6E7a7}
    \begin{gathered}
        \includegraphics[page=4,scale=0.9]{DecayAndFission_figures.pdf}
    \end{gathered}\,.
\end{equation}
We can see that this quiver manifests an $\mathfrak{su}(8) \oplus
\mathfrak{su}(2)$ Coulomb symmetry. Applying the unitary Decay and Fission
algorithm, we see that there are two inequivalent minimal Higgsings of this
quiver, and we depict the full Hasse diagram of the Coulomb branch in Figure
\ref{fig:ClassSHiggsableE6E7a7HasseUnitary}. Since this is an
$\mathfrak{su}(4)$ class $\mathcal{S}$ theory, there is an isomorphic
$\mathfrak{so}(6)$ class $\mathcal{S}$ theory, which is
\begin{equation}
	\mathcal{S}_{\mathfrak{so}_6}\langle C_{0,3} \rangle \{ [1^6], [1^6],[2^2,1^2] \} \,.
\end{equation}
The 3d mirror from this perspective is the following orthosymplectic quiver:
\begin{equation}\label{eqn:ClassSHiggsableE6E7a7SO}
    \begin{gathered}
        \includegraphics[page=6,scale=0.8]{DecayAndFission_figures.pdf}
    \end{gathered}\,.
\end{equation}
We can similarly identify the enhanced $\mathfrak{su}(8) \oplus
\mathfrak{su}(2)$ flavour symmetry from the balanced nodes in this quiver.

\afterpage{
\begin{landscape}
\pagestyle{empty}
\begin{figure}[p]
\begin{subfigure}[b]{0.48\linewidth}
        \centering
        \includegraphics[page=5,width=1.0\linewidth]{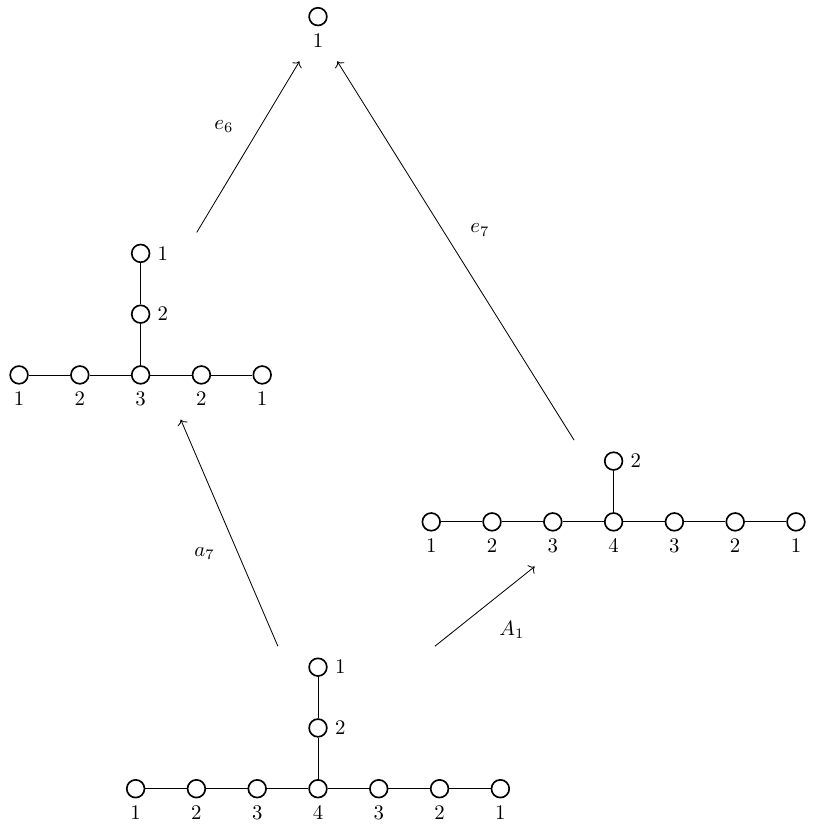}
        \caption{}
        \label{fig:ClassSHiggsableE6E7a7HasseUnitary}
\end{subfigure}
\hspace{0.5cm}
\begin{subfigure}[b]{0.48\linewidth}
        \centering
        \includegraphics[page=7,width=1.0\linewidth]{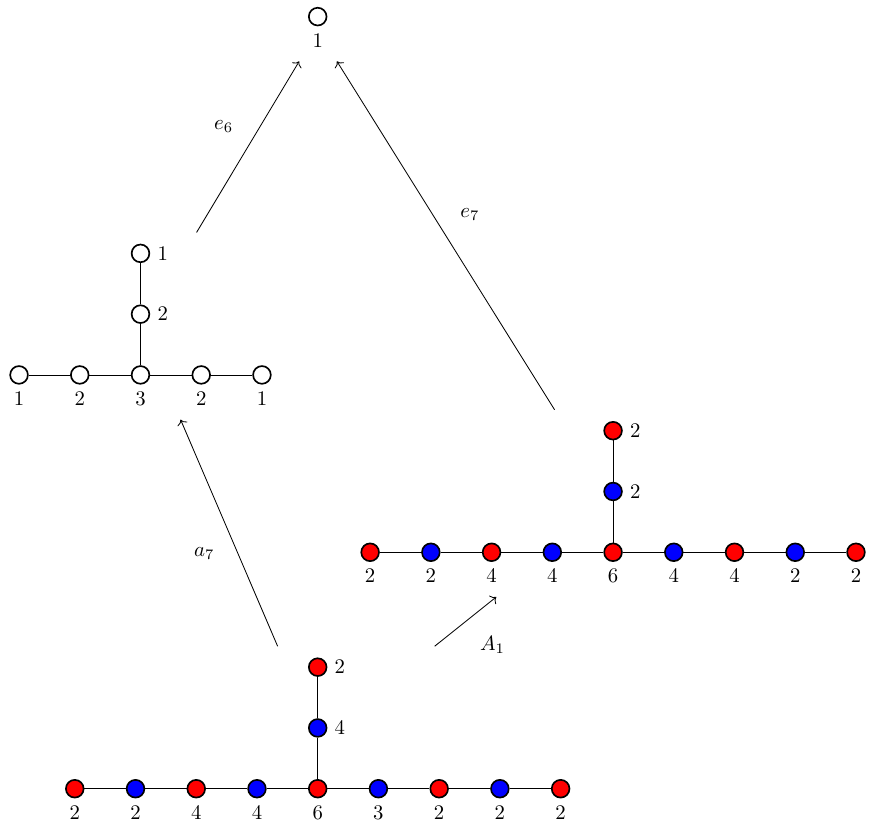}
        \caption{}
        \label{fig:ClassSHiggsableE6E7a7HasseOrtho}
    \end{subfigure}
    \caption{The Higgs branch Hasse diagram of $\mathcal{S}_{\mathfrak{su}_4}\langle C_{0,3} \rangle \{ [1^4], [1^4],[2,1^2] \} \cong \mathcal{S}_{\mathfrak{so}_6}\langle C_{0,3} \rangle \{ [1^6], [1^6],[2^2,1^2] \}$ derived via Decay and Fission in (a) through the unitary 3d mirror and in (b) through the orthosymplectic 3d mirror.}
\end{figure} %side-by-side on rotated page
\end{landscape}}

We now wish to discuss the Orthosymplectic Decay and Fission algorithm from
Section \ref{sec:Algorithm} as applied to the quiver in equation
\eqref{eqn:ClassSHiggsableE6E7a7SO}. The first transition to consider is a
unitarisation step:
\begin{equation}
	\begin{gathered}
		\resizebox{0.3\linewidth}{!}{\begin{tikzpicture}
    \node[sonode,label=below:{$2$}] at (0,0) (1) {};
    \node[spnode,label=below:{$2$}] [right=0.7cm of 1] (2) {};
    \node[sonode,label=below:{$4$}] [right=0.7cm of 2] (3) {};
    \node[spnode,label=below:{$4$}] [right=0.7cm of 3] (4) {};
    \node[sonode,label=below:{$6$}] [right=0.7cm of 4] (5) {};
    \node[spnode,label=below:{$4$}] [right=0.7cm of 5] (4r) {};
    \node[sonode,label=below:{$4$}] [right=0.7cm of 4r] (3r) {};
    \node[spnode,label=below:{$2$}] [right=0.7cm of 3r] (2r) {};
    \node[sonode,label=below:{$2$}] [right=0.7cm of 2r] (1r) {};
    \node[spnode,label=right:{$4$}] [above=0.7cm of 5] (2t) {};
    \node[sonode,label=right:{$2$}] [above=0.7cm of 2t] (1t) {};

\draw (1)--(2) (2)--(3) (3)--(4) (4)--(5) (5)--(4r) (4r)--(3r) (3r)--(2r) (2r)--(1r) (5)--(2t) (2t)--(1t);
\end{tikzpicture}}
	\end{gathered} \qquad \rightarrow \qquad
    \begin{gathered}
		\resizebox{0.2\linewidth}{!}{\begin{tikzpicture}
    \node[node,label=below:{$1$}] at (0,0) (1) {};
    \node[node,label=below:{$2$}] [right=0.7cm of 1] (2) {};
    \node[node,label=below:{$3$}] [right=0.7cm of 2] (3) {};
    \node[node,label=below:{$2$}] [right=0.7cm of 3] (2r) {};
    \node[node,label=below:{$1$}] [right=0.7cm of 2r] (1r) {};
    \node[node,label=right:{$2$}] [above=0.7cm of 3] (2t) {};
    \node[node,label=right:{$1$}] [above=0.7cm of 2t] (1t) {};

\draw (1)--(2) (2)--(3) (3)--(2r) (2r)--(1r) (3)--(2t) (2t)--(1t);
\end{tikzpicture}}
	\end{gathered} \quad\,.
\end{equation}
As we can see, we land directly on the quiver in the shape of an affine $E_6$
Dynkin diagram from Figure \ref{fig:ClassSHiggsableE6E7a7HasseUnitary}, which
also appears identically from the unitary Decay and Fission algorithm. We note that this is the only fission operation which produces a (product of) quivers which are all sufficiently good according to Section \ref{sec:Algorithm}. The only
other non-trivial Orthosymplectic Decay and Fission of the quiver in equation
\eqref{eqn:ClassSHiggsableE6E7a7SO} is
\begin{equation}
	\begin{gathered}
		\resizebox{0.3\linewidth}{!}{\begin{tikzpicture}
    \node[sonode,label=below:{$2$}] at (0,0) (1) {};
    \node[spnode,label=below:{$2$}] [right=0.7cm of 1] (2) {};
    \node[sonode,label=below:{$4$}] [right=0.7cm of 2] (3) {};
    \node[spnode,label=below:{$4$}] [right=0.7cm of 3] (4) {};
    \node[sonode,label=below:{$6$}] [right=0.7cm of 4] (5) {};
    \node[spnode,label=below:{$4$}] [right=0.7cm of 5] (4r) {};
    \node[sonode,label=below:{$4$}] [right=0.7cm of 4r] (3r) {};
    \node[spnode,label=below:{$2$}] [right=0.7cm of 3r] (2r) {};
    \node[sonode,label=below:{$2$}] [right=0.7cm of 2r] (1r) {};
    \node[spnode,label=right:{$4$}] [above=0.7cm of 5] (2t) {};
    \node[sonode,label=right:{$2$}] [above=0.7cm of 2t] (1t) {};

\draw (1)--(2) (2)--(3) (3)--(4) (4)--(5) (5)--(4r) (4r)--(3r) (3r)--(2r) (2r)--(1r) (5)--(2t) (2t)--(1t);
\end{tikzpicture}}
	\end{gathered} \qquad \rightarrow \qquad
    \begin{gathered}
		\resizebox{0.3\linewidth}{!}{\begin{tikzpicture}[scale=0.6]
    \node[sonode,label=below:{$2$}] at (0,0) (1) {};
    \node[spnode,label=below:{$2$}] [right=0.7cm of 1] (2) {};
    \node[sonode,label=below:{$4$}] [right=0.7cm of 2] (3) {};
    \node[spnode,label=below:{$4$}] [right=0.7cm of 3] (4) {};
    \node[sonode,label=below:{$6$}] [right=0.7cm of 4] (5) {};
    \node[spnode,label=below:{$4$}] [right=0.7cm of 5] (4r) {};
    \node[sonode,label=below:{$4$}] [right=0.7cm of 4r] (3r) {};
    \node[spnode,label=below:{$2$}] [right=0.7cm of 3r] (2r) {};
    \node[sonode,label=below:{$2$}] [right=0.7cm of 2r] (1r) {};
    \node[spnode,label=right:{$2$}] [above=0.7cm of 5] (2t) {};
    \node[sonode,label=right:{$2$}] [above=0.7cm of 2t] (1t) {};

\draw (1)--(2) (2)--(3) (3)--(4) (4)--(5) (5)--(4r) (4r)--(3r) (3r)--(2r) (2r)--(1r) (5)--(2t) (2t)--(1t);
\end{tikzpicture}}
	\end{gathered} \quad\,,
\end{equation}
where the $USp(4)$ node on the vertical tail decayed to
$USp(2)$. It is well-known that the quiver on the right has a
Coulomb branch which is the closure of the minimal nilpotent orbit of
$\mathfrak{e}_7$, exactly the same as the affine-$E_7$-shaped quiver that
appears in Figure \ref{fig:ClassSHiggsableE6E7a7HasseUnitary}. We have drawn
the corresponding orthosymplectic Coulomb branch Hasse diagram in Figure
\ref{fig:ClassSHiggsableE6E7a7HasseOrtho}, which we see has identical structure 
to the unitary case, as required by the $\mathfrak{su}(4) \cong \mathfrak{so}(6)$ 
isomorphism. 

We can perform a similar analysis of all of the class $\mathcal{S}$ theories of type $\mathfrak{su}(4) \cong \mathfrak{so}(6)$ where the flavour symmetry is enhanced. The unitary 3d mirrors for each such case are listed in Figure \ref{fig:all_enc}, and the corresponding orthosymplectic 3d mirrors can be directly worked out from the equivalence of $\mathfrak{su}(4)$ and $\mathfrak{so}(6)$ nilpotent orbits given in Table \ref{tbl:A3vsD3}. In all cases, the Orthosymplectic Decay and Fission algorithm of Section \ref{sec:Algorithm} produces the same Coulomb branch Hasse diagram as that of the unitary Decay and Fission algorithm applied to the unitary 3d mirrors. The Higgs branch of class $\mathcal{S}$ theories has a generic structure: first, the transitions are given by partial puncture closure, until there exists an enhanced flavour symmetry; after the enhancement, the transitions are as one of the quivers in Figure \ref{fig:all_enc}. As such, we have demonstrated that the structure of the Higgs branch of all $\mathfrak{so}(6)$ class $\mathcal{S}$ theories (with only regular untwisted punctures) can be recovered from the Orthosymplectic Decay and Fission algorithm presented in Section \ref{sec:Algorithm}.

\section{Walking the Path to the Rules}\label{sec:Rules}

Although as explained in Section \ref{sec:Class_S} we heavily leveraged on the predictions following from class $\mathcal{S}$ and the algebra isomorphism $\mathfrak{su}_4\cong \mathfrak{so}_6$ to lay down the rule of an Orthosymplectic Decay and Fission algorithm, some more general insights follow from venturing deeper in the physics behind each of the rules stated in Section \ref{sec:Algorithm}.

\subsection{Symmetry through Monopoles}

Already at the beginning of Section \ref{sec:Algorithm}, we defined the \emph{balance} as a quantity locally controlling the conformal dimension of the theories' monopole operators in relation to the unitary bound. The definition given in \eqref{eqn:balance} stems from a general formula that takes into account the type of gauge theory under consideration in the presence of fundamental matter. As already noticed by the original authors \cite{Gaiotto:2008ak}, for a significant number of orthosymplectic gauge theories, the goodness condition $b\ge0$ does not suffice to ensure the goodness of the entire theory. When the quiver theory is comprised of only gauge nodes, there exists a redundant $\mathbb{Z}_2$ symmetry acting trivially on the product of the gauge groups, which we always choose to ungauge \cite{Bourget:2020xdz}. This choice is reflected in the flux lattice of monopole operators that, together with their usual contribution in a chamber of a $\mathbb{Z}$ valued lattice of dimension dictated by the product of the rank of the various gauge groups, also receive a contribution from a chamber in a lattice valued on the shifted half integers $\mathbb{Z}+\frac{1}{2}$.
Explicitly, the balance condition of \eqref{eqn:balance} is locally sensitive to the contribution of $\mathbb{Z}$-valued fluxes; thus, whereas for quiver theories with special orthogonal nodes of rank $k\ge4$ this condition suffices to assess the global goodness of a theory, for $k=3$, i.e.\ $SO(6)$ gauge nodes, this condition fails as the minimal monopole falling below the unitarity bound corresponds to fluxes valued in the half-integer lattice. These kinds of ``bad" contributions are not localised on the single gauge node but are spread along the whole quiver theory. It is possible to circumvent this problem by having an empirical definition of balance that mimics the inclusion of these half-integers valued monopoles by readjusting the contribution of a neighbouring symplectic node towards the number of flavours. Exploiting the aforementioned algebra isomorphism and class $\mathcal{S}$, such an empirical statement about the goodness of the theory regardless of the quiver's shape can be derived.

The theories $\mathcal{S}_{\mathfrak{su}_4}\langle C_{0,n} \rangle \{ [3,1], \cdots, [3,1] \} $ and $\mathcal{S}_{\mathfrak{so}_6}\langle C_{0,n} \rangle \{ [3^2], \cdots, [3^2] \} $, in \eqref{eqn:4with1and6with2}, are isomorphic by construction for every $n$.
\begin{equation}\label{eqn:4with1and6with2}
    \begin{gathered}
    \includegraphics[page=3,scale=0.8]{DecayAndFission_figures.pdf}
    \end{gathered}
\end{equation}
The unitary theory dictates that for $n>8$, both theories must be good, for $n=8$ balanced, and for $n\le7$ bad\footnote{Actually for $n=7$ the unitary theory is ugly, but since this definition does not translate to general orthosymplectic theories we incorporate it into the bad notion. Moreover, ugly theories do not appear as Decay and Fission products of good/balanced unitary theories. Hence, it is sensible to exclude the orthosymplectic counterpart from appearing in the Orthosymplectic Decay and Fission algorithm.}. Reproducing the same behaviour from the balance notion for special-orthogonal theories of \ref{eqn:balance} requires that all the stand-alone $USp(2)$ in the orthosymplectic construction must be effectively counted as contributing $\frac{5}{8}$-s towards $N_f$. 
The same reasoning can be applied to the theories $\mathcal{S}_{\mathfrak{su}_4}\langle C_{0,n} \rangle \{ [2^2], \cdots, [2^2] \} $ and $\mathcal{S}_{\mathfrak{so}_6}\langle C_{0,n} \rangle \{ [3,1^3], \cdots, [3,1^3] \} $ to assess that a $USp(2)$ between an $SO(6)$ and an $SO(2)$ gauge nodes contributes to $N_f$ as $\frac{5}{4}$. 
The just performed analysis sheds light behind \texttt{Rule 1} of our algorithm, justifying it as an empirical readjustment of \eqref{eqn:balance} to be sensitive to all the monopoles that may fall below the unitarity bound.

The role of half-integer lattice flux-valued monopoles extends beyond the scope of $SO(6)$ gauge theories: it is a known phenomenon that flavour symmetries can enhance due to the presence of extra contribution stemming from monopoles valued in this shifted lattice. Classically, the Coulomb branch flavour symmetry algebra\footnote{Coherently with a choice of global structure for the gauge algebra that determines the group nature, we denote the flavour algebra by the associated group even though we are not determining the presence of discrete actions.} in an orthosymplectic quiver is read from a chain of $p$ balanced nodes to be:
\begin{itemize}
    \item $SO(p+3)$ if the chain starts and ends with an $SO(2)$ gauge node.
    \item $SO(p+2)$ if the chain starts or ends with an $SO(2)$ gauge node.
    \item $SO(p+1)$ in all the other cases. 
\end{itemize}
This symmetry arises from monopoles valued in the integer lattice and corresponds, in general, to the full symmetry of the theory. In specific cases, the half-integer-valued monopoles recombine into a spinor representation of the Coulomb branch symmetry read from the aforementioned rule and can interplay with the integer-valued monopoles to enhance the symmetry. We can keep track of this possibility by noticing that, coherently with a spinor representation's dimension, half-integer valued monopoles contribute to the adjoint action of the flavour symmetry (i.e.\ to the flavour current) at most $2^d$, where $d$ is the number of $SO(even)$ gauge nodes in the quiver. The group theory of the problem requires finding an embedding of $G$ of the form $\prod_i SO(r_i)$, where the spinor representation of the obtained algebras recombine with the adjoint to form the adjoint representation of the original group $G$. We tabulated all such occurrences\footnote{The case of $G_2\rightarrow SU(2)\times SU(2) \leadsto 14=3+3+2^3$ is a special embedding therefore it does not entail a naive recombination of spinors.} in Table \ref{tab:Flavour_Enhancement}.
\begin{table}[ht]
\renewcommand{\arraystretch}{1.25}
    \centering
    \begin{tabular}{c|c|l}\toprule
        $G$ & $\prod_i SO(r_i)$ & $\mathbf{adj}$ Decomposition  \\ \midrule
        $SU(8)$ & $SO(6) \times SO(6) \times SO(2)$ & $63=15+15+1+2^5$\\
        $SU(6)$ & $SO(6) \times SO(3) \times SO(2)$ & $35=15+3+1+2^4$\\
        $SU(5)$ & $SO(6) \times SO(2)$ & $24=15+1+2^3$\\
        $SU(3)$ & $SO(3) \times SO(2)$ & $8=3+1+2^2$\\
        $F_4$ & $SO(9)$ & $52=36+2^4$\\
        $E_6$ & $SO(10) \times SO(2)$ & $78=45+1+2^5$\\
        $E_7$ & $SO(12) \times SO(3)$ & $133=66+3+2^6$\\
        $E_8$ & $SO(16)$ & $248=120+2^7$\\ \bottomrule
    \end{tabular}
    \caption{List of flavour symmetry groups $G$ that can be realised as enhancement in orthosymplectic quiver gauge theories.}
    \label{tab:Flavour_Enhancement}
\end{table}

All these examples, apart from the $E_8$ and the $F_4$ ones, can be realised in the controlled $\mathcal{S}_{\mathfrak{su}_4}\langle C_{0,n} \rangle$ framework. The $E_8$ case is known from the 6d $(1,0)$ E-string theory \cite{Minahan:1997ct} and its magnetic quiver \cite{Cabrera:2019dob}. Whereas an $F_4\times F_4$ example was realised in \cite{Sperling:2021fcf} by studying balanced simply-laced forked-quivers of orthosymplectic type. Studying the unitary class $\mathcal{S}$ Hasse diagram we found the list of admitted decays found in Table \ref{tab:admitted decays}, and thus justify \texttt{Rule 3a}.

\subsection{Vacuum of bad theories}
We emphasised in multiple instances that we are reticent towards bad theories. The main reason supporting our attitude lies in the nature of their vacuum\footnote{The word ``vacuum'' has been used in the sense of \cite{Assel:2017jgo} where it stands for the global collection of all the local geometries, i.e\ Coulomb branch moduli space of vacua, admitted near each singular point in the space.} structure: the Coulomb branch moduli space of vacua, different from good theories where it supports one and only singular locus, admits a set of (possibly continuous) singular loci. The understanding of the global geometry of this space is hindered by the impossibility of an explicit counting of the theory's monopole operators due to the IR $R$-symmetry being different from the UV one. Nevertheless, it is still possible to probe the local geometry of the Coulomb branch moduli space near each of their singular loci. In \cite{Assel:2018exy,Assel:2017jgo}, a first approach has been made in this direction, identifying, in the unitary case, a Seiberg-like duality that maps the local geometry near each singular point to the Coulomb branch moduli space of a good unitary theory plus some decoupled hypermultiplets responsible for the flat directions in the bad theory. This result was also recovered from index-like computations in \cite{Giacomelli:2023zkk,Giacomelli:2024laq}, where the index of bad theories behaves as a sum of distributions, each identifying a particular singular locus of the bad theory. 

Shifting our attention to bad orthosymplectic theories, already a simple case such as the $USp(2N)$ SQCD reveals itself to be more subtle with respect to its unitary counterpart. Symplectic gauge theories, as well as special-orthogonal ones, do not admit an ugly notion, and their moduli space of vacua drifts from the good single-cone-behaviour already when $b=-1$, i.e.\ when the $USp(2N)$ theory is paired with $N_f=2N$ fundamental hypermultiplets. In this case, the vacuum consists of two singular loci that are quantum mechanically separated, but swapped by a $\mathfrak{G}_2 \cong \mathbb{Z}_2$ global symmetry on the Coulomb branch \cite{Assel:2018exy}, without any flat direction. The origin of this discrete symmetry is explained when interpreting the bad symplectic SQCD theory as a $T_\rho(SO(4N))$ theory: there exist two in-equivalent nilpotent orbits, mapped under a $\mathbb{Z}_2$ symmetry, $O_I , O_{II}$ in the nilpotent cone of $SO(4N)$ cone which are associated with the same (very-)even partition $\rho=[(2N)^2]$ of $4N$. Studying the local IR SCFT living at each of the singular points in the moduli space results in it being the source of a Higgs branch that admits a unitary 3d mirror \cite{Assel:2018exy, Ferlito:2016grh}. For more severe badnesses, i.e.\ $b<-1$ and $N_f < 2N$, the vacuum behaviour shown by $USp(2N)$ SQCDs matches the classic expectations for bad theories: there exist multiple singular points where the local geometry looks like a good theory with extra flat directions.

This can be extended to general $T_\rho(SO(2N))$ theories. Leveraging on the well-definiteness of the nilpotent cone of $\mathfrak{so}_{2N}$ algebras as a singular space with the singularity localised in a single locus, any slice to it retains the singularity structure: thus, for any even partition $\rho$ of $2N$ the singularity which is the Coulomb branch moduli space of the associated $T_\rho(SO(2N))$ stays well behaved. The mathematical definition of these spaces signals that the appearance of bad $USp(2m)$ nodes with $N_f=2m$ flavours in  $T_\rho(SO(2N)$ theories does not and needs not to disrupt the goodness\footnote{Here meant as having a Coulomb branch moduli space constituting of a single singular locus, and not in the balance sense of \eqref{eqn:balance} that pertains the conformal dimension of the theory's monopole operators.} of the theory. A pathway to open a physical understanding of this phenomenon can be opened thinking that the $\mathbb{Z}_2$ symmetry exchanging the local two singular points of the bad symplectic gauge node, in reality, identifies them, despite their quantum mechanical in-equivalence. Another example is to consider the 3d mirror pair $T(SO(2n+1))$ and $T(USp(2n))$ established in \cite{Gaiotto:2008ak} which have as maximal branches the maximal/regular nilpotent orbit closures of $\mathfrak{so}(2n+1)$ and $\mathfrak{usp}(2n)$, respectively. The Higgs branch of $T(USp(2n))$ is a well defined hyper-K\"ahler cone. Therefore, the Coulomb branch of $T(SO(2n+1))$ is also a well defined hyper-K\"ahler cone with one most singular point despite the fact that all its symplectic gauge nodes have balance $b=-1$. Consequently, we generically admit bad nodes with $b=-1$ in our Hasse Diagrams, as stated in \texttt{Rule 2}. Notice, however, that although an explicit analysis has been carried out for symplectic gauge nodes, an alternative one for even special-orthogonal nodes is lacking, and we leave it for future analysis.

Equipped with this new understanding, we can push the study of $T_\rho(SO(2N))$ theories to also justify the decay pattern predicted by \texttt{Rule 3b}. The natural inclusion ordering on the closure of nilpotent orbits induces an ordering in the $T_\rho(SO(2N))$ theories on the set of partitions $\rho$-s associated with the orbits $O$-s. In fact, starting from a partition $\rho_B$ such that the resulting $T_{\rho_B}(SO(2N))$ theory has $m$ gauge nodes of balance $b=-1$\footnote{From the construction of \cite{Gaiotto:2008ak}, it is clear that the minimum balance for a gauge node achievable in a $T_\rho$ theory is $b=-1$.}, it is possible to flow to the next special partition\footnote{Special partitions are such that they are mapped into themselves after having applied twice the Barbasch-Vogan map, which takes partitions of $G$ to special partitions of its GNO dual $G^\vee$. This requirement is physically equivalent to demanding $(G^\vee)^\vee=G$ \cite{Goddard:1976qe}: we can go from the electric to the magnetic phase of a theory, i.e.\ trading monopoles for multiplets in the 3d language and back, trading again multiplets for monopoles, consistently.} $\rho_>$ in the aforementioned partial ordering. The $T_{\rho_>}(SO(2N))$ theory has the same structure of $T_{\rho_B}(SO(2N))$ but having $n<m$ of the previous bad nodes with rank lowered by one, i.e.\ the bad node in $T_{\rho_B}(SO(2N))$ has now become balanced. This transition is characterized by a $c_n$ transverse slice, i.e.\ the closure of the minimal nilpotent orbit of $USp(2n)$, with $n$ given by the number of $-1$ balance nodes that are now balanced. In Appendix~\ref{sec:Appendix_Trho}, the  algorithm is showcased in the context of $T_\rho^\sigma(SO(2N))$ theories.
We require said $c_m$ pattern to be transported to more general orthosymplectic quiver theories consistently with observations, leading to \texttt{Rule 3b}. 

\begin{table}[p]
\renewcommand{\arraystretch}{1.25}
    \centering
\begin{tabular}{cc}\toprule
     Slice & Quiver  \\ \midrule
     $d_{n \ge 3}$ & $\begin{gathered}
 \underbrace{\begin{tikzpicture}
         \node[sonode, label=below:{\footnotesize $2$}] at (0,0) (c1) {};
         \node[spnode, label=below:{\footnotesize $2$}, right of=c1] (d1) {};
         \node[tnode, right of=d1, label={[yshift=-12.5pt]\footnotesize $\cdots$}] (dots) {};
         \node[sonode, label=below:{\footnotesize $2$},right of=dots]  (dn-1) {};
         \node[spnode, label=below:{\footnotesize $2$},right of=dn-1]  (cn) {};
         \node[sonode, label=below:{\footnotesize$ 2$}, right of=cn] (dn) {};
         \draw (c1)--(d1)--(dots)--(dn-1)--(cn)--(dn);
     \end{tikzpicture}}_{n-1 \, SO(2) \text{ nodes and } n-2 \, USp(2) \text{nodes}}   
     \end{gathered}$\\
     $b_{n \ge 3}$ & $\begin{gathered}
 \underbrace{\begin{tikzpicture}
         \node[sonode, label=below:{\footnotesize $2$}] at (0,0) (c1) {};
         \node[spnode, label=below:{\footnotesize $2$}, right of=c1] (d1) {};
         \node[tnode, right of=d1, label={[yshift=-12.5pt]\footnotesize $\cdots$}] (dots) {};
         \node[sonode, label=below:{\footnotesize $2$},right of=dots]  (dn-1) {};
         \node[spnode, label=below:{\footnotesize $2$},right of=dn-1]  (cn) {};
         \draw (c1)--(d1)--(dots)--(dn-1)--(cn);
     \end{tikzpicture}}_{n-1 \, SO(2) \text{ nodes and } n-1 \, USp(2) \text{nodes}}   
     \end{gathered}$\\
%     $c_n$ & $\begin{gathered}\underbrace{\begin{tikzpicture}
%         \node[spnode, label=below:{\footnotesize $2$}] at (0,0) (c1) {};
%         \node[sonode, label=below:{\footnotesize $\mathfrak{d}_0$}, right of=c1] (d1) {};
%         \node[tnode, right of=d1, label={[yshift=-12.5pt]\footnotesize $\cdots$}] (dots) {};
%         \node[spnode, label=below:{\footnotesize $2$},right of=dots]  (cn) {};
%         \draw (c1)--(d1)--(dots)--(cn);
%     \end{tikzpicture}}_{n \, 2 \text{ nodes}} \end{gathered}$\\
     $a_7$ & $\begin{gathered}
     \begin{tikzpicture}
        \node[sonode, label=below:{\footnotesize $2$}] at (0,0) (d1) {};    
        \node[spnode, label=below:{\footnotesize $2$}, right of=d1] (c1) {};
        \node[sonode, label=below:{\footnotesize $2$}, right of=c1] (d2) {};
        \node[spnode, label=below:{\footnotesize $0$},right of=d2]  (c0) {};
        \node[sonode, label=below:{\footnotesize $2$}, right of=c0] (d3) {};
        \node[spnode, label=below:{\footnotesize $0$},right of=d3]  (c01) {};
        \node[sonode, label=below:{\footnotesize $2$}, right of=c01] (d4) {};
        \node[spnode, label=below:{\footnotesize $2$},right of=d4]  (c2) {};
        \node[sonode, label=below:{\footnotesize $2$}, right of=c2] (d5) {};
         \draw (d1)--(c1)--(d2)--(c0)--(d3)--(c01)--(d4)--(c2)--(d5);
     \end{tikzpicture}
     \end{gathered}$\\
     $a_5$ & $\begin{gathered}
     \begin{tikzpicture}
        \node[sonode, label=below:{\footnotesize $2$}] at (0,0) (d1) {};    
        \node[spnode, label=below:{\footnotesize $2$}, right of=d1] (c1) {};
        \node[sonode, label=below:{\footnotesize $2$}, right of=c1] (d2) {};
        \node[spnode, label=below:{\footnotesize $0$},right of=d2]  (c0) {};
        \node[sonode, label=below:{\footnotesize $2$}, right of=c0] (d3) {};
        \node[spnode, label=below:{\footnotesize $0$},right of=d3]  (c01) {};
        \node[sonode, label=below:{\footnotesize $2$}, right of=c01] (d4) {};
         \draw (d1)--(c1)--(d2)--(c0)--(d3)--(c01)--(d4);
     \end{tikzpicture}
     \end{gathered}$\\
     $a_4$ & $\begin{gathered}
     \begin{tikzpicture}
        \node[sonode, label=below:{\footnotesize $2$}] at (0,0) (d1) {};    
        \node[spnode, label=below:{\footnotesize $2$}, right of=d1] (c1) {};
        \node[sonode, label=below:{\footnotesize $2$}, right of=c1] (d2) {};
        \node[spnode, label=below:{\footnotesize $0$},right of=d2]  (c0) {};
        \node[sonode, label=below:{\footnotesize $2$}, right of=c0] (d3) {};
         \draw (d1)--(c1)--(d2)--(c0)--(d3);
     \end{tikzpicture}
     \end{gathered}$\\
     $a_2$ & $\begin{gathered}
     \begin{tikzpicture}
        \node[sonode, label=below:{\footnotesize $2$}] at (0,0) (d1) {};    
        \node[spnode, label=below:{\footnotesize $0$}, right of=d1] (c0) {};
        \node[sonode, label=below:{\footnotesize $2$}, right of=c0] (d2) {};
         \draw (d1)--(c0)--(d2);
     \end{tikzpicture}
     \end{gathered}$\\
     $a_1$ or $A_k$ or $D_k$ & $\begin{gathered}
     \begin{gathered}
              \begin{tikzpicture}
        \node[sonode, label=below:{\footnotesize $2$}] at (0,0) (d1) {};
     \end{tikzpicture}
     \end{gathered} \text{ or }     \begin{gathered}
              \begin{tikzpicture}
        \node[spnode, label=below:{\footnotesize $2$}] at (0,0) (d1) {};
     \end{tikzpicture}
     \end{gathered} 
     \end{gathered} $\\
          $e_7$ & $\begin{gathered}
     \begin{tikzpicture}
        \node[sonode, label=below:{\footnotesize $2$}] at (0,0) (d1) {};    
        \node[spnode, label=below:{\footnotesize $2$}, right of=d1] (c1) {};
        \node[sonode, label=below:{\footnotesize $4$}, right of=c1] (d2) {};
        \node[spnode, label=below:{\footnotesize $4$},right of=d2]  (c0) {};
        \node[sonode, label=below:{\footnotesize $6$}, right of=c0] (d3) {};
        \node[spnode, label=below:{\footnotesize $4$},right of=d3]  (c01) {};
        \node[sonode, label=below:{\footnotesize $4$}, right of=c01] (d4) {};
        \node[spnode, label=below:{\footnotesize $2$},right of=d4]  (c2) {};
        \node[sonode, label=below:{\footnotesize $2$}, right of=c2] (d5) {};
        \node[spnode, label=right:{\footnotesize $2$},above of=d3]  (c1t) {};
        \node[sonode, label=right:{\footnotesize $2$}, above of=c1t] (d1t) {};
         \draw (d1)--(c1)--(d2)--(c0)--(d3)--(c01)--(d4)--(c2)--(d5);
         \draw (d3)--(c1t)--(d1t);
     \end{tikzpicture}
     \end{gathered}$\\ 
          $e_8$ & $\begin{gathered}
     \begin{tikzpicture}
        \node[sonode, label=below:{\footnotesize $2$}] at (0,0) (d1) {};    
        \node[spnode, label=below:{\footnotesize $2$}, right of=d1] (c1) {};
        \node[sonode, label=below:{\footnotesize $4$}, right of=c1] (d2) {};
        \node[spnode, label=below:{\footnotesize $4$}, right of=d2] (c2) {};
        \node[sonode, label=below:{\footnotesize $6$}, right of=c2] (d3) {};
        \node[spnode, label=below:{\footnotesize $6$},right of=d3]  (c3) {};
        \node[sonode, label=below:{\footnotesize $8$}, right of=c3] (d4) {};
        \node[spnode, label=below:{\footnotesize $4$},right of=d4]  (c3s) {};
        \node[sonode, label=below:{\footnotesize $6$}, right of=c3s] (d3s) {};
        \node[spnode, label=below:{\footnotesize $4$},right of=d3s]  (c2s) {};
        \node[sonode, label=below:{\footnotesize $4$}, right of=c2s] (d2s) {};
        \node[spnode, label=below:{\footnotesize $2$},right of=d2s]  (c1s) {};
        \node[sonode, label=below:{\footnotesize $2$}, right of=c1s] (d1s) {};
        \node[spnode, label=right:{\footnotesize $2$},above of=d4]  (c1t) {};
         \draw (d1)--(c1)--(d2)--(c2)--(d3)--(c3)--(d4)--(c3s)--(d3s)--(c2s)--(d2s)--(c1s)--(d1s);
         \draw (d4)--(c1t);
     \end{tikzpicture}
     \end{gathered}$\\      
     \bottomrule
\end{tabular}
\caption{Table of admitted decay structures from orthosymplectic quiver theories. Each entry in the table is associated with the closure of the minimal nilpotent orbit of the manifest said symmetry in the quiver that is supposed to be Higgsed via the decay process.}
\label{tab:admitted decays}
\end{table}

\subsection{Unitary within Orthosymplectic}\label{unitarization}

In dealing with orthosymplectic quiver theories, it is crucial to encompass the possibility of partial higgsing directions leading to unitary theories. As remarked in Section~\ref{sec:Algorithm}, \texttt{Rule 4a} and \texttt{Rule 4b}, play the fundamental role of explaining the appearance of unitary quiver gauge theories as fission or unitarisation products of their orthosymplectic ancestors. The formal justification of said procedure lies in the representation theory of orthosymplectic algebras, each of which admits a Levi decomposition as a sum of subalgebras of the form:
\begin{equation}
    \mathfrak{so}_{2N} \rightarrow \mathfrak{so}_{2N-2K} \oplus \mathfrak{su}_K \oplus \mathfrak{u}_1 \qquad , \qquad \mathfrak{usp}_{2N} \rightarrow \mathfrak{usp}_{2N-2K} \oplus \mathfrak{su}_K \oplus \mathfrak{u}_1 \,.
\end{equation}
Under this decomposition, a nilpotent orbit $O$ of the parent algebra fractionate and induces the nilpotent orbits $O_{1^{\text{st}}}$ and $O_{2^{\text{nd}}}$ in the first and second addendum in the decomposition. In terms of physics, it would be improper to talk about this decomposition in terms of $T_\rho(G)$ theories: the reason, we soon thoroughly illustrate, lies in the presence of the $G$ flavour node. Guided by unitary Decay and Fission and other works in this direction \cite{Mansi:2023faa,Lawrie:2023uiu}, we can interpret unitary fission processes as separations of stacks of D$p$-branes along a transverse direction spanned by D$(p+2)$ branes. The analogous picture for orthosymplectic quiver theories requires Type II superstring theory constructions of a D$p$-D$(p+2)$-NS5 brane system in the presence of an additional $Op$-orientifold plane. Differently from the unitary case, the orientifold is spatially localised; thus, a separation in stacks of D$p$-branes followed by a careful infinite distance limit leads to having only one of the stacks remaining on top of the orientifold and giving rise to an orthosymplectic theory. Oppositely, the stack that has been pulled away realises a unitary theory since it does not sense the presence of the orientifold anymore. In the case of $T_\rho(G)$ realised in a Type IIB D3-D5-NS5 system, a transverse pulling away action on D3-branes would not be able to break the flavour symmetry determined by D5-branes, as the D5-s would be unmovable and do not share any spatial direction with the NS5-branes more than the one also shared with D3-branes.

There is yet one process that needs to be understood: unitarisation, which corresponds to having $N=K$ in the subalgebra splitting. The phenomenology of this choice corresponds to the extreme action of pulling away all the allowed branes from the orientifold plane: here, the term allowed includes the possibility of having fractional branes remaining stuck on the orientifold plane and not contributing to physical gauge degrees of freedom.

An analogous Higgsing pattern compatible with the phenomena just described has been encountered in \cite{Lawrie:2024zon}, where studying $(2,0)$ six-dimensional SCFTs, in Type $D_N$ there exist complex structure deformations, the analogous of Higgsings in F-theory construction, leading to a product of theories of Type $D_{N-k}$ and Type $A_{k-1}$ for $1\le k \le n$. Type $D_N$ theories admit a magnetic quiver constituted by a single $SO(2N)$ gauge node with antisymmetric matter, whereas the magnetic quiver for the Type $A_{N-1}$ theory has a single $U(N)$ with an adjoint matter field. Therefore, the F-theory predictions of the Hasse Diagram are mimicked in \texttt{Rule 4a} and \texttt{Rule 4b} of the Orthosymplectic Decay and Fission.  
Arguments from F-theory in support of these rules also stem from the possibility of turning special-orthogonal algebras, corresponding to fibrations of Type $I^*_n$ in F-theory constructions, to (special-)unitary ones, corresponding to fibrations of Type $I_n$ in F-theory constructions. The analogue for symplectic algebra a has been discussed in \cite{Mansi:2023faa} and substitutes the $I^*_n$ fibers with $I_n$ ones but with a different monodromy with respect to the (special-)unitary ones. This geometrical phenomenon is the electric analogue of the fission and unitarisation process described via magnetic quivers.

\section{Beyond \texorpdfstring{\boldmath{$\mathfrak{su}(4) \cong \mathfrak{so}(6)$}}{su4 = so6} Class \texorpdfstring{\boldmath{$\mathcal{S}$}}{S}}\label{sec:Benchmark}

The development of the Orthosymplectic Decay and Fission algorithm that we presented in Section \ref{sec:Algorithm} was influenced by an exhaustive analysis of the Higgs branch renormalization group flows of class $\mathcal{S}$ of type $\mathfrak{su}(4) \cong \mathfrak{so}(6)$. The key which unlocked this analysis was the Lie algebra isomorphism, which allowed us to use the unitary Decay and Fission algorithm to predict the required structure obtained from the Orthosymplectic Decay and Fission algorithm. 

In this section, we demonstrate that the algorithm holds beyond these class $\mathcal{S}$ examples. We consider both examples of higher rank class $\mathcal{S}$ theories, and examples of 6d $(1,0)$ SCFTs whose Higgs branch is given by a known orthosymplectic 3d mirror/magnetic quiver and for which the Higgs branch renormalization group flows can be understood via the complex structure deformations of the singularity realizing the 6d SCFT in F-theory. In these example, we observe that the Orthosymplectic Decay and Fission algorithm on the magnetic quiver/3d mirror produces the expected result for the structure of the Higgs branch of the higher dimensional theory.

In fact, there are many more constructions that lead to Coulomb branches which have simultaneous unitary and orthosymplectic quiver descriptions. For example, in \cite{Hanany:2023tvn,Bennett:2024llh}, subgroups of the Coulomb symmetry of unitary and orthosymplectic quivers have been gauged, and quiver descriptions for the resulting theories has been proposed. In this way, new orthosymplectic quiver theories whose moduli space coincides with a unitary one have been derived, see \cite[Tables 2 and 3]{Bennett:2024llh}. It is straightforward to apply the Orthosymplectic Decay and Fission algorithm to these examples and see that they agree with the unitary Decay and Fission. Hence, providing another non-trivial consistency check of the rules proposed here. Yet another source of 3d $\mathcal{N}=4$ quivers with both unitary and orthosymplectic descriptions is \cite{Akhond:2020vhc}; we leave this additional construction as a future testing ground for the Orthosymplectic Decay and Fission algorithm.

\subsection{Higgsing the D-type Orbi-instanton}

We now turn to the study of 3d $\mathcal{N}=4$ orthosymplectic quivers that are magnetic quivers for the Higgs branch of certain 6d $(1,0)$ SCFTs. Related to the anomaly cancellation constraints, 6d $(1,0)$ SCFTs are believed to be classified via a construction in terms of F-theory compactified on non-compact elliptically-fibered Calabi--Yau threefolds satisfying certain restrictive conditions \cite{Heckman:2013pva,DelZotto:2014hpa,Heckman:2015bfa}. The effective theories that live on subloci of the Higgs branch can be determined by studying the space of complex structure deformations of the Calabi--Yau threefold engineering the 6d $(1,0)$ SCFT in question. Therefore, the Orthosymplectic Decay and Fission algorithm proposed in Section \ref{sec:Algorithm} can be geometrically tested. In fact, applying the Orthosymplectic Decay and Fission algorithm allows for both new predictions about Higgs branch renormalization group flows of six-dimensional SCFTs, as well as recovering and generalizing fission processes predicted in \cite{Bao:2024eoq}. 

We consider a class of 6d $(1,0)$ SCFTs which not only have a geometric construction in F-theory, but also possess a realization in M-theory. Consider a stack of $N$ M5-branes probing a $\mathbb{C}^2/\Gamma$, with $\Gamma$ a finite subgroup of $SU(2)$, orbifold singularity and contained inside of an M9-plane, as depicted in Figure \ref{fig:M-theory_OrbiInstanton}. The SCFT living on the worldvolume of the M5-branes is known as the rank $N$ $(\mathfrak{e}_8, \mathfrak{g})$ orbi-instanton \cite{DelZotto:2014hpa}, where $\mathfrak{g}$ is the simple and simply-laced Lie algebra associated to $\Gamma$ via the McKay correspondence. To describe such a theory, it is necessary to specify a flat connection on the $E_8$-bundle associated with the M9-plane on the $S^3/\Gamma$ boundary of the orbifold; the trivial flat connection leads to an $\mathfrak{e}_8$ global symmetry factor, whereas other choices generally have a reduced global symmetry.

\afterpage{
\begin{figure}[t]
\begin{multicols}{2}
\centering
    \begin{tikzpicture}
    \node[] at (3.5,0) (label) {\footnotesize $\mathbb{C}^2/\Gamma_{ADE}$};
    \draw[dashed] (0,0)--(label)--(7,0);
    \draw[blue] (7,-1.5)--(7,1.5) node[above] () {M9};

    \node at (7,0) [red,right=0.3] () {\footnotesize $N$ M5};
    \draw[red] (6.8,0.2)--(7.2,-0.2);
    \draw[red] (6.8,-0.2)--(7.2,0.2);

    \end{tikzpicture}
    
    \vfill
    
    \begin{equation*}
        \begin{aligned}
        \resizebox{0.4\textwidth}{!}{
                \begin{tabular}{c|c}\toprule
              & Spatial Orientation  \\ \midrule
             {\color{red}M5} & $x^0,x^1,x^2,x^3,x^4,x^5$ \\ 
             {\color{blue}M9} & $x^0,x^1,x^2,x^3,x^4,x^5,x^7,x^8,x^9,x^{10}$ \\ 
             $\mathbb{C}^2/\Gamma_{ADE}$ & $x^0,x^1,x^2,x^3,x^4,x^5,x^6$ \\ \bottomrule
    \end{tabular} 
    }
        \end{aligned}
    \end{equation*}

\end{multicols}
    \caption{The M-theory configuration for a rank $N$ $(\mathfrak{e}_8, \mathfrak{g}_{ADE})$ orbi-instanton theory.}
    \label{fig:M-theory_OrbiInstanton}
\end{figure}}

When $\mathfrak{g} = \mathfrak{so}(2n + 8)$, there exists an alternative construction of the rank $N$ $(\mathfrak{e}_8, \mathfrak{g})$ orbi-instanton in Type IIA. The orbifold becomes a stack of $n+4$ full D6-branes on top of an $O6^-$-orientifold-plane, the M5-branes are replaced by NS5-branes that fractionate on the orientifold, and the M9-plane is reduced to an $O8^-$-orientifold-plane paired with $8$ full D8-branes plus an $ON^0$ plane living at the intersection of the two orientifold planes \cite{Hanany:1999sj}. For $n=2k$, a particular choice of flat connection for the $E_8$-bundle associated with the M9-plane exists such that the Type IIA brane engineering takes the following form:
\begin{align}
\raisebox{-.5\height}{
\includegraphics[page=1]{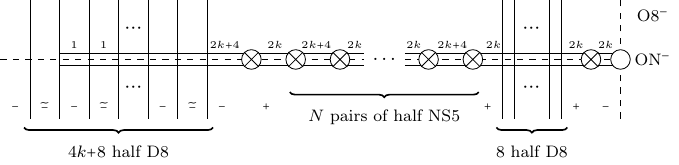}
}\,.
\end{align}
The 6d $(1,0)$ SCFT engineered by this brane configuration can be captured by writing the following generalized quiver:
\begin{align}
\raisebox{-.5\height}{
\includegraphics[page=2]{orbi-instanton_figures.pdf}
}\,.
\end{align}
In 6d SQFTs, there are vector multiplets, here associated with the written Lie algebra, and tensor multiplets, one for each simple gauge algebra and associated with self-dual strings whose string charge is given by the corresponding bold integer. There are also hypermultiplets, however the hypermultiplet spectrum is generally fixed by anomaly cancellation after the vector and tensor spectrum if specified. The SCFT is obtained by taking all the gauge couplings simultaneously to infinity. In Calabi--Yau geometry, the string charge associated to the tensor multiplets are the self-intersection numbers of smooth rational curves, and the geometry giving rise to the SCFT is obtained via simultaneous contraction of each of these curves. For a detailed introduction to the language of tensor branch curve configurations in F-theory compactifications to six dimensions, we refer the reader to the review \cite{Heckman:2018jxk}.

We can now apply the technique of passing to the magnetic phase of the Type IIA brane system to extract a magnetic quiver for the Higgs branch of the 6d $(1,0)$ SCFT (i.e., where all the gauge couplings in the tensor branch description are taken to infinity) \cite{Cabrera:2019dob,Sperling:2021fcf}. The magnetic brane system is: 
\begin{align}
\raisebox{-.5\height}{
\includegraphics[page=3]{orbi-instanton_figures.pdf}
}\,,
\end{align} 
where the Higgs branch moduli near the intersection of the orientifolds are arranged as follows
\begin{align}
\raisebox{-.5\height}{
\includegraphics[page=4]{orbi-instanton_figures.pdf}
}
\qquad \longrightarrow \qquad 
\raisebox{-.5\height}{
\includegraphics[page=5]{orbi-instanton_figures.pdf}
}\,.
\end{align}
On the right, we have colour-coded the nodes in the resulting quiver to highlight which stacks of D6-branes they correspond to. The complete magnetic quiver for the Higgs branch of the 6d $(1,0)$ SCFT, which is the 3d $\mathcal{N}=4$ object we are interested in, is thus read off as
\begin{align}
\raisebox{-.5\height}{
\includegraphics[page=6,width=0.85\linewidth]{orbi-instanton_figures.pdf}
} \,.
\end{align}
We note that for $N=0$ this coincides with the magnetic quiver derived in \cite{Sperling:2021fcf}. As we can determine from the balance of the nodes in this quiver, the global symmetry is generically (for $N > 0$) given by
\begin{align}
\mathfrak{f} = \mathfrak{so}(4k+8) \oplus \mathfrak{so}(8) \oplus \mathfrak{so}(8)
\,.
\end{align}
For the rank zero D-type orbi-instanton, the 6d $(1,0)$ SCFT is in fact a product theory, and correspondingly the global symmetry is enhanced. The product is two copies of the 6d $(1,0)$ SCFTs known as rank one $(\mathfrak{so}(2k), \mathfrak{so}(2k))$ conformal matter; it would be interesting to apply the Orthosymplectic Decay and Fission algorithm to this magnetic quiver (which is not itself a product) and recover the product nature of the Higgs branch.

From the perspective of the M-theory realization of the orbi-instanton theories, a Higgs branch deformation that separates $\ell$ of the $N$ M5-branes out of the M9-brane is expected. As a benchmark of the Orthosymplectic Decay and Fission algorithm from Section \ref{sec:Algorithm}, we would like to be able to reproduce this Higgs branch RG flow from the magnetic quiver for the Higgs branch. For simplicity we consider the case where $k = 0$; then, the SCFTs is specified by the following curve configuration at the generic point of the tensor branch:\footnote{It would be excessively cumbersome to provide a careful introduction for this notation to capture 6d $(1,0)$ SCFTs in this paper. We instead refer the reader to \cite{Heckman:2018jxk,Baume:2021qho,Baume:2023onr} for comprehensive reviews.}
\begin{equation}\label{eqn:D-type_OI}
    [\mathfrak{so}_{8}]\underbrace{1\stackon{$4$}{$\mathfrak{so}_8$}\cdots}_{N-1} \stackon{$1\stackon{$4$}{$\mathfrak{so}_8$}$}{\stackon{$1$}{$[\mathfrak{so}_{8}]$}}1[\mathfrak{so}_{8}] \,.
\end{equation}
The magnetic quiver for the Higgs branch then takes the following form:
\begin{equation}\label{eqn:orbiinsta_before_higgs}
\begin{gathered}
    \includegraphics[page=10,scale=0.75]{DecayAndFission_figures.pdf}
\end{gathered} \,.
\end{equation}
This is a star-shaped orthosymplectic quiver, and it is straightforward to apply the Orthosymplectic Decay and Fission algorithm. While the resulting Coulomb branch Hasse diagram is rather large and intricate, it is easy to show that, for each $0 < \ell \leq N$ there exists a fission to the following product of quivers:
\begin{equation}\label{eqn:orbiinstantoafterfission}
\begin{gathered}
    \includegraphics[page=11,scale=0.75]{DecayAndFission_figures.pdf} 
\end{gathered}\,.
\end{equation}
The unitary quiver appearing in the product after fission  is nothing other than a magnetic quiver for the Higgs branch of the 6d $(1,0)$ SCFT engineered as a stack of $\ell$ M5-branes, not probing any orbifold, inside of an M9-plane \cite{Cabrera:2019izd,Lawrie:2023uiu}; this is the rank $\ell$ E-string theory. In the F-theory language, this transition corresponds to the Higgsing 
\begin{equation}
        [\mathfrak{so}_{8}]\underbrace{1\stackon{$4$}{$\mathfrak{so}_8$}\cdots}_{N-1} \stackon{$1\stackon{$4$}{$\mathfrak{so}_8$}$}{\stackon{$1$}{$[\mathfrak{so}_{8}]$}}1[\mathfrak{so}_{8}] 
        \quad \rightarrow \quad 
        [\mathfrak{so}_{8}]\underbrace{1\stackon{$4$}{$\mathfrak{so}_8$}\cdots}_{N-1-\ell} \stackon{$1\stackon{$4$}{$\mathfrak{so}_8$}$}{\stackon{$1$}{$[\mathfrak{so}_{8}]$}}1[\mathfrak{so}_{8}]
        \quad \sqcup \quad [\mathfrak{e}_8]\underbrace{\,12\cdots 2\,}_{\ell}[\mathfrak{su}_2] \,\,.
\end{equation}
Although this feature is something expected from all rank $N$ $(\mathfrak{e}_8, \mathfrak{g})$ orbi-instanton theories \cite{Lawrie:2023uiu,Mansi:2023faa,Heckman:2015ola}, this was the first time that an explicit example of this phenomenon was carried out beyond the $\mathfrak{g}=\mathfrak{su}(K)$ case.\footnote{Intriguingly, there is a growing body of evidence that torus compactifications (with Stiefel--Whitney and outer-automorphism twists) followed by mass deformations of orbi-instanton SCFTs provides a higher-dimensional origin for all 4d $\mathcal{N}=2$ SCFTs, see, for example, \cite{Martone:2021drm,Heckman:2022suy,Giacomelli:2024dbd,Giacomelli:2024ycb}. While we remain agnostic on such a conjecture, it highlights that understanding the Higgs branches of orbi-instanton SCFTs can have far-reaching implications.}

\subsection{Higher-rank Class \texorpdfstring{$\mathcal{S}$}{S}: Beyond Partial Puncture Closure}

General class $\mathcal{S}_\mathfrak{g}\langle C_{g,n} \rangle \{ O_1, \cdots, O_n \} $ theories with $\mathfrak{g}=\mathfrak{so}_{2k}$ and their Higgsings have been studied in \cite{Chacaltana:2011ze}. In the literature, there has been control over such theories leveraging the association between certain Higgsings, nilpotent orbits, and partial puncture closure in class $\mathcal{S}$. Such interplays allow us to predict Higgs branch RG flows of the 4d $\mathcal{N}=2$ SCFTs that are localized on a puncture via the corresponding nilpotent data, leading in the extreme case to the full close of the puncture itself, which changes the Riemann surface to $C_{g,n-1}$. 

However, there are more general Higgsings in class $\mathcal{S}$ going beyond just partial puncture closure. More general Higgsing directions take the class $\mathcal{S}$ of type $\mathfrak{g}$ to (possibly a product of) class $\mathcal{S}$ of type $\mathfrak{g}'$ for $\mathfrak{g}'\subset \mathfrak{g}$ a Levi subalgebra of $\mathfrak{g}$. Such possibilities were fully explored only for $\mathfrak{g}=\mathfrak{su}(n)$, where exploiting the unitary-nature of the 3d mirrors for the associated class $\mathcal{S}$ theories \cite{Benini:2010uu}, the Decay and Fission algorithm \cite{Bourget:2024mgn,Bourget:2023dkj} demonstrates the transition
\begin{equation}
    \mathcal{S}_{\mathfrak{su}_n}\langle C_{g,n} \rangle \{ O_1, \cdots, O_n \} 
    \; \rightarrow \;
    \mathcal{S}_{\mathfrak{su}_{n-k}}\langle C_{g',n} \rangle \{ O'_1, \cdots, O'_n \} \times \mathcal{S}_{\mathfrak{su}_{k}}\langle C_{g'',n} \rangle \{ O''_1, \cdots, O''_n \}\,.
\end{equation}
Since our algorithm implements Decay and Fission for orthosymplectic quiver theories (without additional antisymmetric matter), we can explore the full Higgs branch of 
\begin{equation}
    \mathcal{S}_{\mathfrak{so}_{2n}}\langle C_{0,n} \rangle \{ O_1, \cdots, O_n \} \,,
\end{equation}
and appreciate the Higgs branch renormalization group flow corresponding to the decomposition of $\mathfrak{so}(2n)$ into the following Levi subalgebras:
\begin{equation}
    \mathfrak{so}({2n}) \rightarrow \mathfrak{so}({2n-2k}) \oplus \mathfrak{su}(k) \,.
\end{equation}
In this section, we highlight an interesting example of the type-changing Higgs branch RG flows expected to exist from the Orthosymplectic Decay and Fission algorithm.

For simplicity, we consider only at a subset of all the possible Higgs branch renormalization group flows  of the theory
\begin{equation}
    \mathcal{S}_{\mathfrak{so}_{12}}\langle C_{0,4} \rangle \{ [3^2,1^{6}], [5,3,1^4],[9,3],[5^2,1^2]  \} \,,
\end{equation}
which are sufficient to demonstrate the most notable features of higher rank class $\mathcal{S}$. The 3d mirror for this theory is the following star-shaped orthosymplectic quiver:
\begin{equation}\label{eqn:QuiverClassSso10}
\begin{gathered}
        \includegraphics[page=18]{DecayAndFission_figures.pdf}
\end{gathered} \,\,\,.
\end{equation}
We note that all the gauge nodes are either balanced or good, leading to an explicit non-Abelian $\mathfrak{so}(6) \oplus \mathfrak{so}(4) \oplus \mathfrak{so}(4)$ global symmetry.\footnote{We have not analyzed extra Abelian factors not emerging from balanced subsets.}

The Orthosymplectic Decay and Fission algorithm applied to the quiver in equation \eqref{eqn:QuiverClassSso10} indicates the existence of a fission process determined by the following algebras splitting
\begin{equation}
\begin{aligned}
    & \mathfrak{so}({12}) \rightarrow \mathfrak{so}({8}) \oplus \mathfrak{u}({2}) \,,\\
    & \mathfrak{usp}({8}) \rightarrow \mathfrak{usp}({6}) \oplus \mathfrak{u}({1}) \,, \\
    & \mathfrak{usp}({6}) \rightarrow \mathfrak{usp}({4}) \oplus \mathfrak{u}({1}) \,, \\
    & \mathfrak{usp}({2}) \rightarrow \mathfrak{\varnothing} \oplus \mathfrak{u}({1}) \,, 
\end{aligned}
\end{equation}
where the fission happens on the central $SO(12)$ gauge node and the four surrounding gauge nodes only. The end result of this fission is the following product theory:
\begin{equation}\label{eqn:producClaasSso8andClassSsu2}
\begin{gathered}
    \includegraphics[page=19,width=0.85\linewidth]{DecayAndFission_figures.pdf}
\end{gathered} \,\,.
\end{equation}
This replicates the following Higgs branch renormalization group flow in class $\mathcal{S}$:
\begin{equation}\label{eqn:kakuna}
\begin{gathered}
\begin{aligned}
    \mathcal{S}_{\mathfrak{so}_{12}}\langle C_{0,4} \rangle \{ [3^2,1^{6}], &[5,3,1^4],[9,3],[5^2,1^2]  \} \longrightarrow \\
    &\mathcal{S}_{\mathfrak{so}_{8}}\langle C_{0,3} \rangle \{ [1^{8}], [3^2,1^2],[3,1^5] \} \times \mathcal{S}_{\mathfrak{su}_{2}}\langle C_{0,4} \rangle \{ [1^{2}], [1^2],[1^2],[1^2] \} \,,
\end{aligned}  
\end{gathered}
\end{equation}
which itself stems from the existence of an $\mathfrak{so}(8) \oplus \mathfrak{su}(2)$ Levi subalgebra in $\mathfrak{so}(12)$.

Let us now further apply the Orthosymplectic Decay and Fission  algorithm to the 3d mirror of the $\mathcal{S}_{\mathfrak{so}_{8}}\langle C_{0,3} \rangle \{ [1^{8}], [3^2,1^2],[3,1^5] \}$ class $\mathcal{S}$ theory appearing in equation \eqref{eqn:kakuna}. We point out that this theory also appears in the class $\mathcal{S}$ analysis of \cite{Chacaltana:2011ze}; hence, we have two independent ways to verify that the global symmetry is $\mathfrak{so}(10)\oplus \mathfrak{usp}(4) \oplus \mathfrak{u}(1)$. The $\mathfrak{so}(10)$ symmetry can be understood from the magnetic quiver by noticing that the central note is balanced; therefore, minimal nilpotent Higgsing of the moment map of this symmetry is an action that cannot be reproduced by only partial puncture closure in the class $\mathcal{S}$ picture. From the magnetic quiver, it corresponds to decaying the set of balanced nodes generating this symmetry via Table \ref{tab:admitted decays}. This operation recovers the quiver in equation \eqref{eqn:ClassSHiggsableE6E7a7SO}; this justifies the class $\mathcal{S}$ Higgs branch transition:
\begin{equation}
    \mathcal{S}_{\mathfrak{so}_{8}}\langle C_{0,3} \rangle \{ [1^{8}], [3^2,1^2],[3,1^5] \} \; \xrightarrow[]{d_5}  \;\mathcal{S}_{\mathfrak{so}_6}\langle C_{0,3} \rangle \{ [1^6], [1^6],[2^2,1^2] \} \,.
\end{equation}
We note that the dimension of the conjectured transverse slice ($\mathrm{dim}_\mathbb{H}\, d_n= 2n-3$, where $d_n$ again denotes the closure of the minimal nilpotent orbit of $\mathfrak{so}(2n)$) correctly reproduces the change in Coulomb branch dimension of the 3d quiver.

Alternatively, we can decay the $\mathfrak{so}(4)$ symmetry involving the balanced $SO(12)$ node of equation \eqref{eqn:QuiverClassSso10} and reduce by $1$ the rank of the bad $USp$-type gauge node resulting from the process. The outcome of this procedure is the 3d mirror of the $\mathfrak{so}(10)$ class $\mathcal{S}$ theory $\mathcal{S}_{\mathfrak{so}_{10}}\langle C_{0,4} \rangle \{ [3,1^{7}], [3^2,1^4],[5,3,1^2],[7,3] \} $, which is
\begin{equation}\label{eqn:dratini}
    \begin{gathered}
        \includegraphics[page=20]{DecayAndFission_figures.pdf}
    \end{gathered} \,.
\end{equation} 
This 3d mirror exhibits an $\mathfrak{so}(7)$ factor in the global symmetry from the balanced nodes on the left tail, agreeing with expectations from the Hall--Littlewood index of the class $\mathcal{S}$ theory. 

We can perform a further sequence of decays on the quiver in equation \eqref{eqn:dratini}. Fully closing the puncture associated with the stand-alone $USp(2)$ node, i.e., decaying this node, renders the central $SO(10)$ balanced which, together with the balanced $USp(6)$, give rise to an $\mathfrak{so}(3)$ symmetry that can be further decayed, leading to the 3d mirror of the $\mathfrak{so}(8)$ class $\mathcal{S}$ theory, $\mathcal{S}_{\mathfrak{so}_{8}}\langle C_{0,3} \rangle \{ [1^{8}], [3^2,1^2],[2^2,1^4] \} $:
\begin{equation}
    \begin{gathered}
        \includegraphics[page=21]{DecayAndFission_figures.pdf}
    \end{gathered} \,.
\end{equation} 
Finally, a further decay of the bad $USp(6)$ node on the right tail leads to the orthosymplectic quiver in equation \eqref{eqn:producClaasSso8andClassSsu2}. As we have discussed, this theory can be further decayed to the orthosymplectic quiver shown in equation \eqref{eqn:ClassSHiggsableE6E7a7SO}. 

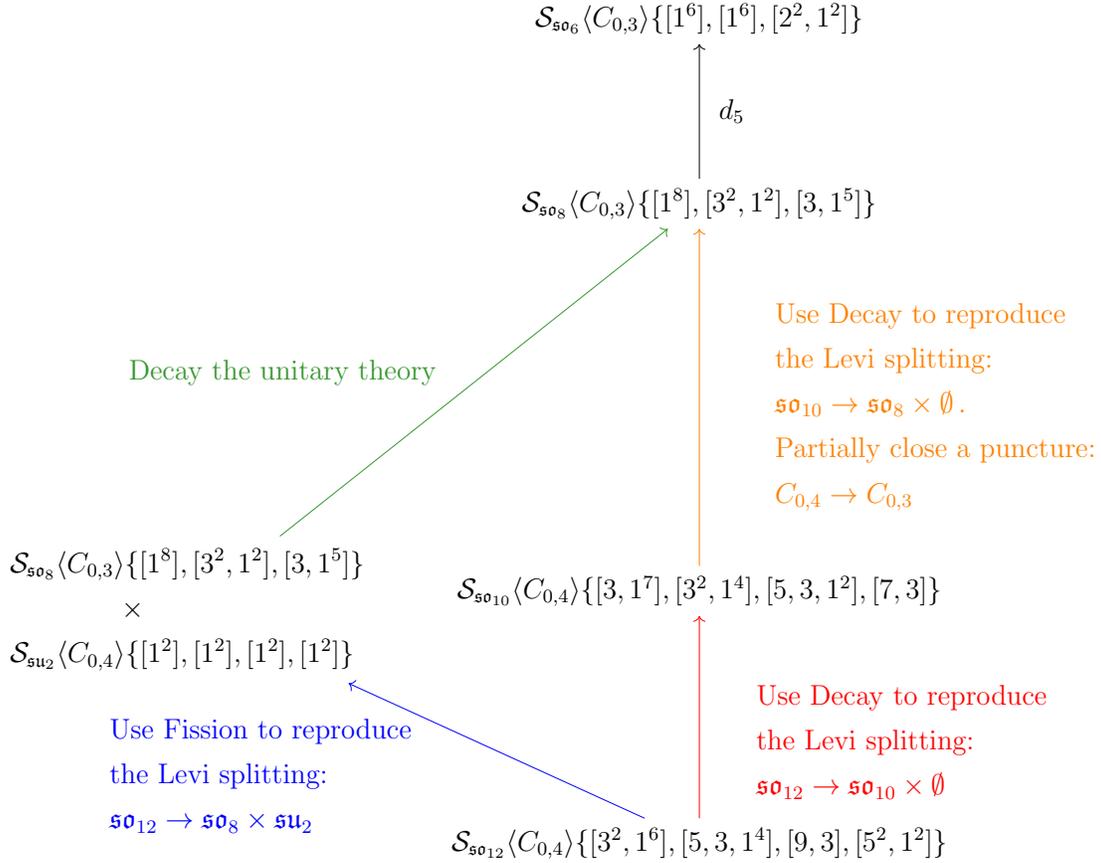
\begin{figure}[t]
    \centering
    $\begin{gathered}
    \resizebox{0.9\linewidth}{!}{
        \begin{tikzpicture}
            \node at (0,0) (tso12) { $\mathcal{S}_{\mathfrak{so}_{12}}\langle C_{0,4} \rangle \{ [3^2,1^{6}], [5,3,1^4],[9,3],[5^2,1^2]  \}$};

\node [above left= 2cm and 1cm of tso12] (tproduct) { $\begin{gathered}
    \begin{aligned}
        &\mathcal{S}_{\mathfrak{so}_{8}}\langle C_{0,3} \rangle \{ [1^{8}], [3^2,1^2],[3,1^5] \}\\
        & \qquad \qquad\times \\
        &\mathcal{S}_{\mathfrak{su}_{2}}\langle C_{0,4} \rangle \{ [1^{2}], [1^2],[1^2],[1^2] \}
    \end{aligned}
\end{gathered}$};

                \node [above=3cm of tso12] (tso10) { $\mathcal{S}_{\mathfrak{so}_{10}}\langle C_{0,4} \rangle \{ [3,1^{7}], [3^2,1^4],[5,3,1^2],[7,3] \}$};
                \node [above=5cm of tso10] (tso8) { $\mathcal{S}_{\mathfrak{so}_{8}}\langle C_{0,3} \rangle \{ [1^{8}], [3^2,1^2],[3,1^5] \}$};
                \node [above=2cm of tso8] (tso6) { $\mathcal{S}_{\mathfrak{so}_6}\langle C_{0,3} \rangle \{ [1^6], [1^6],[2^2,1^2] \}$};

        \draw[->,blue] (tso12)--(tproduct); \node[blue] () at (-6.5,1) {$\begin{gathered}
            \begin{aligned}
                &\text{Use Fission to reproduce}\\
                &\text{the Levi splitting:}\\
                &\mathfrak{so}_{12}\rightarrow\mathfrak{so}_8\times \mathfrak{su}_2
            \end{aligned}
        \end{gathered}$};
        \draw[->,red] (tso12)--(tso10); \node[red] () at (3,1.5) {$\begin{gathered}
            \begin{aligned}
                &\text{Use Decay to reproduce}\\
                &\text{the Levi splitting:}\\
                &\mathfrak{so}_{12}\rightarrow\mathfrak{so}_{10}\times \emptyset
            \end{aligned}
        \end{gathered}$};
        \draw[->,orange] (tso10)--(tso8); \node[orange] () at (3.5,6.5) {$\begin{gathered}
            \begin{aligned}
                &\text{Use Decay to reproduce}\\
                &\text{the Levi splitting:}\\
                &\mathfrak{so}_{10}\rightarrow\mathfrak{so}_{8}\times \emptyset\,.\\
                &\text{Partially close a puncture:}\\
                &C_{0,4}\rightarrow C_{0,3}
            \end{aligned}
        \end{gathered}$};
        \draw[->,myGreen] (tproduct)--(tso8); \node[myGreen] () at (-6,7) {$\begin{gathered}
            \begin{aligned}
                &\text{Decay the unitary theory}
            \end{aligned}
        \end{gathered}$};
        \draw[->,black] (tso8)--(tso6) node[midway,label=right:{$d_5$}] () {};
        \end{tikzpicture}
        }
    \end{gathered}$
    \caption{Some of the Higgs branch RG flows between class $\mathcal{S}$ theories of different type conjectured to exist from the application of the Orthosymplectic Decay and Fission algorithm applied to the 3d mirror of the class $\mathcal{S}$ theory at the bottom. As explained in the text, the orange arrow does not represent a minimal transition, whereas all other arrows are minimal.}
    \label{fig:finalfig}
\end{figure}

We have depicted these sequences of decays and fissions in Figure \ref{fig:finalfig}. This result is particularly interesting as this Higgs branch structure, in particular the existence of at least two separate Higgs branch renormalization group flows between the class $\mathcal{S}$ theories
\begin{equation}
    \mathcal{S}_{\mathfrak{so}_{12}}\langle C_{0,4} \rangle \{ [3^2,1^{6}], [5,3,1^4],[9,3],[5^2,1^2] \,,
\end{equation}
and
\begin{equation}
    \mathcal{S}_{\mathfrak{so}_6}\langle C_{0,3} \rangle \{ [1^6], [1^6],[2^2,1^2] \} \,,
\end{equation}
which are not manifest from the class $\mathcal{S}$ perspective. We leave the construction of the explicit Higgsing on the purely class $\mathcal{S}$ side for future work, however such a pattern matches the expected behavior from the notions of simultaneously deletion in class $\mathcal{S}$ theories proposed in \cite{DKL}. We remark that this result was, before our algorithm, inaccessible from the perspective of the 3d mirror.

\section{Discussion}\label{sec:discussion}

Quiver gauge theories in three dimensions have proven themselves a fruitful and enriching object of study both for understanding their intrinsic physical phenomena and also as a portal to access information difficult to extract from higher dimensional theories. In this spirit, a long-standing problem has been the understanding of the Higgs branch and the associated renormalization group flows and Higgsing patterns of supersymmetric quantum field theories in $d \geq 4$. A portal on the sought information was opened, following the development of the magnetic quiver technology \cite{Cabrera:2018jxt,Cabrera:2019izd}, via the study of Coulomb branches of 3d $\mathcal{N}=4$ quiver gauge theories. While multiple efforts have been proven successful in extracting partial Higgsing patterns \cite{Bourget:2019aer} from unitary quiver gauge theories \cite{Cabrera:2018ann, Bourget:2023dkj, Bourget:2024mgn}, analogous results have been only partially found for orthosymplectic theories \cite{Cabrera:2019dob}. 

In our work, we propose an algorithm that predicts the Higgsing pattern for simply-laced orthosymplectic quiver theories (with edge multiplicity one). Multiple features, both expected and unexpected, emerge from the algorithm. Similar to what was shown in the unitary Decay and Fission case, an orthosymplectic theory can only Higgs to a theory with a gauge group of lower ranks or fission to a product of theories. Unexpectedly, we found that an orthosymplectic quiver theory can undergo fission into an orthosymplectic theory and a unitary one, or in the extreme case, a purely unitary theory when it unitarises. For instance, using the known existence of a 3d mirror, we can now predict the full Higgs branch Hasse diagram for type $\mathfrak{so}({2n})$ class $\mathcal{S}$ theories. Similarly, we demonstrate that certain 6d $(1,0)$ D-type orbi-instanton SCFTs have Higgs branch RG flows that force them to split into specific product theories, consistent with the idea of separating stacks of M5-branes in their construction via M-theory.

The algorithm developed in this paper applies to all orthosymplectic quivers. Therefore, equipped with this technology, we can expand our knowledge of the Higgs mechanism across dimensions. In general, orthosymplectic quivers that are magnetic quivers of higher dimensional SCFTs do not have known unitary counterparts. Therefore, our algorithm now provides a method of Higgsing along the Higgs branch for these SCFTs. In this spirit, a direction to pursue is extending the algorithm to encompass more exotic matter, such as $2$-rank antisymmetric hypermultiplets, and quivers with non-simply-laced edges which are key features of known magnetic quivers for the Higgs branches of certain higher dimensional SCFTs. Moreover, our work lays the ground for a more detailed justification of the algorithm from a purely mathematical study of the landscape of symplectic singularities that can be built via Coulomb branches of orthosymplectic quivers.

\section*{Acknowledgements}
We would like to thank Jiakang Bao, Sam Bennett, Riccardo Comi, Jacques Distler, Amihay Hanany, Monica Jinwoo Kang, Guhesh Kumaran, Deshuo Liu, and Sara Pasquetti for interesting discussions and useful comments.
The authors thank the Mainz Institute for Theoretical Physics for hospitality at the 2024 workshop on the ``Geometry of SCFTs'' during the initial phase of this work. C.L.~and L.M.~acknowledge support from DESY (Hamburg, Germany), a member of the Helmholtz Association HGF; C.L.~and L.M.~also acknowledges the Deutsche Forschungsgemeinschaft under Germany's Excellence Strategy - EXC 2121 ``Quantum Universe'' - 390833306 and the Collaborative Research Center - SFB 1624 ``Higher Structures, Moduli Spaces, and Integrability'' - 506632645. Z.Z.~is supported by the ERC Consolidator Grant \# 864828 ``Algebraic Foundations of Supersymmetric Quantum Field Theory'' (SCFTAlg). 
The work of M.S.~is supported by the Austrian Science Fund (FWF), START project ``Phases of quantum field theories: symmetries and vacua'' STA 73-N [grant DOI: 10.55776/STA73]. M.S.~also acknowledge support from the Faculty of Physics, University of Vienna.

\appendix

\section{Higgsing 3d \texorpdfstring{\boldmath{$\mathcal{N}=4$}}{N=4} \texorpdfstring{\boldmath{$T^{\sigma}_{\rho}(G)$}}{Trhosigma(G)} theories}\label{sec:Appendix_Trho}

In this paper we have proposed an Orthosymplectic Decay and Fission algorithm for extracting the structure of the Coulomb branch of orthosymplectic quivers. So far, the quivers we have studied, inspired by class $\mathcal{S}$, are star-shaped unframed/flavourless orthosymplectic quivers. These are theories that are well known to be magnetic quivers for the Higgs branch/3d mirrors of class $\mathcal{S}$ theories and other higher dimensional SCFTs. 

In this appendix, we would like to go beyond this paradigm and consider the Orthosymplectic Decay and Fission as applied to \emph{framed} orthosymplectic quivers, that is, those with flavour nodes. Examples of such theories are the $T_\rho^\sigma(G)$ from \cite{Gaiotto:2008ak}, where $G$ is orthosymplectic. 

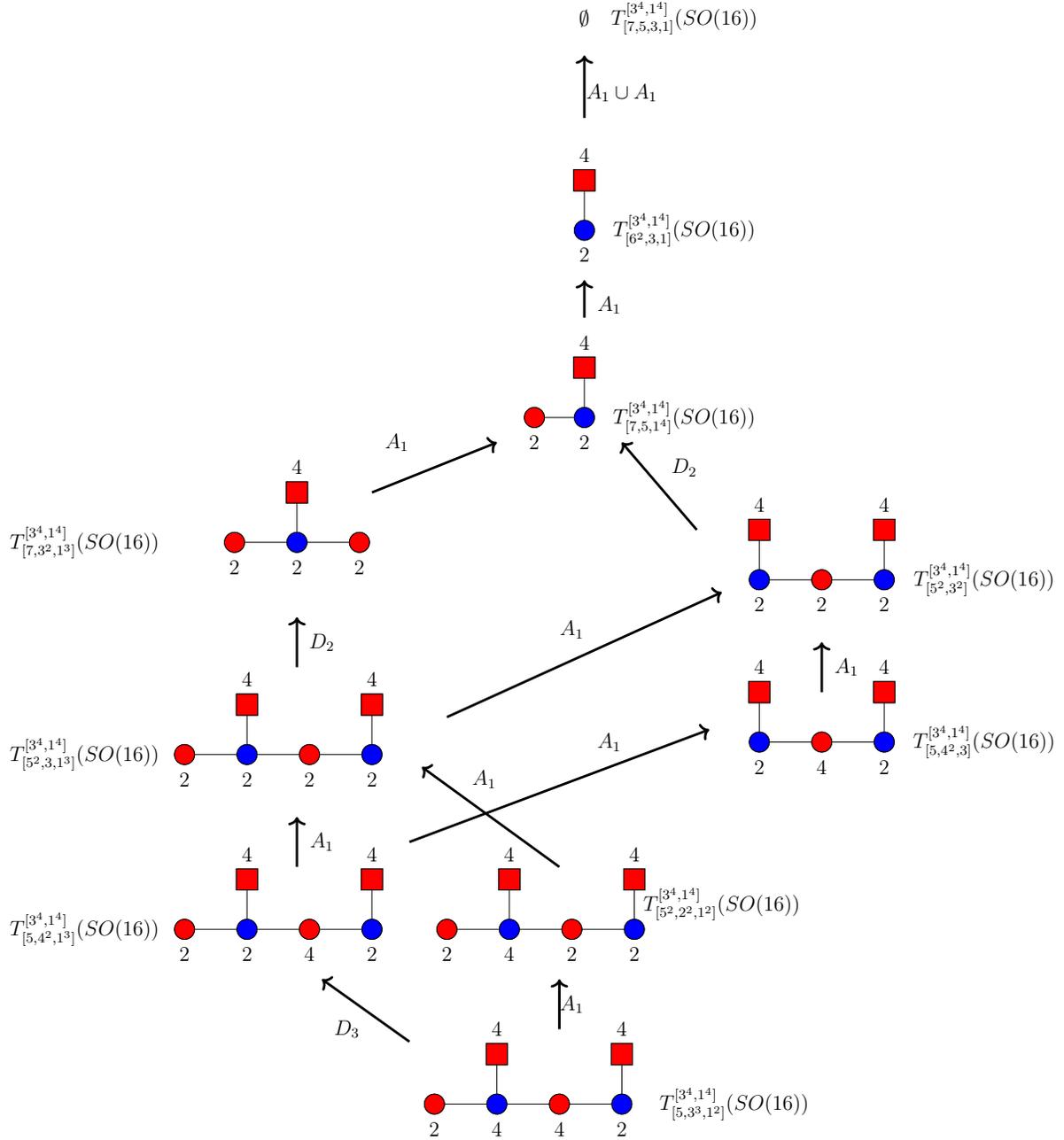
\begin{figure}[p]
    \centering
    $\begin{gathered}
    \scalebox{0.75}{\begin{tikzpicture}
	\begin{pgfonlayer}{nodelayer}
		\node [style=gauge3] (0) at (5.5, -6.75) {};
		\node [style=gauge3] (1) at (6.75, -6.75) {};
		\node [style=gauge3] (2) at (8, -6.75) {};
		\node [style=gauge3] (3) at (9.25, -6.75) {};
		\node [style=flavour2] (4) at (9.25, -5.75) {};
		\node [style=flavour2] (5) at (6.75, -5.75) {};
		\node [style=none] (6) at (9.25, -5.25) {4};
		\node [style=none] (7) at (6.75, -5.25) {4};
		\node [style=redgauge] (8) at (5.5, -6.75) {};
		\node [style=redgauge] (9) at (8, -6.75) {};
		\node [style=bluegauge] (10) at (9.25, -6.75) {};
		\node [style=bluegauge] (11) at (6.75, -6.75) {};
		\node [style=redflavour] (12) at (9.25, -5.75) {};
		\node [style=redflavour] (13) at (6.75, -5.75) {};
		\node [style=none] (14) at (5.5, -7.25) {2};
		\node [style=none] (15) at (6.75, -7.25) {4};
		\node [style=none] (16) at (8, -7.25) {4};
		\node [style=none] (17) at (9.25, -7.25) {2};
		\node [style=gauge3] (33) at (0.5, 0.25) {};
		\node [style=gauge3] (34) at (1.75, 0.25) {};
		\node [style=gauge3] (35) at (3, 0.25) {};
		\node [style=gauge3] (36) at (4.25, 0.25) {};
		\node [style=flavour2] (37) at (4.25, 1.25) {};
		\node [style=flavour2] (38) at (1.75, 1.25) {};
		\node [style=none] (39) at (4.25, 1.75) {4};
		\node [style=none] (40) at (1.75, 1.75) {4};
		\node [style=redgauge] (41) at (0.5, 0.25) {};
		\node [style=redgauge] (42) at (3, 0.25) {};
		\node [style=bluegauge] (43) at (4.25, 0.25) {};
		\node [style=bluegauge] (44) at (1.75, 0.25) {};
		\node [style=redflavour] (45) at (4.25, 1.25) {};
		\node [style=redflavour] (46) at (1.75, 1.25) {};
		\node [style=none] (47) at (0.5, -0.25) {2};
		\node [style=none] (48) at (1.75, -0.25) {2};
		\node [style=none] (49) at (3, -0.25) {2};
		\node [style=none] (50) at (4.25, -0.25) {2};
		\node [style=gauge3] (51) at (1.5, 4.5) {};
		\node [style=gauge3] (52) at (2.75, 4.5) {};
		\node [style=gauge3] (53) at (4, 4.5) {};
		\node [style=flavour2] (54) at (2.75, 5.5) {};
		\node [style=none] (55) at (2.75, 6) {4};
		\node [style=redgauge] (56) at (1.5, 4.5) {};
		\node [style=redgauge] (57) at (4, 4.5) {};
		\node [style=bluegauge] (58) at (2.75, 4.5) {};
		\node [style=redflavour] (59) at (2.75, 5.5) {};
		\node [style=none] (60) at (1.5, 4) {2};
		\node [style=none] (61) at (2.75, 4) {2};
		\node [style=none] (62) at (4, 4) {2};
		\node [style=gauge3] (63) at (12, 3.75) {};
		\node [style=gauge3] (64) at (13.25, 3.75) {};
		\node [style=gauge3] (65) at (14.5, 3.75) {};
		\node [style=flavour2] (66) at (14.5, 4.75) {};
		\node [style=flavour2] (67) at (12, 4.75) {};
		\node [style=none] (68) at (14.5, 5.25) {4};
		\node [style=none] (69) at (12, 5.25) {4};
		\node [style=redgauge] (70) at (13.25, 3.75) {};
		\node [style=bluegauge] (71) at (14.5, 3.75) {};
		\node [style=bluegauge] (72) at (12, 3.75) {};
		\node [style=redflavour] (73) at (14.5, 4.75) {};
		\node [style=redflavour] (74) at (12, 4.75) {};
		\node [style=none] (75) at (12, 3.25) {2};
		\node [style=none] (76) at (13.25, 3.25) {2};
		\node [style=none] (77) at (14.5, 3.25) {2};
		\node [style=none] (78) at (5, -5.5) {};
		\node [style=none] (79) at (3.25, -4.25) {};
		\node [style=none] (80) at (10, -5.75) {};
		\node [style=none] (81) at (14.25, -0.75) {};
		\node [style=none] (84) at (2.75, 2) {};
		\node [style=none] (85) at (2.75, 3) {};
		\node [style=none] (86) at (3.75, -5.25) {$D_3$};
		\node [style=none] (89) at (3.25, 2.5) {$D_2$};
		\node [style=gauge3] (90) at (7.5, 7) {};
		\node [style=gauge3] (91) at (8.5, 7) {};
		\node [style=flavour2] (92) at (8.5, 8) {};
		\node [style=none] (93) at (8.5, 8.5) {4};
		\node [style=redgauge] (94) at (7.5, 7) {};
		\node [style=bluegauge] (95) at (8.5, 7) {};
		\node [style=redflavour] (96) at (8.5, 8) {};
		\node [style=none] (97) at (7.5, 6.5) {2};
		\node [style=none] (98) at (8.5, 6.5) {2};
		\node [style=gauge3] (99) at (8.5, 10.75) {};
		\node [style=flavour2] (100) at (8.5, 11.75) {};
		\node [style=none] (101) at (8.5, 12.25) {4};
		\node [style=bluegauge] (102) at (8.5, 10.75) {};
		\node [style=redflavour] (103) at (8.5, 11.75) {};
		\node [style=none] (104) at (8.5, 10.25) {2};
		\node [style=none] (105) at (4.25, 5.5) {};
		\node [style=none] (106) at (6.75, 6.5) {};
		\node [style=none] (107) at (9.25, 6.5) {};
		\node [style=none] (108) at (10.75, 4.75) {};
		\node [style=none] (109) at (8.5, 9) {};
		\node [style=none] (110) at (8.5, 9.75) {};
		\node [style=none] (111) at (4.75, 6.5) {$A_1$};
		\node [style=none] (112) at (10.5, 6) {$D_2$};
		\node [style=none] (113) at (9, 9.25) {$A_1$};
		\node [style=gauge3] (114) at (0.5, -3.25) {};
		\node [style=gauge3] (115) at (1.75, -3.25) {};
		\node [style=gauge3] (116) at (3, -3.25) {};
		\node [style=gauge3] (117) at (4.25, -3.25) {};
		\node [style=flavour2] (118) at (4.25, -2.25) {};
		\node [style=flavour2] (119) at (1.75, -2.25) {};
		\node [style=none] (120) at (4.25, -1.75) {4};
		\node [style=none] (121) at (1.75, -1.75) {4};
		\node [style=redgauge] (122) at (0.5, -3.25) {};
		\node [style=redgauge] (123) at (3, -3.25) {};
		\node [style=bluegauge] (124) at (4.25, -3.25) {};
		\node [style=bluegauge] (125) at (1.75, -3.25) {};
		\node [style=redflavour] (126) at (4.25, -2.25) {};
		\node [style=redflavour] (127) at (1.75, -2.25) {};
		\node [style=none] (128) at (0.5, -3.75) {2};
		\node [style=none] (129) at (1.75, -3.75) {2};
		\node [style=none] (130) at (3, -3.75) {4};
		\node [style=none] (131) at (4.25, -3.75) {2};
		\node [style=none] (132) at (2.75, -2) {};
		\node [style=none] (133) at (2.75, -1) {};
		\node [style=none] (134) at (3.25, -1.5) {$A_1$};
		\node [style=none] (135) at (8.5, 15) {$\emptyset$};
		\node [style=none] (136) at (8.5, 13) {};
		\node [style=none] (137) at (8.5, 14.25) {};
		\node [style=none] (138) at (9.25, 13.5) {$A_1 \cup A_1$};
		\node [style=none] (139) at (5.75, 1) {};
		\node [style=none] (140) at (11.25, 3.5) {};
		\node [style=none] (141) at (8.25, 2.75) {$A_1$};
		\node [style=gauge3] (142) at (5.75, -3.25) {};
		\node [style=gauge3] (143) at (7, -3.25) {};
		\node [style=gauge3] (144) at (8.25, -3.25) {};
		\node [style=gauge3] (145) at (9.5, -3.25) {};
		\node [style=flavour2] (146) at (9.5, -2.25) {};
		\node [style=flavour2] (147) at (7, -2.25) {};
		\node [style=none] (148) at (9.5, -1.75) {4};
		\node [style=none] (149) at (7, -1.75) {4};
		\node [style=redgauge] (150) at (5.75, -3.25) {};
		\node [style=redgauge] (151) at (8.25, -3.25) {};
		\node [style=bluegauge] (152) at (9.5, -3.25) {};
		\node [style=bluegauge] (153) at (7, -3.25) {};
		\node [style=redflavour] (154) at (9.5, -2.25) {};
		\node [style=redflavour] (155) at (7, -2.25) {};
		\node [style=none] (156) at (5.75, -3.75) {2};
		\node [style=none] (157) at (7, -3.75) {4};
		\node [style=none] (158) at (8.25, -3.75) {2};
		\node [style=none] (159) at (9.5, -3.75) {2};
		\node [style=none] (160) at (8, -2) {};
		\node [style=none] (161) at (5.25, 0) {};
		\node [style=none] (162) at (6.5, -0.25) {$A_1$};
		\node [style=gauge3] (163) at (12, 0.5) {};
		\node [style=gauge3] (164) at (13.25, 0.5) {};
		\node [style=gauge3] (165) at (14.5, 0.5) {};
		\node [style=flavour2] (166) at (14.5, 1.5) {};
		\node [style=flavour2] (167) at (12, 1.5) {};
		\node [style=none] (168) at (14.5, 2) {4};
		\node [style=none] (169) at (12, 2) {4};
		\node [style=redgauge] (170) at (13.25, 0.5) {};
		\node [style=bluegauge] (171) at (14.5, 0.5) {};
		\node [style=bluegauge] (172) at (12, 0.5) {};
		\node [style=redflavour] (173) at (14.5, 1.5) {};
		\node [style=redflavour] (174) at (12, 1.5) {};
		\node [style=none] (175) at (12, 0) {2};
		\node [style=none] (176) at (13.25, 0) {4};
		\node [style=none] (177) at (14.5, 0) {2};
		\node [style=none] (178) at (13.25, 1.5) {};
		\node [style=none] (179) at (13.25, 2.5) {};
		\node [style=none] (180) at (13.75, 2) {$A_1$};
		\node [style=none] (181) at (5, -1.5) {};
		\node [style=none] (182) at (11, 0.75) {};
		\node [style=none] (183) at (9, 0.5) {$A_1$};
		\node [style=none] (184) at (8, -5.25) {};
		\node [style=none] (185) at (8, -4.25) {};
		\node [style=none] (186) at (8.25, -4.75) {$A_1$};
		\node [style=none] (187) at (10.5, 15) {$T^{[3^4,1^4]}_{[7,5,3,1]}(SO(16))$};
		\node [style=none] (188) at (10.5, 10.75) {$T^{[3^4,1^4]}_{[6^2,3,1]}(SO(16))$};
		\node [style=none] (189) at (10.5, 7) {$T^{[3^4,1^4]}_{[7,5,1^4]}(SO(16))$};
		\node [style=none] (190) at (16.5, 3.75) {$T^{[3^4,1^4]}_{[5^2,3^2]}(SO(16))$};
		\node [style=none] (191) at (16.5, 0.5) {$T^{[3^4,1^4]}_{[5,4^2,3]}(SO(16))$};
		\node [style=none] (192) at (11.25, -2.75) {$T^{[3^4,1^4]}_{[5^2,2^2,1^2]}(SO(16))$};
		\node [style=none] (193) at (11.5, -6.75) {$T^{[3^4,1^4]}_{[5,3^3,1^2]}(SO(16))$};
		\node [style=none] (194) at (-1.5, -3.25) {$T^{[3^4,1^4]}_{[5,4^2,1^3]}(SO(16))$};
		\node [style=none] (195) at (-1.5, 0.25) {$T^{[3^4,1^4]}_{[5^2,3,1^3]}(SO(16))$};
		\node [style=none] (196) at (-1.5, 4.5) {$T^{[3^4,1^4]}_{[7,3^2,1^3]}(SO(16))$};
	\end{pgfonlayer}
	\begin{pgfonlayer}{edgelayer}
		\draw (0) to (1);
		\draw (1) to (3);
		\draw (3) to (4);
		\draw (1) to (5);
		\draw (33) to (34);
		\draw (34) to (36);
		\draw (36) to (37);
		\draw (34) to (38);
		\draw (51) to (52);
		\draw (52) to (54);
		\draw (57) to (58);
		\draw (63) to (65);
		\draw (65) to (66);
		\draw (63) to (67);
		\draw [style=arrowed] (78.center) to (79.center);
		\draw [style=arrowed] (84.center) to (85.center);
		\draw (90) to (91);
		\draw (91) to (92);
		\draw (99) to (100);
		\draw [style=arrowed] (105.center) to (106.center);
		\draw [style=arrowed] (108.center) to (107.center);
		\draw [style=arrowed] (109.center) to (110.center);
		\draw (114) to (115);
		\draw (115) to (117);
		\draw (117) to (118);
		\draw (115) to (119);
		\draw [style=arrowed] (132.center) to (133.center);
		\draw [style=arrowed] (136.center) to (137.center);
		\draw [style=arrowed] (139.center) to (140.center);
		\draw (142) to (143);
		\draw (143) to (145);
		\draw (145) to (146);
		\draw (143) to (147);
		\draw [style=arrowed] (160.center) to (161.center);
		\draw (163) to (165);
		\draw (165) to (166);
		\draw (163) to (167);
		\draw [style=arrowed] (178.center) to (179.center);
		\draw [style=arrowed] (181.center) to (182.center);
		\draw [style=arrowed] (184.center) to (185.center);
	\end{pgfonlayer}
\end{tikzpicture}
}
\end{gathered}$
\caption{The Coulomb branch Hasse diagram of the 3d $\mathcal{N}=4$ SCFT $T^{[3^4,1^4]}_{[5,3^3,1^2]}(SO(16))$ obtained from the Orthosymplectic Decay and Fission algorithm.}
\label{fig:app}
\end{figure}

As an example, we consider the orthosymplectic quiver $T^{[3^4,1^4]}_{[5,3^3,1^2]}(SO(16))$. Applying the Orthosymplectic Decay and Fission algorithm yields the Coulomb branch Hasse diagram that is depicted in Figure \ref{fig:app}. We note that this reproduces the dominance ordering of the partitions associated to the nilpotent orbits $\rho$, as desired. In this figure, the transverse slices can be obtained either through brane dynamics similar to the analysis in \cite{Cabrera:2017njm} or through subtracting the quivers before/after decay and rebalancing the gauge nodes along the same lines as in \cite{Bourget:2023dkj}. 

\bibliography{references}{}
\bibliographystyle{utphys}

\end{document}